\documentclass[twocolumn]{ms}

\usepackage{amsmath}

\begin{document}

\title{From Equipartition to Curvature: The Spectral Evolution of 4FGL Blazars}




\author[0000-0001-6820-717X,gname=Muhammad Shahzad,sname=Anjum]{Muhammad S. Anjum}
\affiliation{College of Physics and Electronic Engineering, Nanyang Normal University, Nanyang, Henan 473061, China}
\email[show]{mshahzadanjum@yahoo.com}

\author[0009-0006-3581-7207,gname=Shujin,sname=Hou]{Shu-Jin Hou}
\affiliation{College of Physics and Electronic Engineering, Nanyang Normal University, Nanyang, Henan 473061, China}
\email[show]{houshujingrb@163.com}

\author[0000-0002-1908-0536,gname=Liang,sname=Chen]{Liang Chen}
\affil{Shanghai Astronomical Observatory, Chinese Academy of Sciences, 80 Nandan Road, Shanghai 200030, China}
\email[]{chenliang@shao.ac.cn}

\author[0000-0002-1324-0893,gname=Zhigang,sname=Li]{Zhigang Li}
\affiliation{College of Physics and Electronic Engineering, Nanyang Normal University, Nanyang, Henan 473061, China}
\email[]{zhigang@nynu.edu.cn}

\author[0000-0002-4455-6946,gname=Minfeng,sname=Gu]{Minfeng Gu}
\affil{Shanghai Astronomical Observatory, Chinese Academy of Sciences, 80 Nandan Road, Shanghai 200030, China}
\email[]{gumf@shao.ac.cn}

\submitjournal{ApJS}


\begin{abstract}
We investigate the evolution of spectral energy distribution (SED) and underlying electron energy distribution (EED) by modeling the nearly simultaneous broadband spectra of selected bright 4FGL blazars, in the context of a combined cooling and stochastic acceleration scenario. We find that one-zone leptonic model with log-parabolic (LP) EED can successfully fit the GeV-TeV emission of blazars. The synchrotron frequency $\nu_s$ of blazars mainly evolves due to variation of electron peak energy $\gamma_{3p}$. The BL Lac objects (BL Lacs) show a negative trend in the $\nu_s- \nu_s L_s$ SED plane, known as blazar sequence, that does not seem to be an artifact of Doppler boosting, but driven by the equipartition constraints. A positive correlation is found between the derived magnetic field $B$ and electron density $n_e$, whereas $n_e$ and $\gamma_{3p}$ negatively relate, as expected in an equipartition scenario. The flat spectrum radio quasars (FSRQs) deviate significantly from such a scenario, indicating their jet parameters should be varying independently. The synchrotron peak frequency $\nu_s$ and its spectral curvature $b_s$ negatively correlate for all blazars, confirming the stochastic particle acceleration in blazar jets. However, blazars do not show the signature of hard-sphere acceleration, indicating that magnetic turbulence in the jets might be soft and physical conditions might be near to steady state, consistent with equipartition. Furthermore, for BL Lacs, the SED curvature $b_s$ and the EED curvature $r$ and nearly meet the theoretical relationship $r=5b_s$, whereas the FSRQs show large deviation due to poor constrain on $b_s$ due to presence of thermal component.
\end{abstract}





\section{\textbf{Introduction}} \label{sec:intro}
Active galaxies host bright nuclei, called active galactic nuclei (AGN), which shine because of accreting supermassive black holes in the center. Blazars are jetted AGN and their emission is dominated by the nonthermal processes in the relativistic plasma jets moving close to our line of sight. Blazars are marked by broadband emission, extreme variability at high energy, and superluminal motion \citep{1978bllo.conf..328B}. The orientation of the AGN plane and jet angle with our line of sight play an important role and describe the observed spectral features of several AGN types \citep{1995PASP..107..803U}. Based on the equivalent width of emission lines from fast-moving clouds in the broad line region (BLR), the blazars can be categorized into flat-spectrum quasars (FSRQs) and BL Lac objects (BL Lacs). However, both classes have similar observed continuum spectra despite their different emission line properties \citep{2011MNRAS.414.2674G}, suggesting similar underlying processes.\\ 

The bulk emission of the jet is produced by nonthermal plasma accelerated in the jet via leptonic or hadronic processes \citep[see, e.g.,][ for a review on physical processes in the jet]{2020Galax...8...72C}. The low-energy hump usually spans radio to X-ray energies and is always associated with leptonic synchrotron emission, whereas the high-energy bump spans from X-rays to very high energy (VHE) $\gamma$-rays and, in the leptonic scenario, is attributed to inverse Compton (IC) scattering of low-energy photons by the same electron population \citep{1981ApJ...243..700K}. In the hadronic scenario, the high energy emission can also arise from a proton synchrotron process \citep{1993A&A...269...67M, 2003APh....18..593M}. The variability of blazars is attributed to the time dependence of the physical processes in the jet \citep{1997A&A...320...19M}. In leptonic models, the high-energy IC emission from the jet can be explained either by the synchrotron self-Compton (SSC) \citep{1992ApJ...397L...5M, 1996ApJ...461..657B} or by the external Compton (EC) mechanism \citep{1993ApJ...416..458D, 1994ApJ...421..153S}. Depending on the location of energy dissipation of the jet from the central black hole, various seed photons can contribute to the observed high energy emission, such as host dust illuminated by the central continuum \citep{2000ApJ...545..107B}. The one-zone leptonic models have been successful in reproducing the observed broadband spectral energy distribution (SED) of blazars \citep[see, e.g.,][]{2009MNRAS.397..985G, 2010MNRAS.402..497G}. Thermal emission can be present at several energies, especially in FSRQs, where the optical-UV emission is usually dominated by thermal emission from a hot thin accretion disk. The synchrotron peak frequency and peak luminosity of the blazar SED are important parameters and are indicators of the physical jet conditions of a source.\\ 

The broadband SED of blazars is dominated by nonthermal emission shows two characteristic bumps in the $\nu - \nu F_\nu$ logspace \citep{1998MNRAS.299..433F}. The peak frequency of low energy synchrotron bump of blazars varies in a wide range of values. The peak frequency of FSRQs falls in the IR band, whereas the peak frequency BL Lacs may fall from IR to X-ray energies \citep{2010ApJ...716...30A}. Based on synchrotron peak frequency, blazars can be classified into three classes; low synchrotron peak (LSP), intermediate synchrotron peak (ISP), and high synchrotron peak (HSP) \citep{2006A&A...445..441N, 2016ApJS..226...20F, 2022ApJS..262...18Y}. The BL Lacs can vary from low- to high-peak sources, whereas almost all FSRQS are practically low-peak sources. Blazars generally show an inverse correlation between synchrotron peak frequency and luminosity or radiation power, a trend known as blazar sequence \citep{1998MNRAS.299..433F}. This observed sequence was explained in terms of radiative cooling scenario in which jet power derives the anti-correlation \citep{1998MNRAS.301..451G, 2008MNRAS.387.1669G}, but criticized as a selection effect \citep{2012MNRAS.420.2899G} or an observational feature due to Doppler boosting \citep{2008A&A...488..867N, 2017ApJ...835L..38F}.\\ 

Blazar SEDs are curved and can be fitted by a higher-order polynomial or log-parabolic (LP) function. \cite{2004A&A...413..489M} found that even in a small energy band, the spectra are significantly curved, suggesting that the source electron energy distribution (EED) must be mildly curved, rather than a power law. Although astrophysical acceleration mechanisms usually produce a power-law EED, the energy dependence of acceleration probability \citep{2004A&A...413..489M} or a random fluctuations of fractional gain in the underlying acceleration mechanisms can produce a LP EED \citep{2011ApJ...739...66T}. Furthermore, in such an acceleration mechanism, the synchrotron peak frequency $\nu_s$ and the synchrotron spectral curvature $b_s$ anti-correlate. Previously it has been found that the peak frequency and curvature parameter are indeed anti-correlated \citep[see, for example][]{2007A&A...466..521T, 2014ApJ...788..179C}. However, the slope is different for various blazar types \citep{2016ApJS..226...20F}. The imprint of this correlation on the emitting EED, i.e., a correlation between EED peak energy and its curvature has not been widely studied and requires the modeling of simulataneous broadband SEDs of a large blazar sample.\\

The spectral evolution of blazar SED is an important but complex topic since the emission and acceleration models have a large number of parameters. It is not clear which parameters drive the spectral behavior of blazars in the rest-frame $\nu L_\nu$ SED. Even the SED of a single source is highly variable over time, and the true origin of the variability is not fully understood. The spectral changes in the $\nu L_{\nu}$ SED can arise either due to EED parameters; such as electron energy and density, or ambient conditions such as the energy density of magnetic field and photon fields, or beaming parameters, such as the Doppler factor, or a combination of all. To investigate the genuine origin of the spectral shifts or variability in the blazar SED, it is important to model the SEDs of a large sample of blazars of the same SED class. A statistical trend in the $\nu_s - \nu_sL_s$ plane, with $\nu_s$ being the synchrotron peak frequency and $\nu_sL_s$ the peak luminosity, produces a systematic evolution of source parameters and might explain how the parameters are linked. Previously, it has been found that HSP BL Lacs show a positive relationship between the synchrotron frequency and luminosity and a negative one between synchrotron frequency and its spectral curvature, which can be explained in a combined scenario of synchrotron cooling and statistical acceleration \citep{2007A&A...466..521T, 2008A&A...478..395M, 2009A&A...501..879T}. However, the observations were limited to X-rays. In this work, we explore the multiwavelenght behavior of blazars.\\

We selected a sample of bright Fermi-detected blazars to study the spectral evolution of the blazars in the $\nu L_\nu$ SED and its imprint on the emitting EED. In the first step, we performed LP parametric fitting of the SEDs to find the synchrotron frequency, flux, luminosity, and spectral curvature to study the behavior of observed SEDs. In the next step, we employ a physical one-zone synchrotron+IC model, in the context of a combined cooling and stochastic scenario, to model the blazar SEDs to investigate how changes in SED and EED are related. We, then, study the relationship between the SED observables and the source parameters of one-zone model to investigate the evolution of blazars in $\nu-\nu L_\nu$ plane and how the spectral curvature evolves against synchrotron frequency. We describe the blazar sample in Section \ref{sec:data}, the blazar model in Section \ref{sec:model}, and our findings and results in Section \ref{sec:results}. Section \ref{sec:summary} provides a discussion and summarizes our results. Throughout this work, we assume the standard cosmological model with $\Omega_{m}=0.32$, $\Omega_{\Lambda}=0.68$, and the $H_{0}=70$ km s$^{-1}$ Mpc$^{-1}$.

\section{\textbf{Sample \& Data}} \label{sec:data}
Spectral fitting and physical modeling of snapshot SEDs of a blazar sample can provide important insights into the emission mechanisms, the origin of their spectral variability, and the evolution of cooling/acceleration processes in the blazar jets. We choose 100 $\gamma$-ray blazars from the Fermi-LAT bright AGN sample (LBAS) with sufficient SED data, corrsponding to the first four months of Fermi-LAT operations, from August 2008 to December 2008 \citep{2009ApJ...700..597A}. The sample includes about 40 BL Lacs and 60 FSRQs. The broadband SEDs of selected LBAS blazars were compiled by \citet{2010ApJ...716...30A}, however, the energy range of the first LAT data release ($0.1-300$ GeV) was relatively small to constrain the high energy peak and shape of several blazars, especially the HSP sources. Recently, the third data release of the fourth Fermi-LAT catalog (4FGL-DR3) released the $\gamma$-ray data of these blazars in a broad energy range ($0.1$ GeV to $50$ TeV) \citep{2022ApJS..263...24A}. Due to a significantly large energy range, the high-energy peak of blazars can be fully constrained, leading to a more reliable estimate of physical jet parameters. We gathered the spectra data of these 100 4FGL blazars from radio to TeV $\gamma$-ray, from the Space Science Data Center (SSDC) of Italian Space Agency (ASI)\footnote{https://www.ssdc.asi.it} \citep{2011arXiv1103.0749S} to build their broadband SEDs. We collected the Planck radio data \citep{2011A&A...536A..15P}, WISE IR data, optical-UV data from SWIFT-UVOT, $0.1-10$ keV X-ray data from the SWIFT-XRT point source catalog \citep[1SWXRT,][]{2013A&A...551A.142D}, and 4FGL-DR3 $\gamma$-ray data \citep{2020ApJS..247...33A}. From IR to X-rays, we ensure that the spectra data is nearly simultaneous, taken within a few weeks to a few months. Occasionally, when available, we use recently released multiband X-ray data from the eROSITA all-sky survey \citep{2024A&A...682A..34M} or historical data from BeppoSAX \citep{2002babs.conf...63G} to provide better constraints to the model parameters. To account for variability and guide the fitting algorithm, we induce a $\simeq 10\%$ systematic error. Incorporating these data, the LP parametric fitting and physical model fitting can be fully constrained, providing an opportunity to compare models with the observed data, to investigate the spectral changes of blazars in the $\nu_s - \nu_s L_{\nu_s}$ plane.

\section{\textbf{The One-Zone Blazar Model}} \label{sec:model}
Firstly, we perform the LP polynomial fitting of Blazar SEDs to estimate the SED observables, such as peak frequency, peak flux and luminosity, and the curvature of the synchrotron/IC components. The high-energy spectra of blazars is generally curved and deviates significantly from a power-law \citep{2004A&A...413..489M}. Recently, Fermi-LAT found several thousands of blazars with log-parabolic spectra \citep{2022ApJS..263...24A}. In a small energy band, the energy spectrum of a blazar shows LP form, given as
\begin{equation}
    F(E)= K \left(\frac{E}{E_0}\right)^{-(a+b \log (E/E_0))},
\end{equation}
where $E_0$ is the reference energy. The peak energy $E_p$ of LP spectrum in terms spectral index $a$ and curvature parameter $b$ becomes
\begin{equation}
    E_p=E_0  10^{\left( \frac{2-a}{2b} \right)}.
\end{equation}
The broadband SED roughly matches the LP function and is a reflection of an underlying LP EED. Such a curved EED is naturally produced in an energy-dependent scenario in which the acceleration efficiency decays with particle energy \citep{2004A&A...413..489M} or a stochastic acceleration scenario in which the fractional energy gain is purely random. The LP function that describes the broadband blazar SED is given as
\begin{equation}
    \log (\nu F_\nu) = \log ( \nu_p F_{\nu_p} ) - b ( \log \nu - \log \nu_p)^2,
\end{equation}
where $\nu_p$ is the peak frequency and $b$ is the curvature of the fitted bump. We fit this polynomial to all blazar SEDs to determine the peak frequency, flux, and curvature of the synchrotron and IC bumps. The peak luminosity is then found to be $L =4\pi d_L^2 (\nu_p F_{\nu_p})$.\\

Secondly, at the pre-fitting, we estimate the approximate one-zone model parameters from the SED observables. The spectral parameters of synchrotron/IC bump reflect the physical conditions inside the jet and help estimate the physical parameters of a one-zone model, such as electron peak energy $\gamma_p$, ambient magnetic field $B$, emission source size $R$, and the Doppler factor $\delta$ \citep[see, e.g.,][]{1996A&AS..120C.503G, 1998ApJ...509..608T, 2018ApJS..235...39C}. For example, the synchrotron peak frequency $\nu_s$ relates $B$ and $\delta$ as
\begin{equation}
    \nu_s = C_0 \gamma_p^2 B' \delta,
\end{equation}
whereas the SSC peak frequency due to these synchrotron photons becomes
\begin{equation}
    \nu_c = \frac{4}{3} \gamma_p'^2 \nu_s.
\end{equation}
The electron peak energy $\gamma_p$ can, thus, be approximated from observed frequencies as
\begin{equation}
    \gamma_p' \simeq \sqrt{\frac{\nu_c}{\nu_s}}.
\end{equation}
The magnetic energy density $U'_B=B'^2/8\pi$ and the synchrotron energy density in the jet frame $U_s'\simeq L_s/4\pi R^2 c\delta^4$ can be constrained from the SSC dominance parameter (the ratio of SCC losses to synchrotron loss rates) as 
\begin{equation}
    \frac{L_c}{L_s}= \frac{U'_s}{U'_B} \simeq \frac{2L_s}{cB'^2R^2\delta^4},
\end{equation}
relating the magnetic field $B'$ and $\delta$. This equation, together with the expressions for $\nu_s$, leads to the estimation of $B$ and $\delta$ as
\begin{equation}
    B' \simeq \frac{R \nu_s^4}{C_0^2 \nu_c^2 L_s } \sqrt{\frac{c L_c}{2}}
\end{equation}
and 
\begin{equation}
    \delta \simeq \frac{C_0 \nu_c L_s}{\nu_s^2 R} \sqrt{\frac{2}{c L_c}}.
\end{equation}
For the EC component, various external photon fields might contribute to IC emission, depending on the location of energy dissipation of the jet, such as from the broadband line region (BLR), a dusty torus (DT), or even cosmic microwave background (CMB). Assuming the DT source photons, the EC frequency can be approximated as
\begin{equation}
    \nu_c = \frac{4}{3} \gamma_p'^2 \Gamma \nu_T \delta, 
\end{equation}
where $\nu_T$ is the frequency of DT seed photons in the AGN frame. The external seed photon energy density in the jet frame becomes
\begin{equation}
    U'_{ext} \simeq \frac{17}{12} \Gamma^2 U_T,
\end{equation}
where $U_T$ is the source photon energy density of the DT field illuminated by the thermal accretion disk, given by
\begin{equation}
    U_T= \frac{\eta L_D}{4\pi R_T^2 c}.
\end{equation}
Here $L_D$ is the central disk luminosity, $R_T$ is the DT radius, and $\eta$ is the fraction of disk luminosity reprocessed by DT. We assume $\eta=0.1$, i.e., a $10\%$ of disk luminosity being reprocessed into the jet. Then the EC dominance parameter becomes
\begin{equation}
    \frac{L_c}{L_s} = \frac{U'_{ext}}{U'_B} \simeq \frac{2 \Gamma^2 L_D}{B^2 R_T^2c}.
\end{equation}

Approximating the jet parameters at the pre-fitting stage, we then use a one-zone model with synchrotron+SSC+EC components to fit the sample SEDs based on the $\chi^2$ minimization algorithm \textit{JetSet} \citep{2020ascl.soft09001T}. The algorithm optimizes the model fit on the SEDs by varying the parameter values around the approximate values, till the fit is converged.\\ 

We assume a one-zone leptonic model with an expanding homogeneous blazar blob of radius $R$, filled with uniform electron distribution $N(\gamma)$ and chaotic magnetic field $B$, moving relativistically with speed $\Gamma$, making an angle $\theta$ with our line of sight. The size of expanding blob $R$ depends on the jet opening angle $\phi$ and distance of the blob from the central black hole $R_b$ as
\begin{equation}
    R= R_b \tan\phi.
\end{equation}
We assume a line-of-sight angle $\theta \simeq 3$ and an opening angle $\phi \simeq 5$ which are typical of blazars, and the distance of the $R_b$ as a free parameter. The size of the torus $R_{T}$ scales with source disk luminosity $L_D$ as
\begin{equation}
    R_T =2\times10^{18} L_{D,45}^{1/2}
\end{equation}
where the disk luminosity $L_{D,45}$ is constrained by fitting a blackbody to the BBB, observed at optical-UV SED. The emitting electron distribution is considered to be LP in the energy ranging from $\gamma_{min}$ to $\gamma_{max}$, given as
\begin{equation}
    N(\gamma)= N_0 \left( \frac{\gamma}{\gamma_0}\right)^{-s-r \log (\gamma/\gamma_0)}
\end{equation}
where $\gamma_0$ is the reference electron energy, and $s$ and $r$ are the spectral index and curvature at $\gamma_0$. The LP EED is essentially a broken power law distribution, broken at several energy points, and is naturally produced due to a balance between a stochastic acceleration and radiative cooling \citep{1962SvA.....6..317K, 2006A&A...448..861M, 2006A&A...453...47K}. The peak energy of LP EED in the $\gamma^3 N(\gamma)$ representation is given by
\begin{equation}
    \gamma_{3p} = \gamma_0 10^{\left(\frac{3-s}{2r}\right)}.
\end{equation}
Thus, the particle energy and curvature are inversely related in stochastic acceleration as
\begin{equation}
    \log \gamma_{3p} =\log \gamma_0 + \left(\frac{3-2s}{2r}\right).
\end{equation}

By one-zone leptonic model fitting of the blazar SEDs assuming LP EED , we derive the physical jet parameters, such as, the magnetic field $B$, blob radius $R$, blob location $R_b$, jet bulk factor $\Gamma$, Doppler factor $\delta=[\Gamma(1-\beta \cos \theta)]^{-1}$, electron density $n_e$, electron energy $\gamma_0$, and spectral curvature $r$. The corresponding components of jet power in the galaxy frame as
\begin{equation}
    L_i \simeq \pi R^2 c\Gamma^2 U_i, 
\end{equation}
where $U_i$ represents the ambient energy density of the magnetic field $U_B$, emitting electrons $U_e$, and cold protons $U_p$, given as
\begin{equation}
    U_B= \frac{B^2}{8\pi}
\end{equation}
\begin{equation}
    U_e=  m_e c^2 \int \gamma n_e(\gamma) d\gamma 
\end{equation}
\begin{equation}
    U_p= m_p c^2 \int n_p(\gamma) d\gamma.
\end{equation}
We use the number density of cold protons equal to electrons $n_p=n_e$. The total jet power of a source, then, becomes $L_{jet}=L_e+L_p+L_B$. The radiation power of the jet is estimated as
\begin{equation}
    L_{rad}\simeq \frac{L_{bol}}{\Gamma^2}
\end{equation}
where $L_{bol}$ is the bolometric luminosity. For IC emission using a full Klein-Nishina cross-section, we calculate the SSC emission due to internal synchrotron photons and EC emission due to external photons from a relatively colder DT with temperature $T_T \sim 10^2$ K, illuminated by a hot disk at a temperature $T_D \sim 10^4$ K with luminosity $L_D$. The disk luminosity $L_D$ is directly constrained by the big blue bump (BBB) in the observed SED, and for sources where the BBB is not observable, we use the optical-UV data as an upper limit. For three nearby HSP BL Lacs (4FGL J1517.7-2422, 4FGL J1653.8+3945, and 2009.4-4849) with $z<0.1$, we find the signature of background host galaxy in their synchrotron spectra and fit the host galaxy template to account for this emission.

\subsection{\textbf{Combined Cooling and Acceleration Scenario}} \label{sec:combined}
The synchrotron SED from a LP EED is nearly a log-parabola with peak frequency
\begin{equation}
    \log \nu_s \propto \log\gamma_0+\frac{3}{10b_s},
\end{equation}
where $b_s$ is the curvature of synchrotron photon spectra and related to electron curvature parameter $r$ as $r=5b_s$ \citep{2006A&A...448..861M}. Thus, modeling the blazar SEDs reveal how the intrinsic electron curvature $r$ evolves in acceleration and cooling conditions. The spectral curvature is an important parameter and reflects the internal conditions of the blazar jet.\\

We discuss the evolution of LP EED curvature $r$ against its peak energy $\gamma_{3p}$ in a combined cooling and stochastic acceleration scenario. In stochastic acceleration models, the injected particles get accelerated by interactions with magnetohydrodynamic waves produced by magnetic turbulence and the process is described as diffusive shock acceleration \citep{2011ApJ...739...66T}. In a combined scenario including all relevant processes, the time evolution of injected electrons is governed by the kinetic equation
\begin{equation}
    \begin{split}
        \frac{\partial N(\gamma,t)}{\partial t}&= \frac{\partial}{\partial t} \left[ \left( C(\gamma,t) - A(\gamma, t) \right) N(\gamma,t)\right] \\
        &+D(\gamma,t) \frac{\partial N(\gamma,t)}{\partial \gamma}-E(\gamma,t)+Q(\gamma,t)
    \end{split}
\end{equation}
where $C(\gamma,t)$ is the cooling parameter, given in terms of ambient energy density in the jet $U_{tot}(\gamma,t)$ as
\begin{equation}
    C(\gamma,t)=\frac{4}{3}\frac{\sigma_T c}{m_ec^2} U_{tot}(t),
\end{equation}
and $A(\gamma,t)$ is the acceleration term defined in terms of diffusion coefficient $D(\gamma,t)$ as
\begin{equation}
    A(\gamma,t)= \frac{2}{\gamma} D(\gamma,t).
\end{equation}
The terms $E(\gamma,t)=N(\gamma,t)/\tau_{esc}$ and $Q(\gamma,t)$ are the escape and injection terms. In a simple acceleration+cooling scenario, the equilibrium energy is determined by competition between acceleration and cooling timescales, i.e., $\tau_c(\gamma) \simeq \tau_a (\gamma)$. The cooling timescale is inversely related to particle energy and ambient energy density in the jet as
\begin{equation}
    \tau_c(\gamma) \simeq \frac{mc^2}{\sigma_T c \gamma U_{tot}},
\end{equation}
where $U_{tot}= U_B + U_s + U_{ext}$ is the total ambient energy density in the jet, available for radiative cooling processes. The acceleration timescale of electrons is described by the diffusion coefficient as
\begin{equation}
    \tau_a (\gamma)= \frac{\gamma^2}{D(\gamma)} \simeq \left(\frac{\gamma}{\gamma_0}\right)^{2-q}, 
\end{equation}
where $q$ is the index of magnetic turbulence spectra $W(k)$ in terms of wave number $k=2\pi/\lambda$, given as
\begin{equation}
    W(k)= \frac{(\delta B)^2}{8\pi} \left(\frac{k}{k_0}\right)^{2-q}.
\end{equation}
The turbulence index $q$ determines the evolution of EED in the combined scenario. For hard-sphere spectrum $q=2$, whereas $q<2$ for the Kolmogorov spectrum or the Kraichnan spectrum. Thus, for a hard-sphere acceleration, the acceleration time becomes independent of particle energy $\gamma$.\\

In diffusive shock acceleration, the fractional energy gain is a random variable and the evolution of EED shows that the curvature $r$ decays with particle energy as long as the acceleration times are shorter than the typical acceleration time \citep{2011ApJ...739...66T}. Thus, for an acceleration dominated regime ($t<t_a$), the curvature $r$ and peak energy of electrons $\gamma_{3p}$ must follow an inverse relation
\begin{equation}
    r \propto \frac{1}{\gamma_{3p}}.
\end{equation}
Such an inverse correlation between curvature and particle energy can also be explained if the acceleration probability is energy dependent, such that $p_i\propto 1/\gamma^a$, where $a$ is a constant showing energy dependence \citep{2004A&A...413..489M}. Although both the stochastic and energy-dependent acceleration mechanisms predict an inverse correlation between peak energy and curvature, the efficiency of both mechanisms could be different. We investigate the impression of stochastic acceleration and radiative cooling mechanisms on the observed blazar SED.

\section{\textbf{Modeling Results}} \label{sec:results}
\begin{figure*}[htb!]
    \centering
    \includegraphics[scale=0.9]{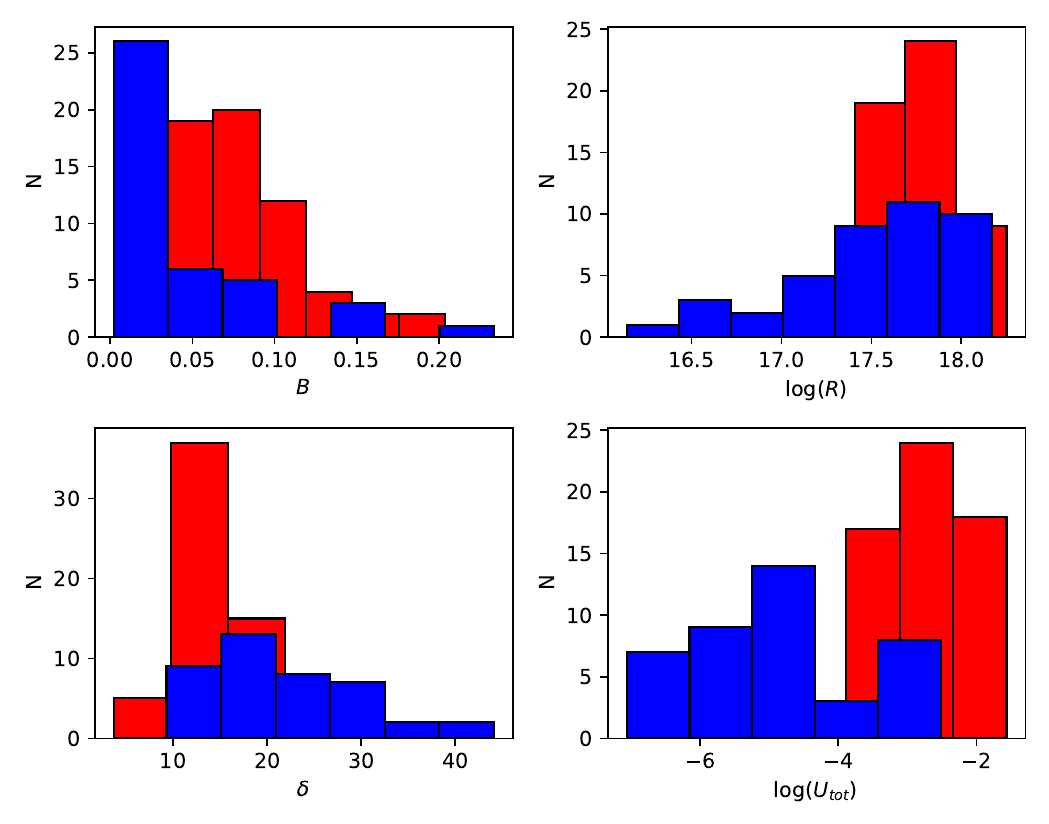}
    \caption{The distributions of physical jet parameters of BL Lacs (blue) and FSRQs (red).}
    \label{fig:HIST-JET}
\end{figure*}

\begin{figure*}[htb!]
    \centering
    \includegraphics[scale=0.9]{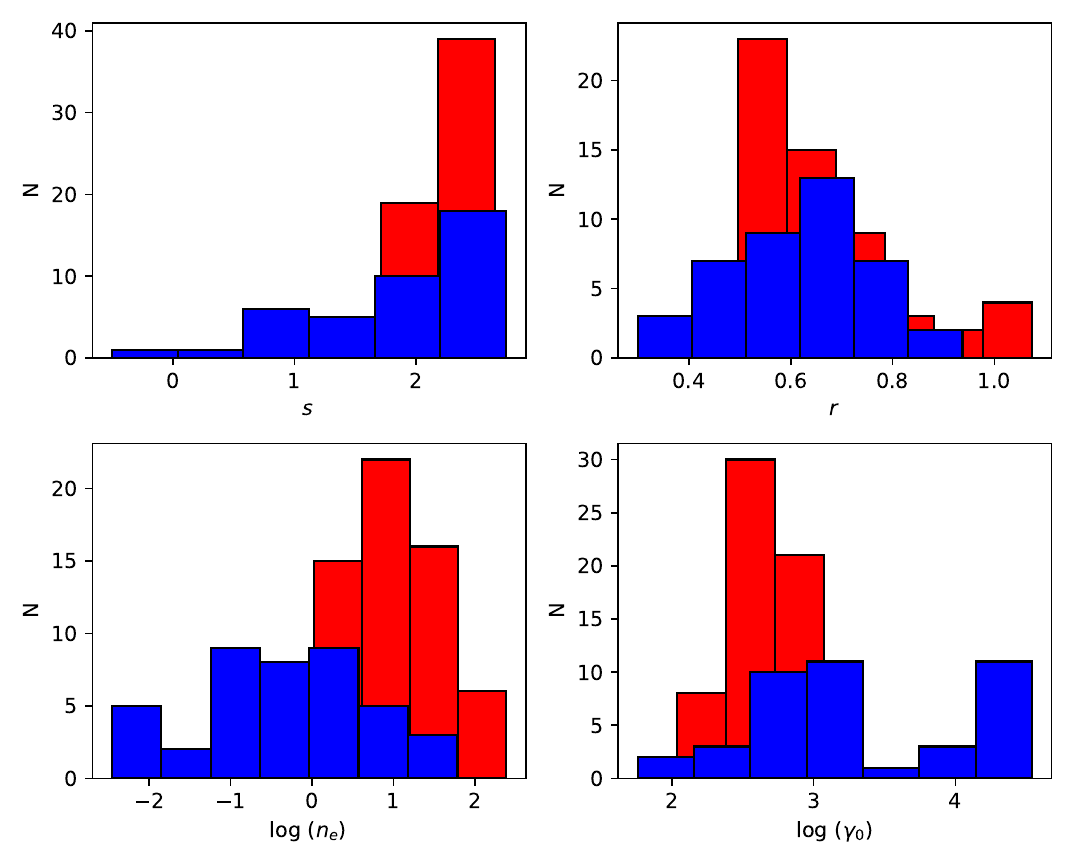}
    \caption{The distributions of source electron energy distribution (EED) parameters for BL Lacs (blue) and FSRQs (red).}
    \label{fig:HIST-EED}
\end{figure*}

\begin{figure*}[htb!]
    \centering
    \includegraphics[scale=0.9]{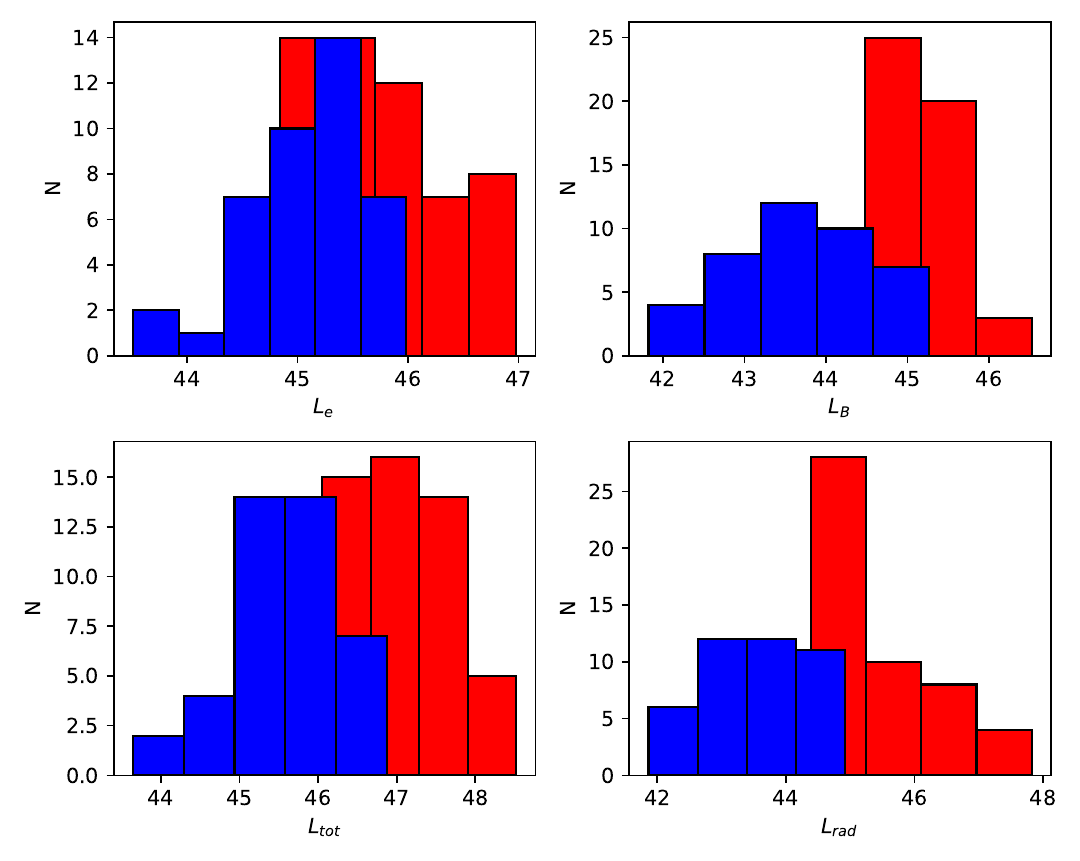}
    \caption{The distributions of jet power for BL Lacs (blue) and FSRQs (red).}
    \label{fig:HIST-JP}
\end{figure*}

\begin{figure}[htb!]
    \centering
    \includegraphics[width=\linewidth]{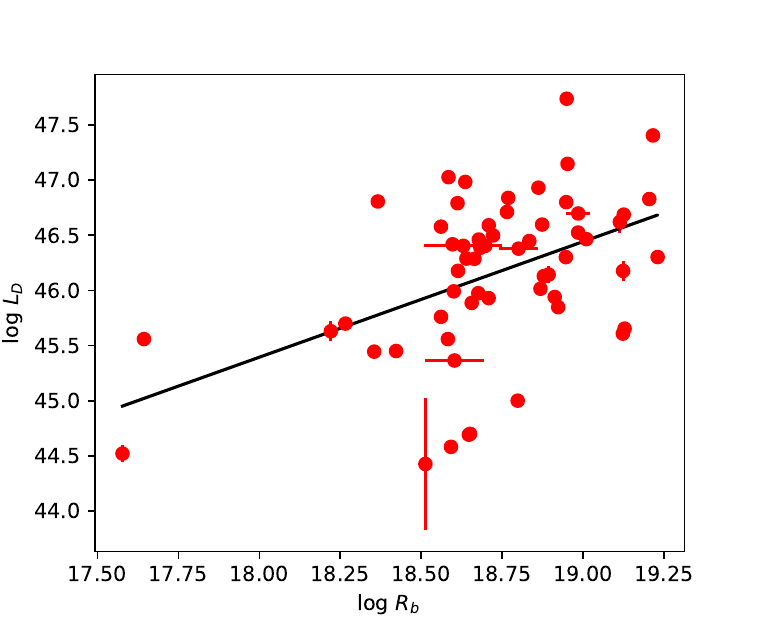}
    \caption{The relationship between dissipation location $R_b$ and the accretion disk luminosity $L_D$ for FSRQs. The solid line shows the best linear regression fit.}
    \label{fig:disk}
\end{figure}

\begin{figure}
    \centering
    \includegraphics[width=0.48\linewidth]{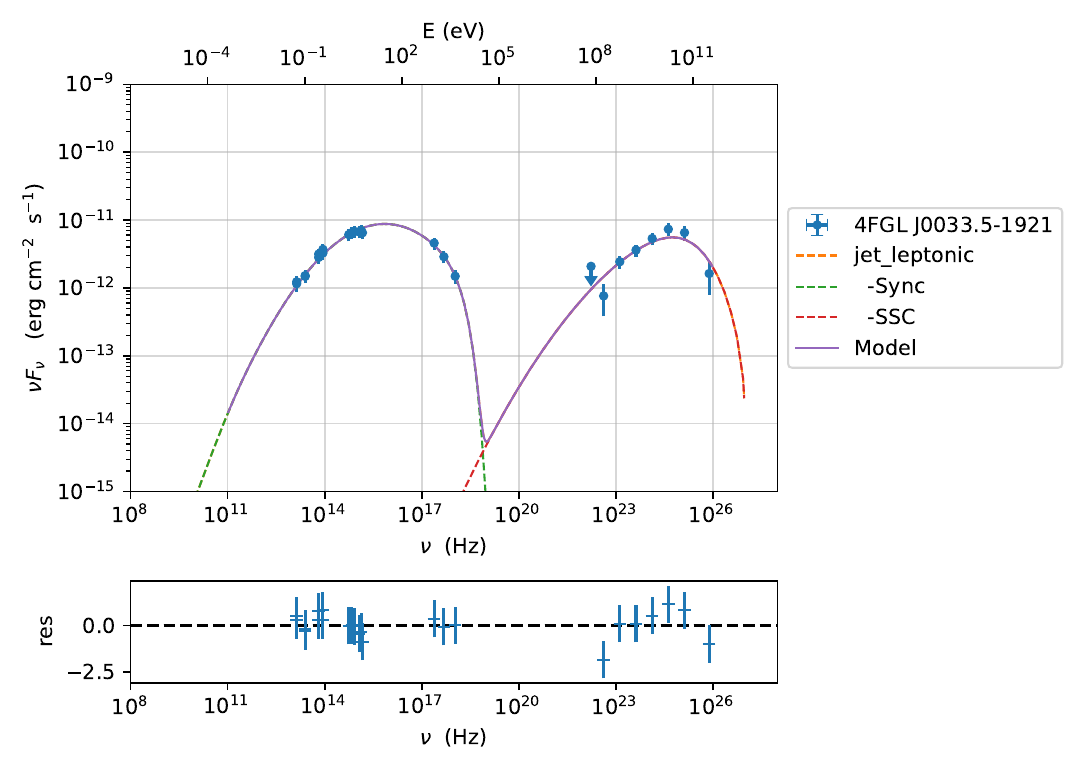}
    \includegraphics[width=0.48\linewidth]{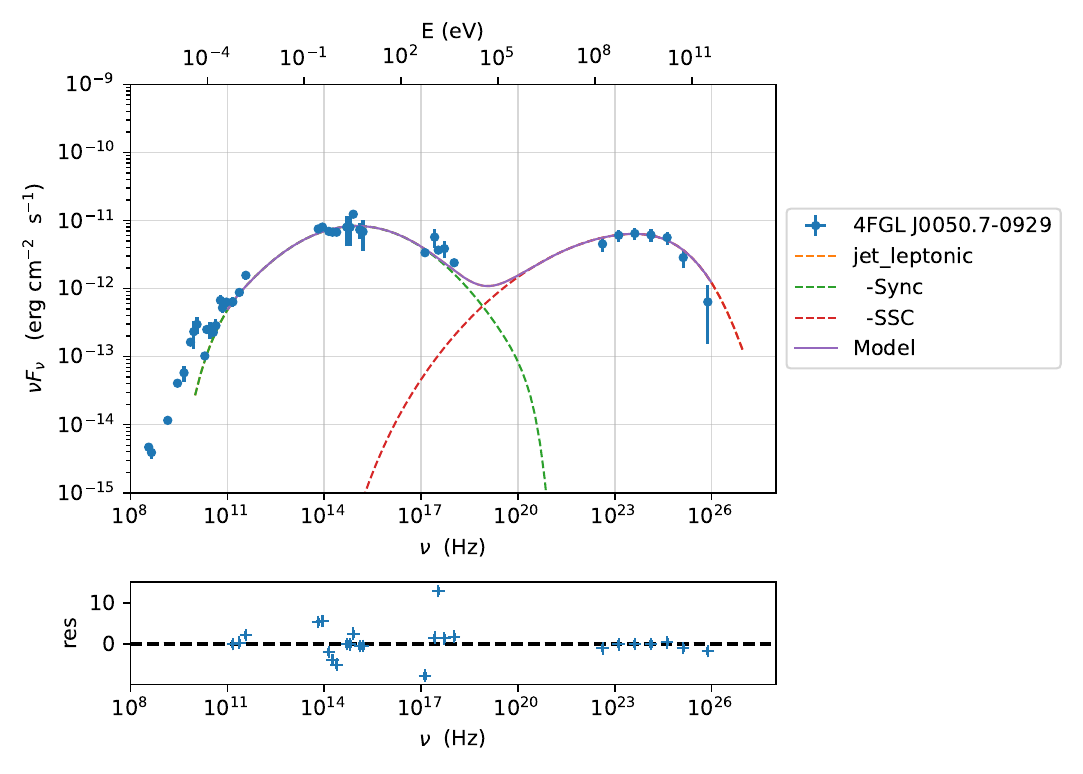}
    \includegraphics[width=0.48\linewidth]{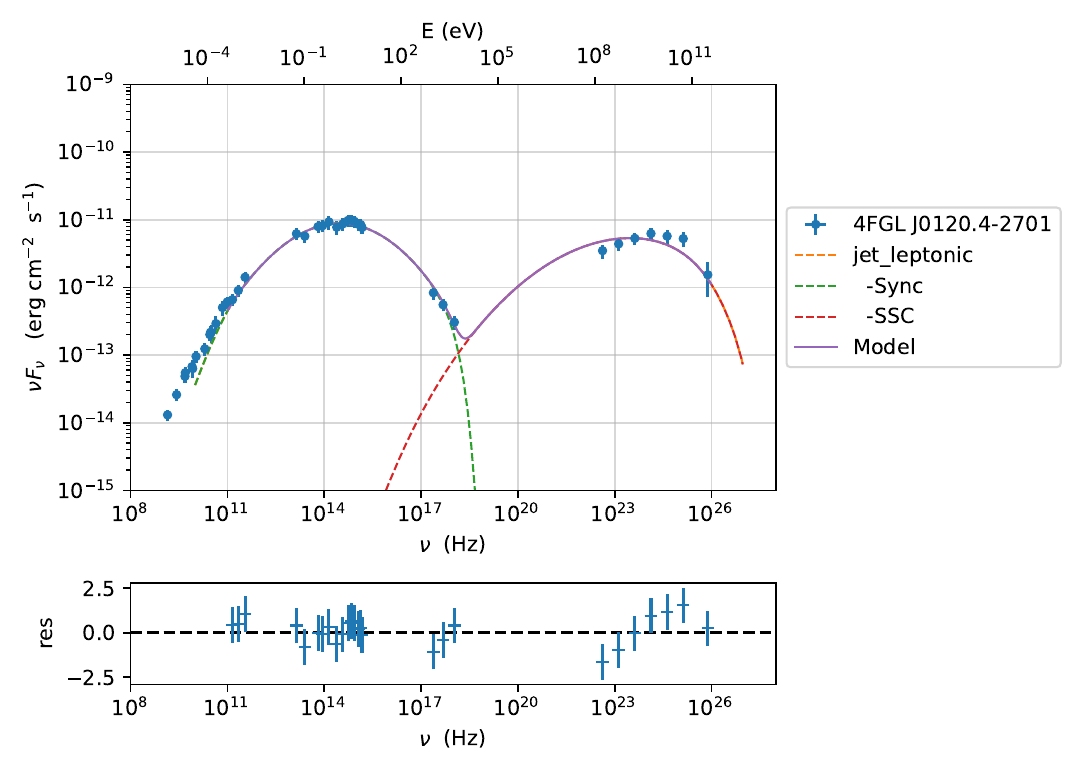}
    \includegraphics[width=0.48\linewidth]{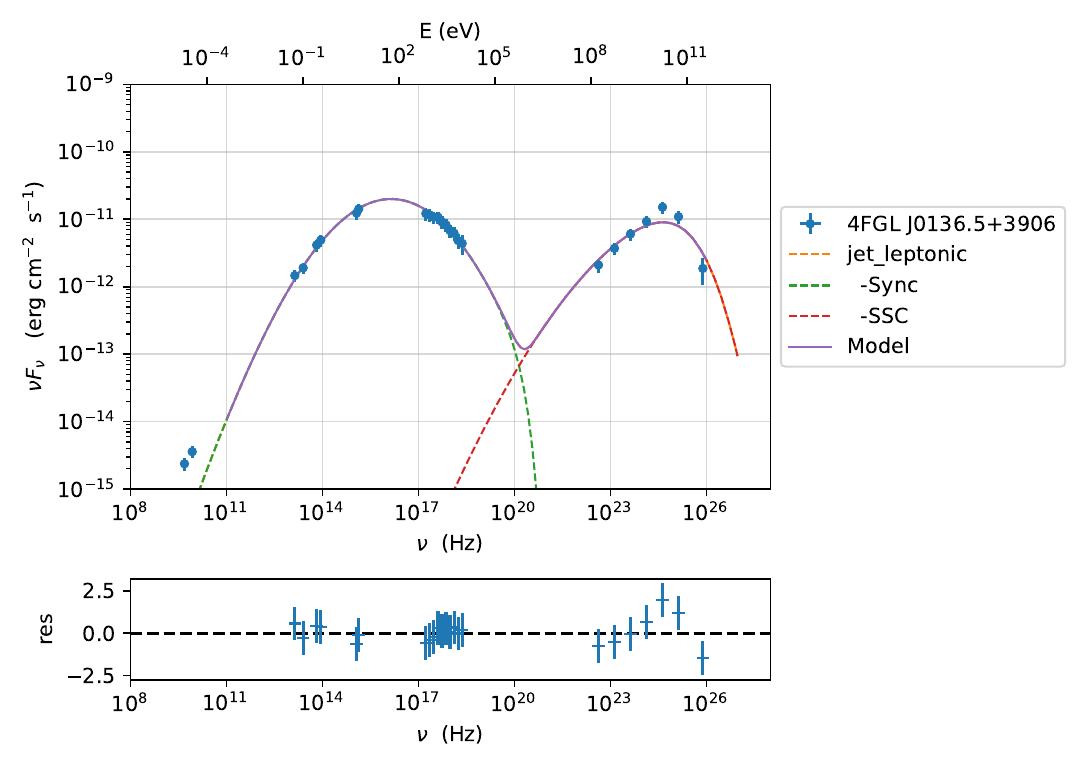}
    \includegraphics[width=0.48\linewidth]{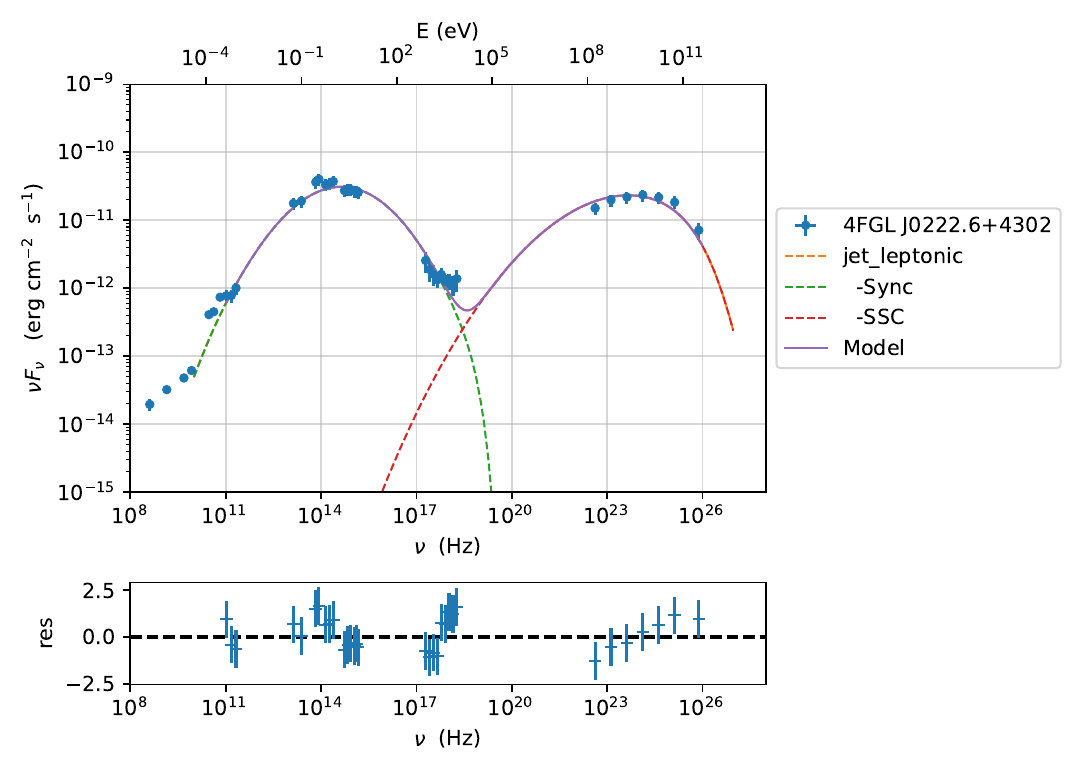}
    \includegraphics[width=0.48\linewidth]{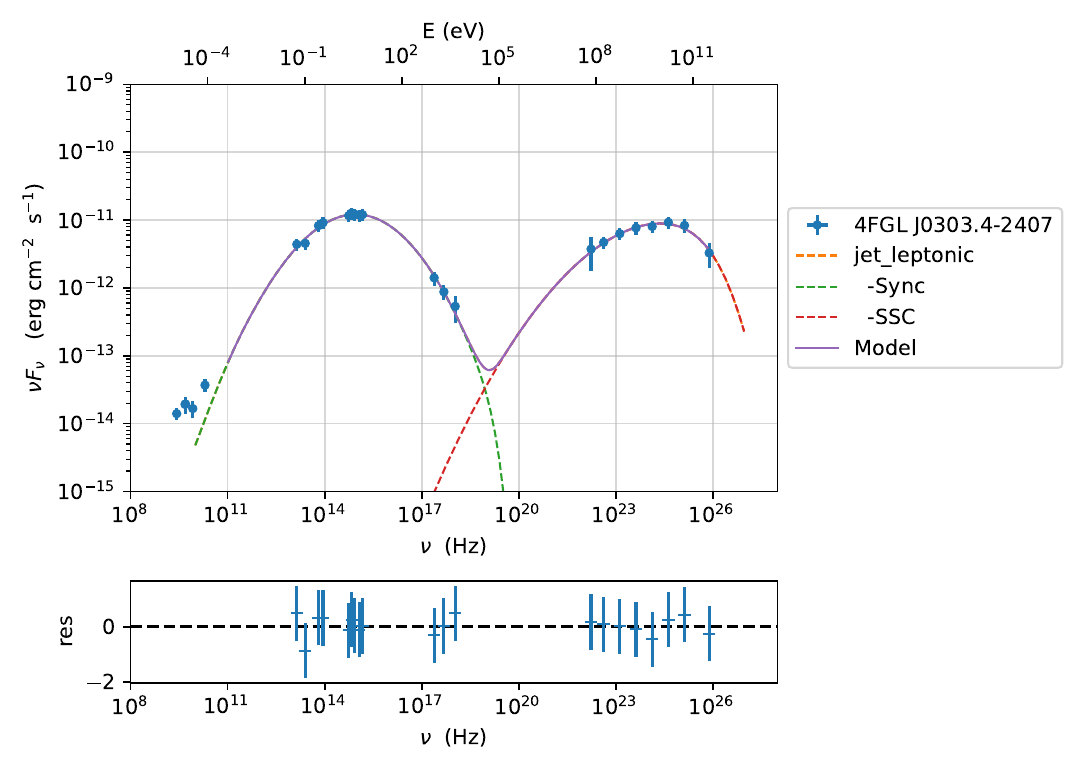}
    \includegraphics[width=0.48\linewidth]{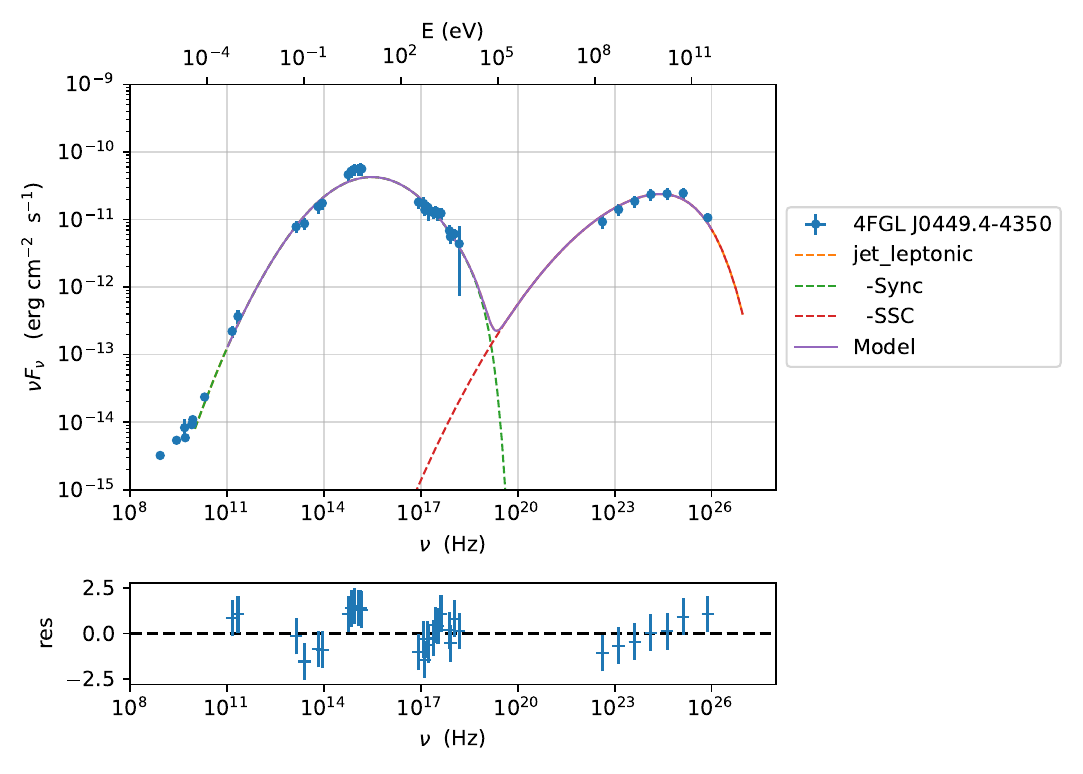}
    \includegraphics[width=0.48\linewidth]{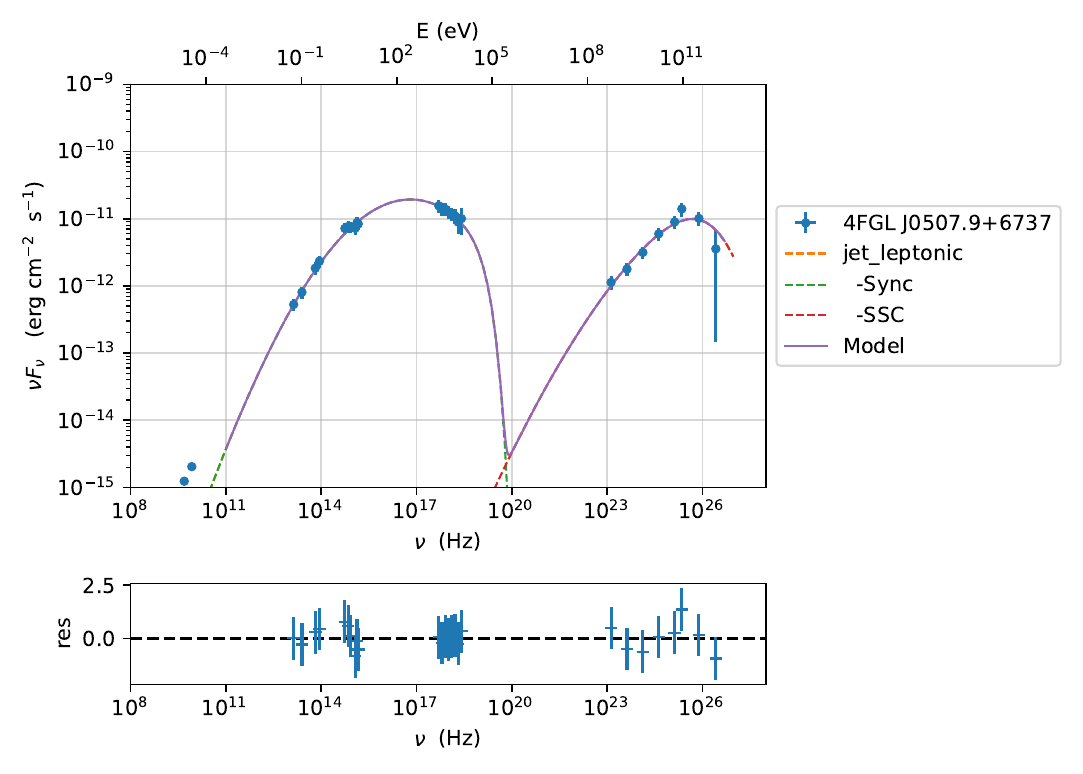}
    \caption{SED Modeling of SSC Blazars.}
    \label{fig:SED-SSC}
\end{figure}

\begin{figure}
\centering
\includegraphics[width=0.48\linewidth]{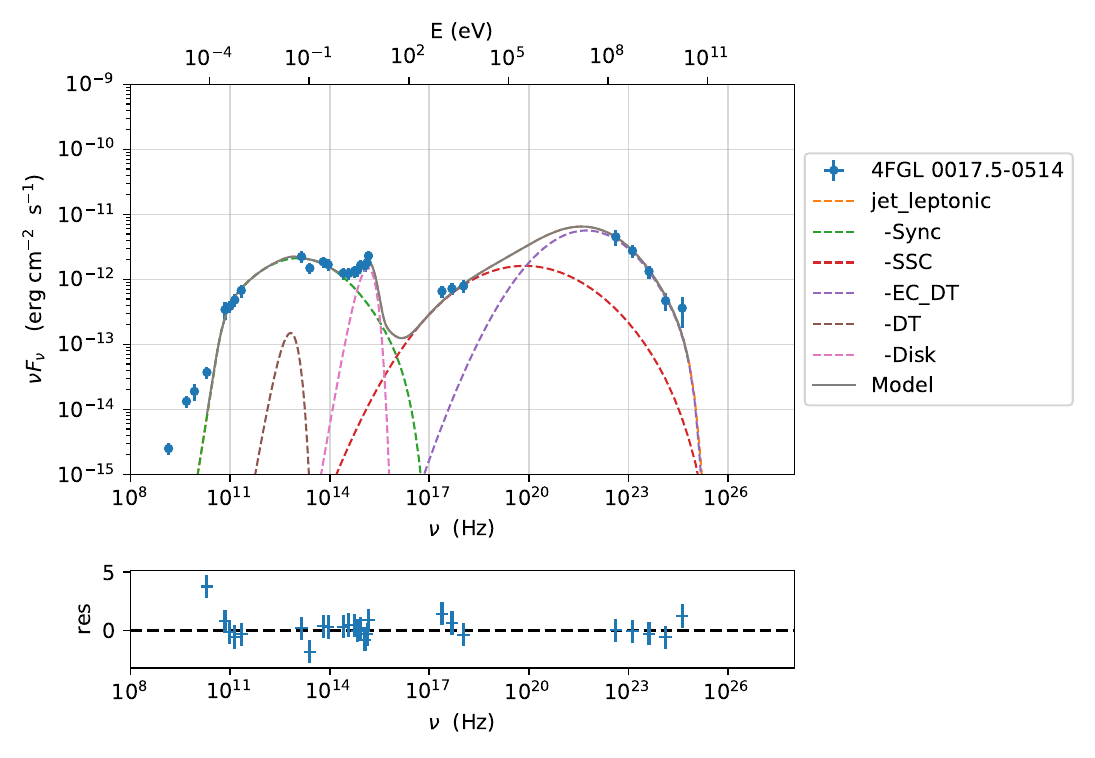}
\includegraphics[width=0.48\linewidth]{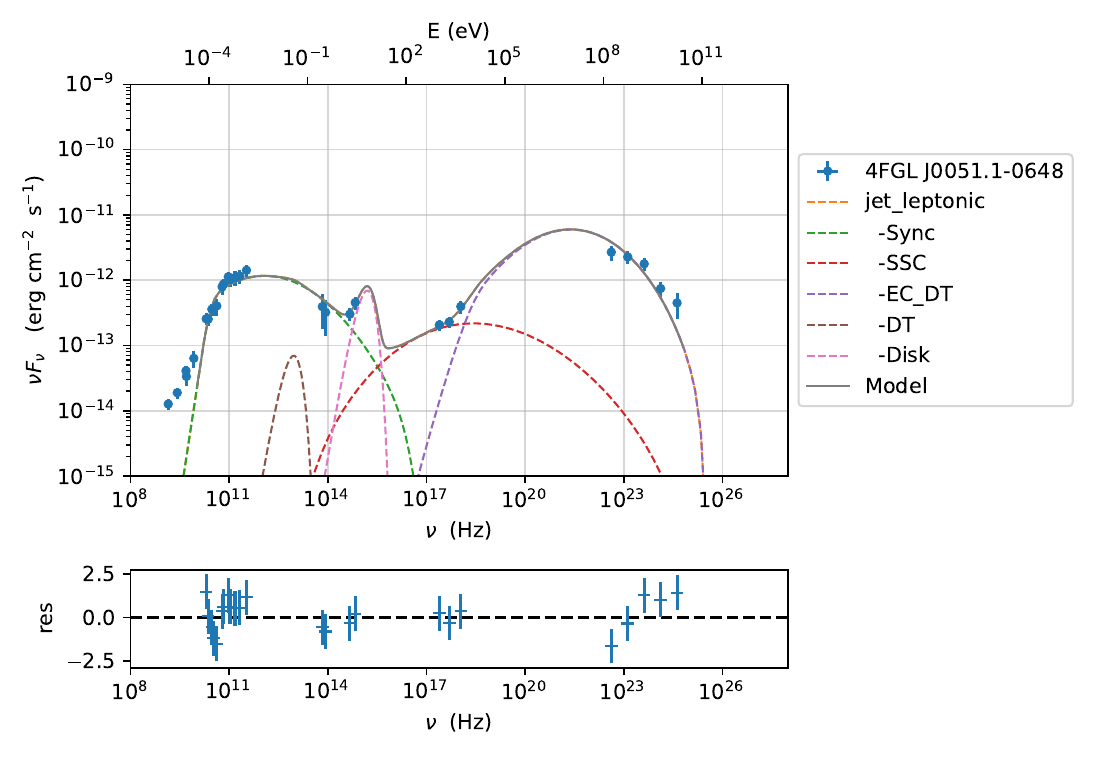}
\includegraphics[width=0.48\linewidth]{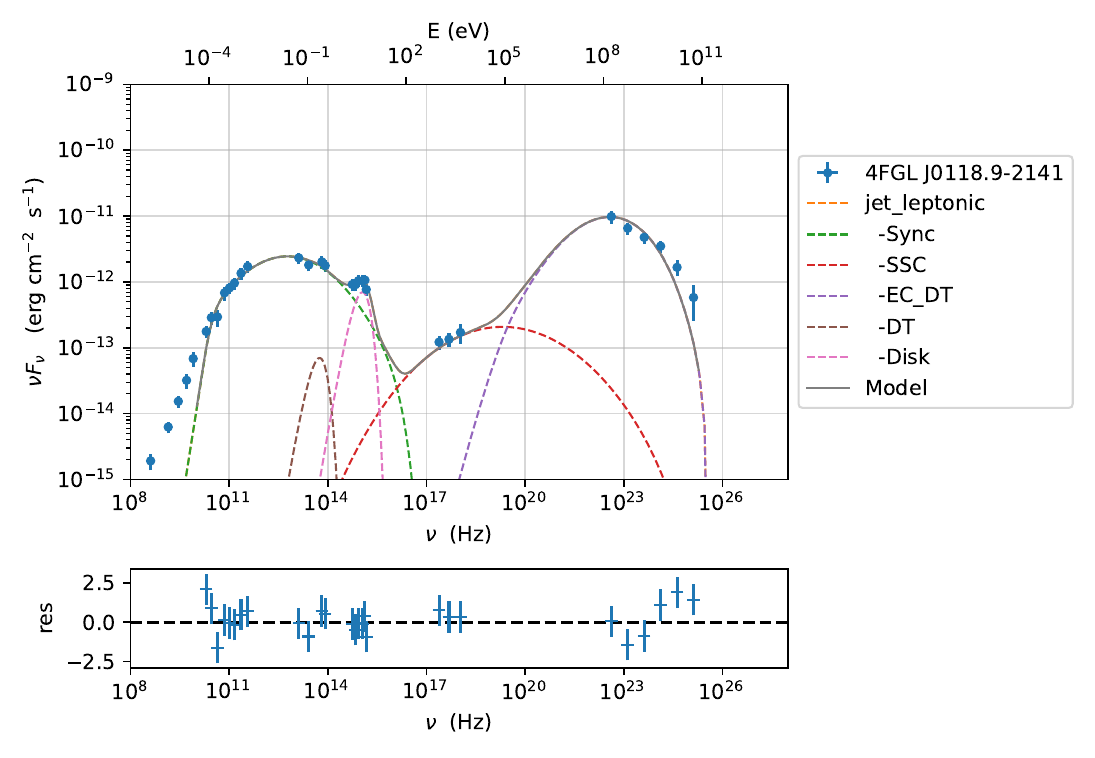}
\includegraphics[width=0.48\linewidth]{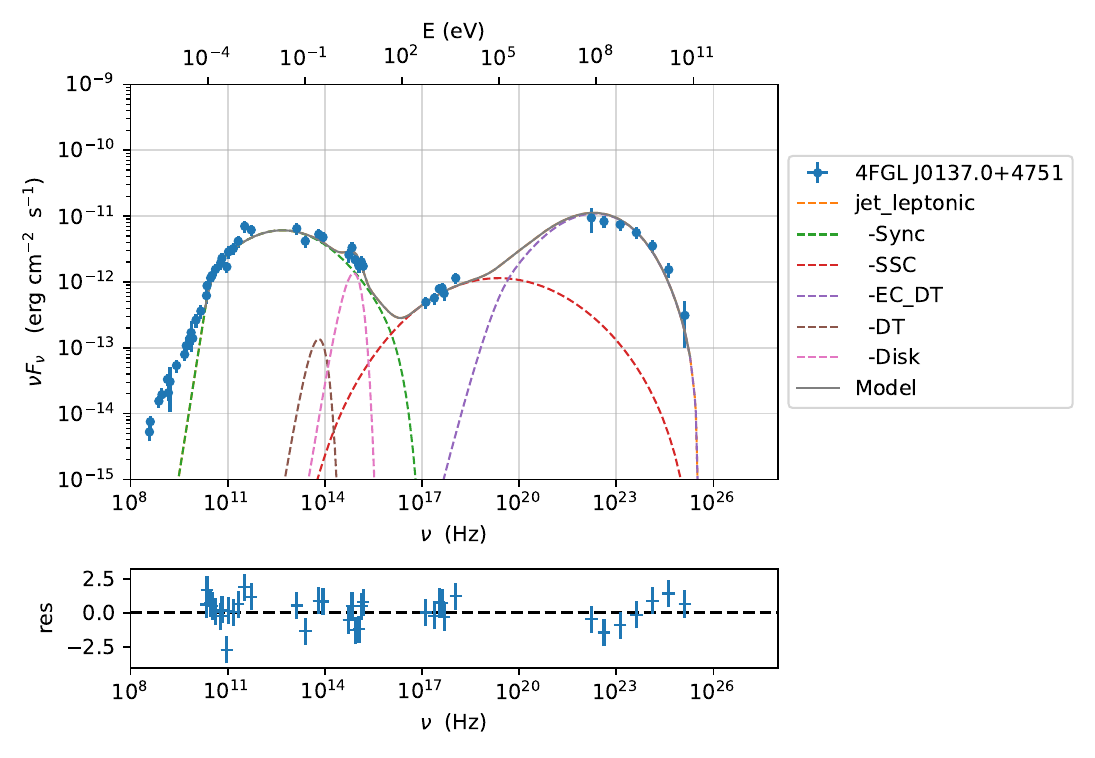}
\includegraphics[width=0.48\linewidth]{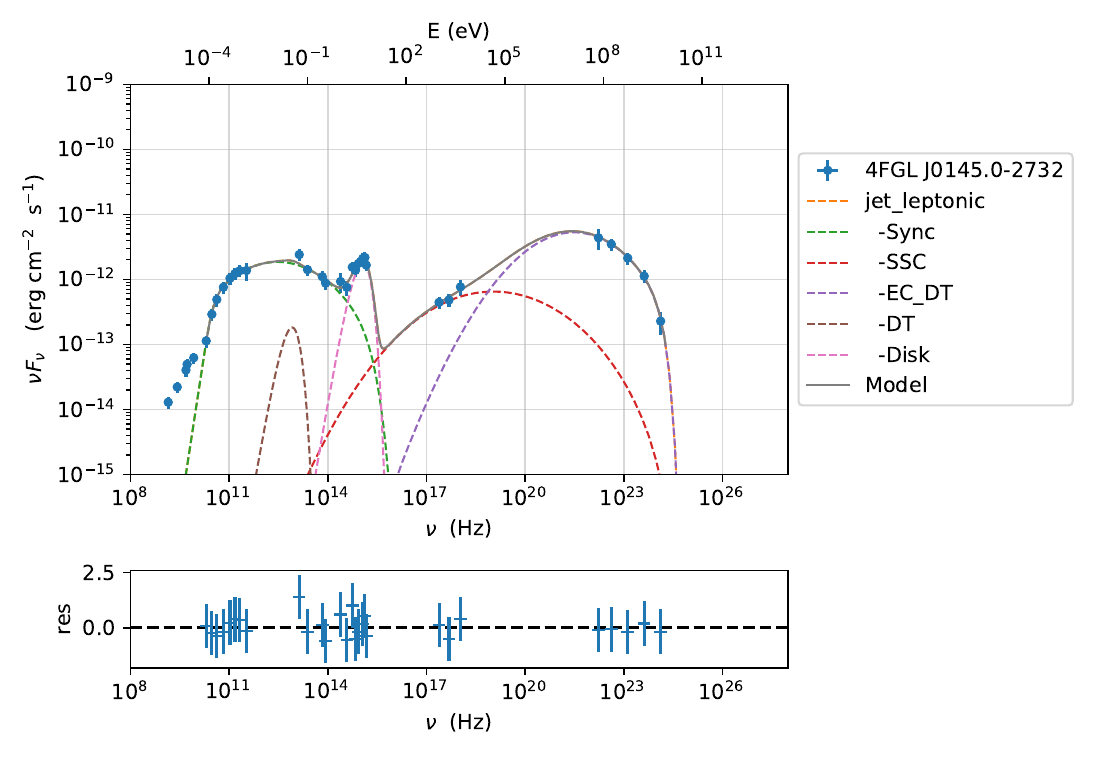}
\includegraphics[width=0.48\linewidth]{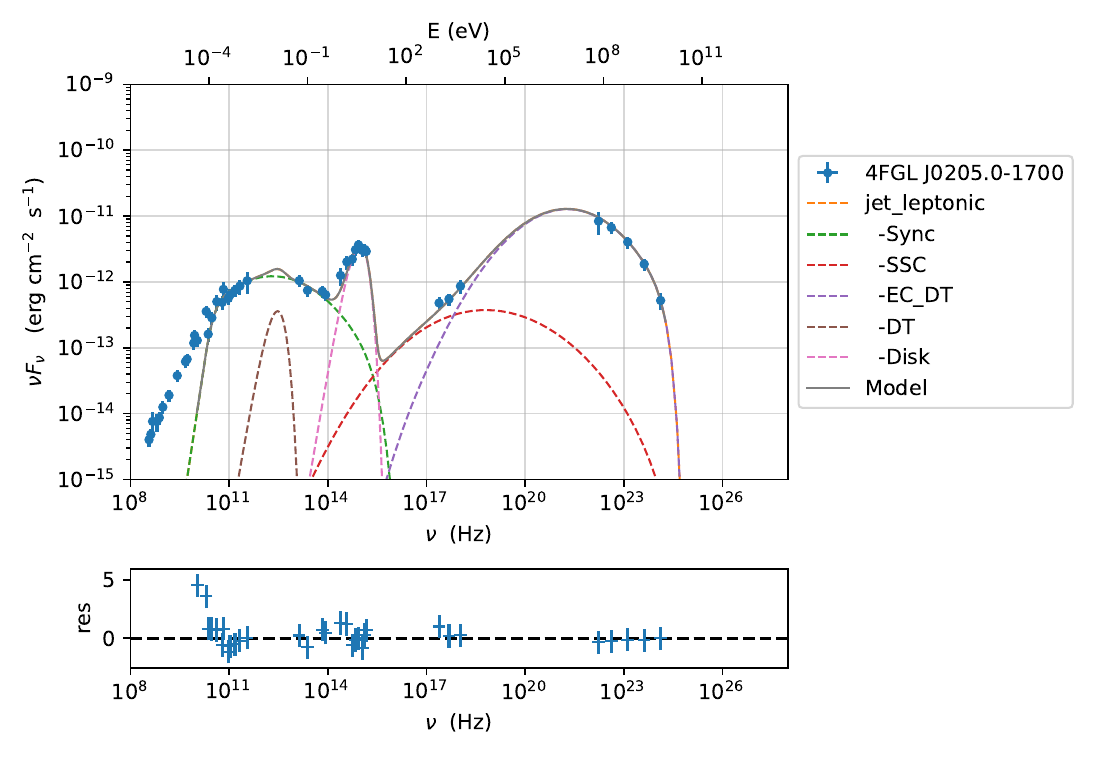} 
\includegraphics[width=0.48\linewidth]{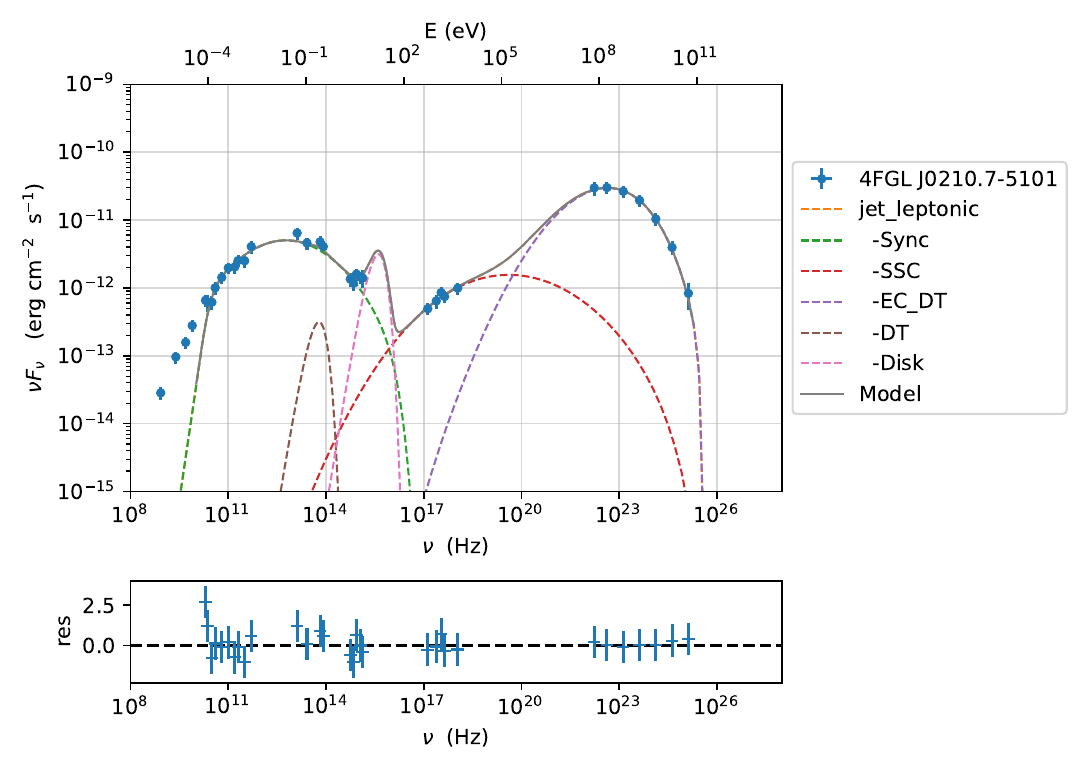}
\includegraphics[width=0.48\linewidth]{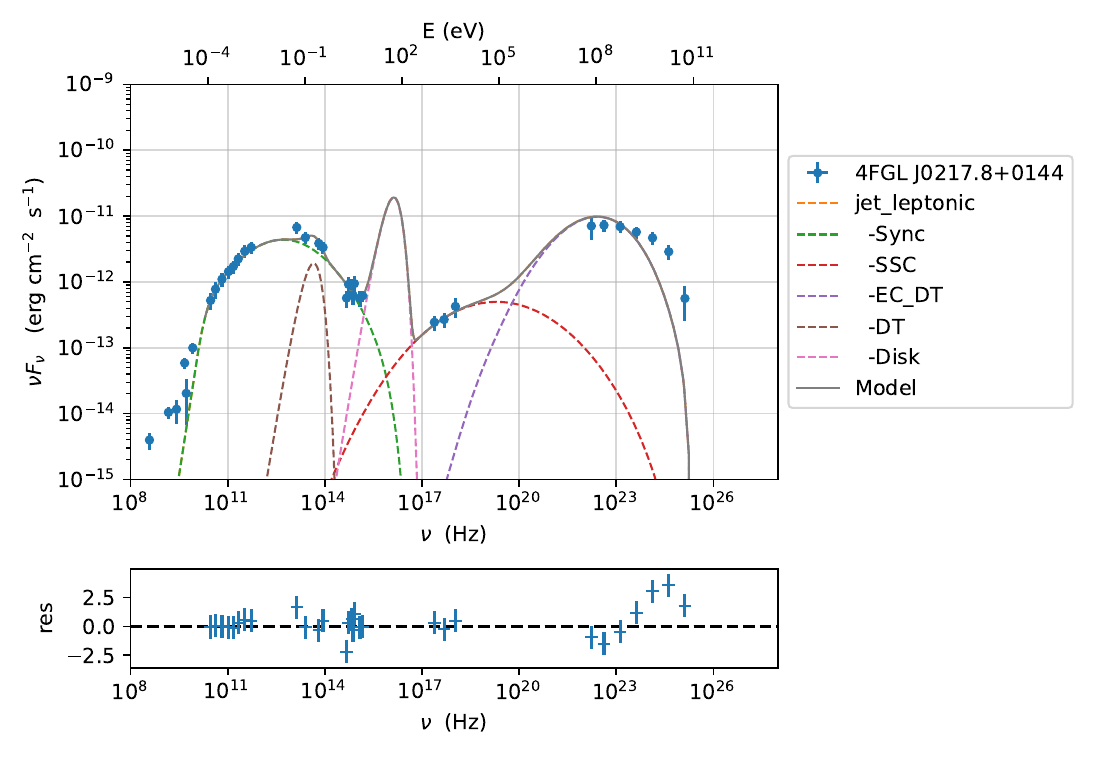}
\caption{SED Modeling of EC Blazars.}
\label{fig:SED-EC}
\end{figure}

We perform a statistical analysis of spectral parameters blazars, derived by LP polynomial fitting and one-zone model fitting, to investigate the spectral evolution of SED and the source EED in the combined scenario. The peak frequency, flux, luminosity, curvature, and the Compton dominance of BL Lacs and FSRQs, derived through polynomial fitting, are reported in Table \ref{tab:LP-BL} and Table \ref{tab:LP-FQ}, respectively. Based on synchrotron peak frequency $\nu_s$, we find that 72 blazars are LSP sources, whereas 28 sources fall into the ISP/HSP category. Then, based on the type of a blazar, we choose the one-zone model with appropriate model components to fit the broadband SEDs using $\chi^2$ minimization technique. We find that the GeV-TeV $\gamma$ ray emission of all ISP/HSP blazars (designated as SSC blazars) can be fitted well with the SSC model satisfactorily, as shown in Figure \ref{fig:SED-SSC}; whereas all LSP blazars (designated as EC blazars) necessarily need an additional EC component, as shown in Figure \ref{fig:SED-EC}. In addition to FSRQs, the EC blazars also include and LSP BL Lacs, as these sources inevitably need the EC component \citep[e.g.,][]{2024MNRAS.528.7587H}. The X-ray emission in the EC blazars is produced mainly by the SSC component, whereas the $\gamma$-ray emission is dominated by the dusty torus EC component (EC-DT). The physical one-zone model parameters of BL Lacs and FSRQs are shown in Table \ref{tab:FIT-BL} and Table \ref{tab:FIT-FQ}, respectively.\\

Interestingly, for some highly $\gamma$-ray dominant FSRQs with highly steep X-ray spectra (such as J0530.9+1332, J1016.0+0512, J1510.8-0542, J1522.1+3144, J2147.1+0931), the EC-DT component completely dominates the SSC emission, either due to strong ambient photon field or a higher collimation of the jet \citep{1997AIPC..410..494S}. This shows that, for highly luminous and Compton dominated FSRQs, the SSC component might be of secondary importance or practically irrelevant at X-ray energies. For a few sources (such as J0217.8+0144, J0221.1+3556, J0238.6+1637, J0334.2-4008, J1147.0-3812, and 4FGL J2203.4+1725), the EC-DT component under-produces the VHE $\gamma-$ rays ($>100$ GeV). It does not affect our jet parameters estimation since the physical parameters are related to SED peaks, however, it indicates that some blazars might need a significant contribution from BLR photons to account for the high energy tail of Fermi-LAT $\gamma$-ray spectra. Our modeling results are consistent with previous studies suggesting that the location of the $\gamma$-ray production should be inside the molecular torus, nearly at the edge of the BLR \citep[see, e.g.,][]{2018MNRAS.477.4749C, 2014ApJS..215....5K, 2024MNRAS.528.7587H}.\\

The distributions of physical jet parameters of blazars are shown in Figure \ref{fig:HIST-JET}. The parameter space is similar to previous blazar studies \citep[see, e.g.,][]{2020ApJS..248...27T, 2020ApJ...898...48A}. On average, the FSRQs possess an order of magnitude higher magnetic field ($B\approx 0.1$ G) as compared to BL Lacs ($B\sim 0.01$ G); however, both sources show a similar jet size $R\sim10^{17}$ cm. The Doppler factor $\delta$ for BL Lacs, on average, is higher ($>20$) than for FSRQs ($<20$). The total ambient energy density $U_{tot}$ for FSRQs is several orders of magnitude higher than BL Lacs due to external DT photon field. Figure \ref{fig:HIST-EED} shows the source EED parameters of best model fitting. The average spectral index $s$ is higher for FSRQs ($s\approx2.5$) than for BL Lacs ($s<2$), however, the curvature parameter $r$ overlap significantly for the two populations. The electron reference energy $\gamma_0$ in BL Lacs varies across several orders of magnitude, whereas for FSRQs it varies in a small range ($100-1000$). The total jet power $L_{tot}$ and radiative power $L_{rad}$ of FSRQs dominates the BL Lacs, as shown in Figure \ref{fig:HIST-JP}, indicating a stronger radiative cooling in FSRQs in the context of blazar sequence. The BL Lacs show average total jet power $L_{tot}< 10^{46}$ erg s$^{-1}$, whereas the FSRQs show $L_{tot}> 10^{47}$ erg s$^{-1}$. For FSRQs sources, the disk luminosity $L_D$ varies in the range $10^{44}-10^{48}$ erg s$^{-1}$ and seems to be positively correlated to the location of the jet blob $R_b$, that varies from subpc to pc scale, as shown in Figure \ref{fig:disk}. This suggests that the location of energy dissipation $R_b$ might be related to accretion power (as $L_{\rm acc} \propto \eta L_D$) and, thus, the $\gamma$-ray power (as $L_\gamma \propto L_D$). Similar results were found by \citep{2012PASJ...64...80Y} who modeled the broadband SEDs of 21 FSRQs.

\subsection{\textbf{Spectral Evolution of Blazars}} \label{sec:evolution}
\begin{figure*}
    \centering
    \includegraphics[width=0.49\linewidth]{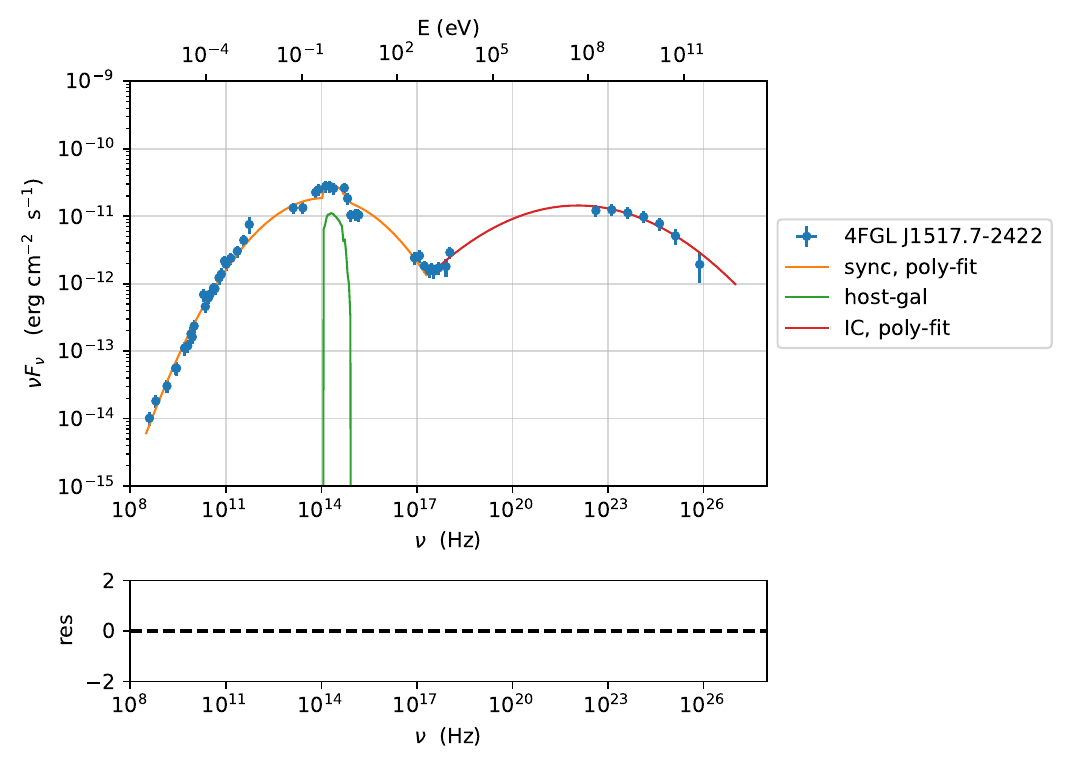}
    \includegraphics[width=0.49\linewidth]{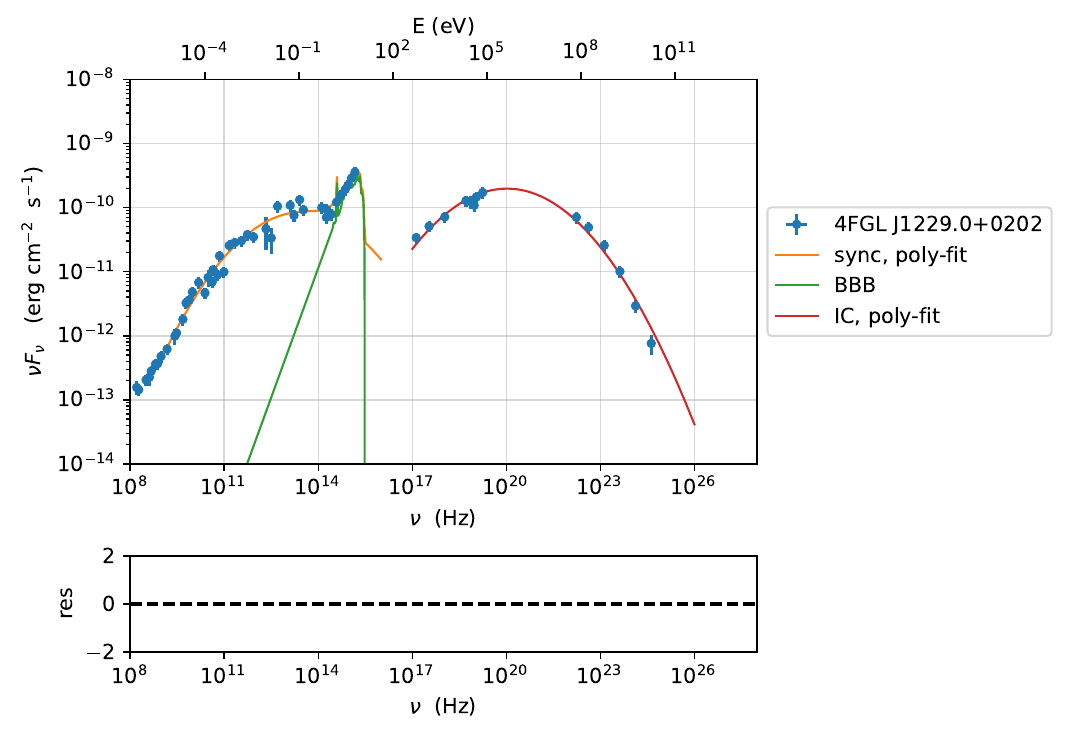}
    \caption{An example of log-parabolic polynomial fitting of blazars. The \textit{left panel} shows the BL Lac source ApLib (4FGL J1517.7-2422), whereas the \textit{right panel} shows the FSRQ 3C273 (4FGL J1229.0+0202).}
    \label{fig:polyfit}
\end{figure*}

 \begin{figure*}[htb!]
    \centering
    \includegraphics[scale=0.9]{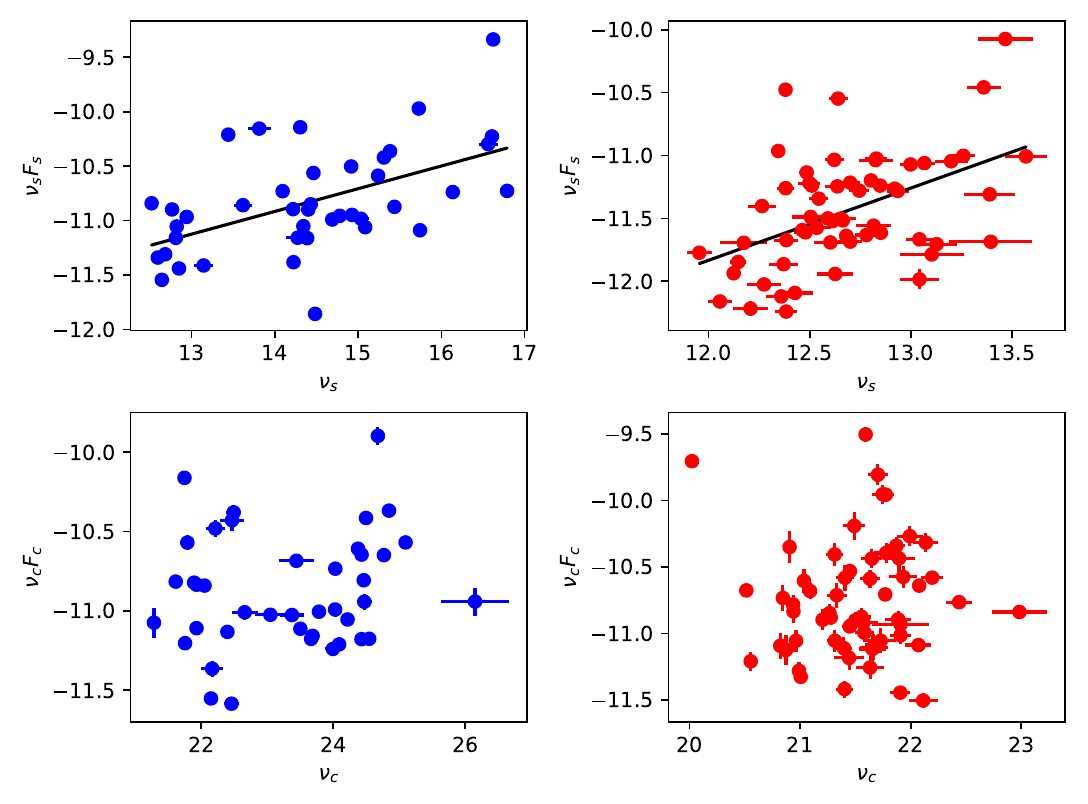}
    \caption{The relationship between spectral peak frequency and peak flux of polynomial. The \textit{upper panels} show the relationship between synchrotron peak frequency $\nu_s$ and synchrotron peak flux $\nu_sF_s$, whereas the \textit{lower panels} show IC peak frequency $\nu_s$ and IC peak flux $\nu_cF_c$. BL Lacs are shown by blue circles and FSRQs are shown by red circles. The solid lines represent the best linear regression fit.}
    \label{fig:nu-F}
\end{figure*}

 \begin{figure*}[htb!]
    \centering
    \includegraphics[scale=0.9]{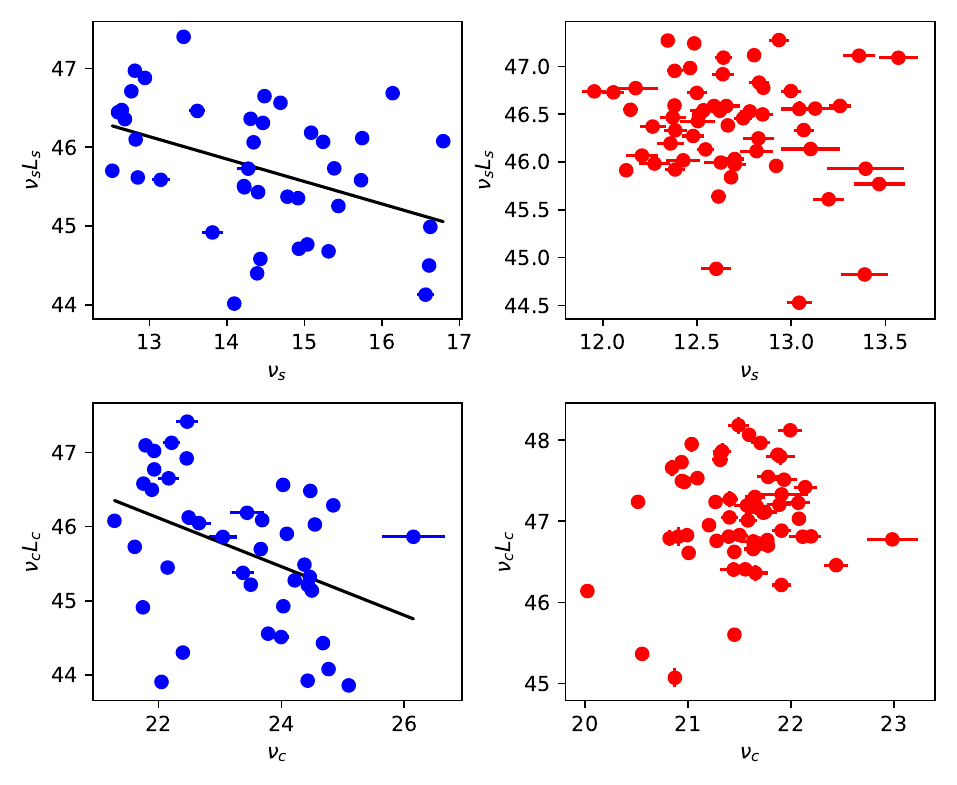}
    \caption{The relationship between spectral frequency and luminosity of polynomial. The \textit{upper panels} show the relationship between synchrotron frequency $\nu_s$ and synchrotron luminosity $L_s$, whereas the \textit{lower panels} show the relationship between IC peak frequency $\nu_c$ and IC peak luminosity $L_c$. BL Lacs are shown by blue circles and FSRQs are shown by red circles. The solid lines represent the best linear regression fit.}
    \label{fig:nu-L}
\end{figure*}

\begin{figure*}[htb!]
    \centering
    \includegraphics[scale=0.9]{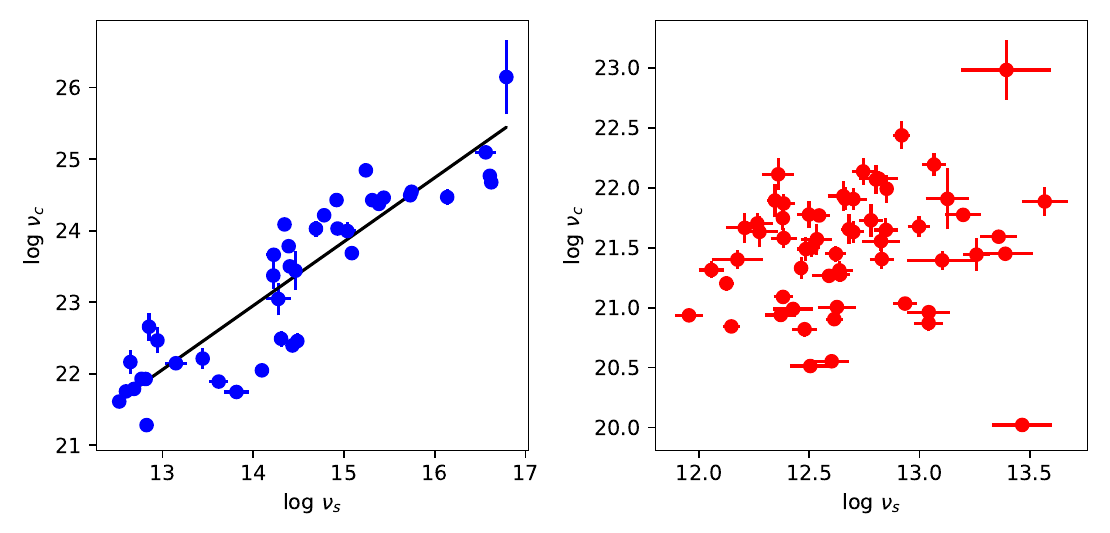}
    \caption{The relationship between synchrotron peak frequency $\nu_s$ and IC peak frequency $\nu_c$. BL Lacs are shown by blue circles and FSRQs are shown by red circles. The solid lines represent the linear regression fit.}
    \label{fig:nu-nu}
\end{figure*}

\begin{figure*}[htb!]
    \centering
    \includegraphics[scale=0.9]{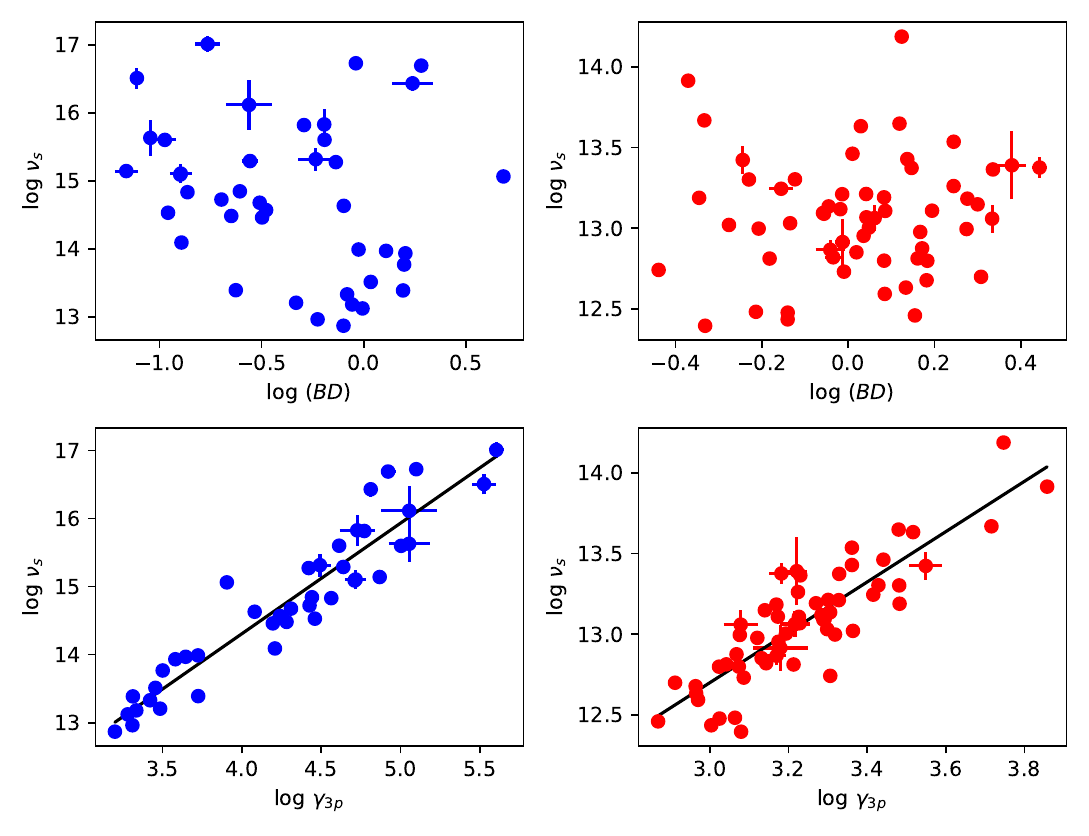}
    \caption{The relationship of synchrotron frequency $\nu_s$ with electron peak energy $\gamma_{3p}$ (\textit{upper panels}) and parameter $B\delta$ (\textit{lower panels}), obtained by one-zone modeling. BL Lacs are shown by blue circles and FSRQs are shown by red circles. The solid lines represent the best linear regression fit.}
    \label{fig:nus}
\end{figure*}

\begin{figure*}[htb!]
    \centering
    \includegraphics[scale=0.9]{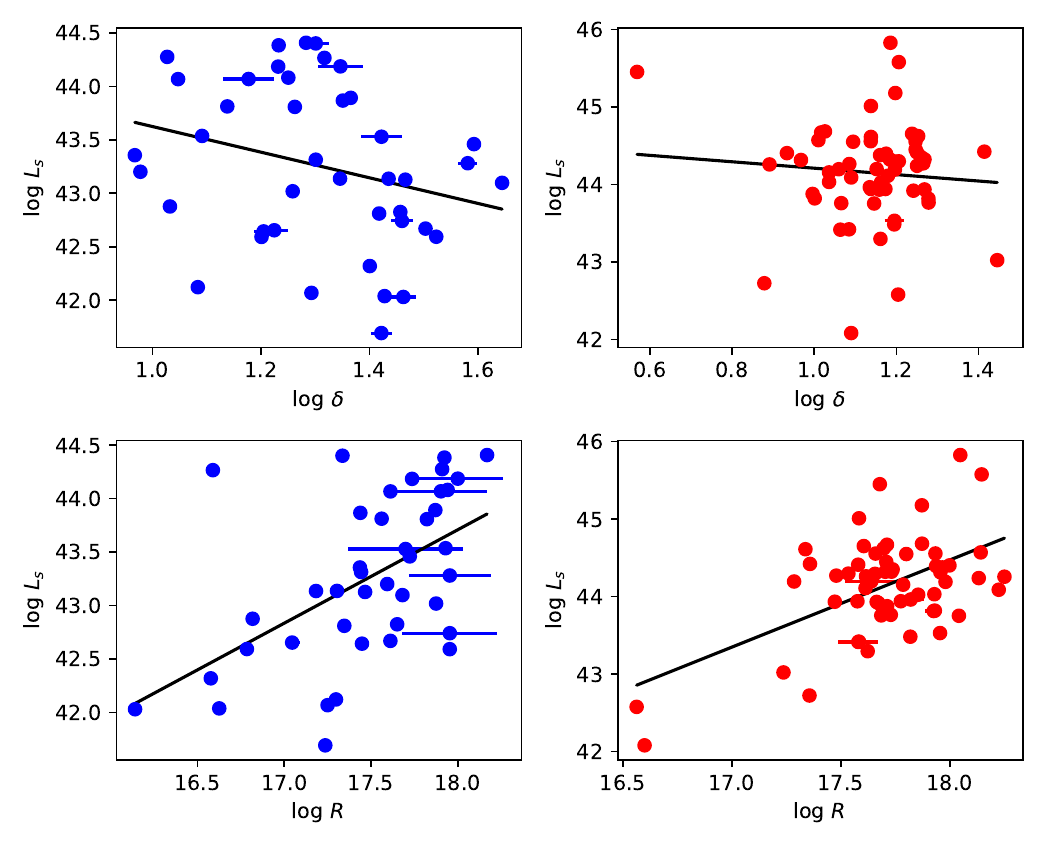}
    \caption{The relationship of synchrotron luminosity $L_s$ with Doppler factor $\delta$ (\textit{upper panels}), and source size $R$ (\textit{lower panels}), obtained by one-zone modeling. BL Lacs are shown by blue circles and FSRQs are shown by red circles. The solid lines represent the best linear regression fit.}
    \label{fig:Ls}
\end{figure*}

\begin{figure*}[htb!]
    \centering
    \includegraphics[scale=0.9]{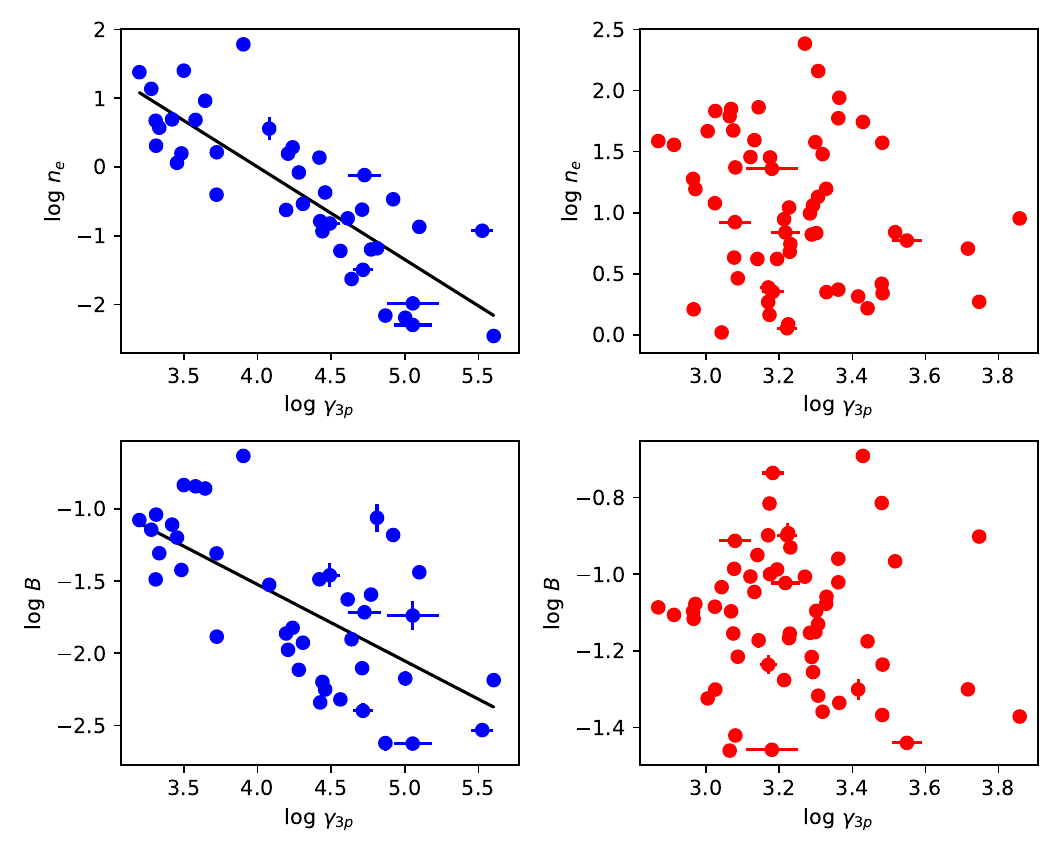}
    \caption{The relationship of particle energy $\gamma_{3p}$ with particle density $n_e$ (\textit{upper panels}), and magnetic field $B$ (\textit{lower panels}). BL Lacs are shown by blue circles and FSRQs are shown by red circles. The solid lines represent the best linear regression fit.}
    \label{fig:gammap}
\end{figure*}

\begin{figure*}[htb!]
    \centering
    \includegraphics[scale=0.9]{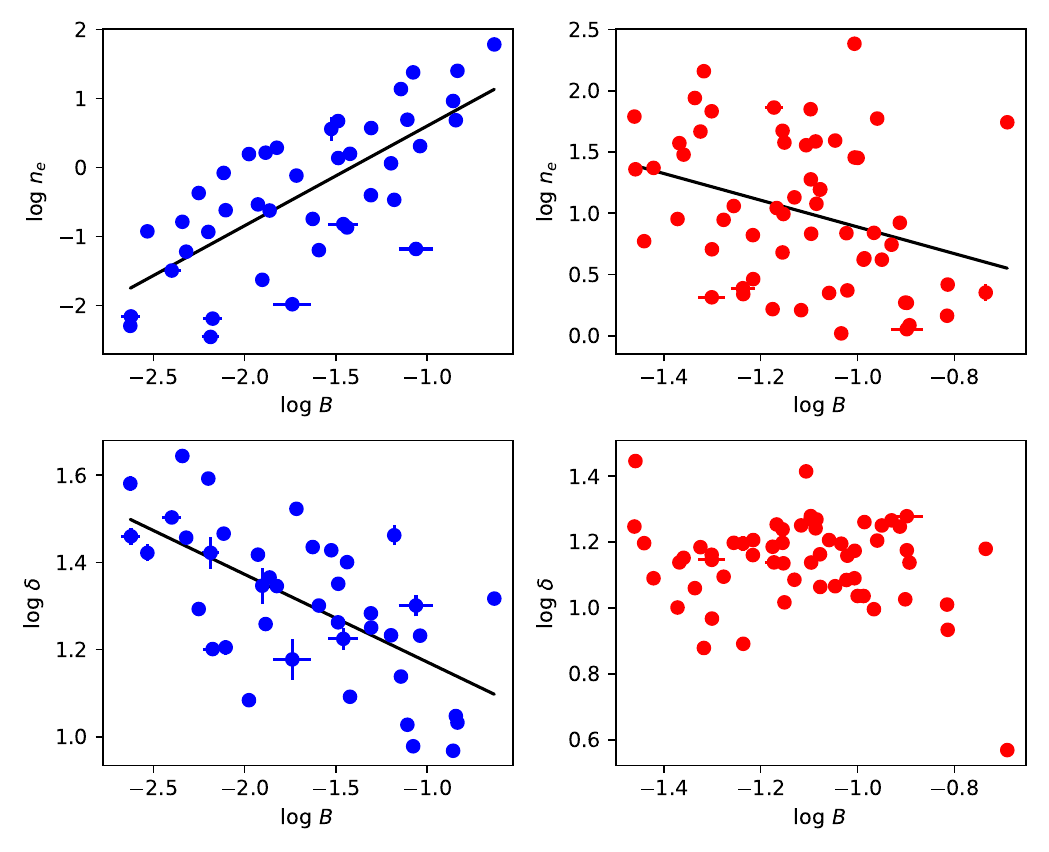}
    \caption{The relationship of magnetic field $B$ with Doppler factor $\delta$ (\textit{upper panels}) and particle density $n_e$ (\textit{lower panels}). BL Lacs are shown by blue circles and FSRQs are shown by red circles. The solid lines represent the best linear regression fit.}
    \label{fig:BDN}
\end{figure*}

The LP polynomial function is relatively milder than an exponential law and fits the blazar spectra adequately. Fitting the blazar SEDs can reveal the spectral variations and underlying parameters that produce these variations. Therefore, based on logarithmic parabolic polynomial fitting using Equation 3, we measure the spectral parameters of both synchrotron and IC components and relate these quantities to physical jet parameters derived through the one-zone model fitting. Since the jet properties of the BL Lacs and FSRQs are similar in each class, studying these classes separately makes their spectral behavior more prominent. To measure the spectral parameters, we fit the polynomial+BBB-template on synchrotron bump of LSP sources, whereas a polynomial+host galaxy template for ISP/HSP sources. While almost all FSRQs shows a BBB, only three nearby BL Lacs show the host galaxy emission. As the curvature measurement is sensitive to fitted frequency window, We fitted the polynomial in same frequency window for BL Lacs and FSRQs to eliminate possible systematic biases. Figure \ref{fig:polyfit} shows two representative cases of LP polynomial fitting.\\

We find that the synchrotron frequency $\nu_s$ of both blazar types seems to be positively correlated with synchrotron flux $\nu_sF_s$, however, the Compton frequency $\nu_c$ and flux $\nu_cF_c$ do not seem to be related, as shown in Figure \ref{fig:nu-F}. This trend is, however, not intrinsic and appear to be an effect of the luminosity distance. The BL Lacs show an inverse relationship in both the $\nu_s-\nu_sL_s$ and $\nu_c-\nu_cL_c$ planes with correlation coefficient $r_p\approx-0.4$ and a chance probability $p\sim10^{-3}$, as shown in Figure \ref{fig:nu-L}. The FSRQs almost show no correlation in these SED planes. Similarly, the peak frequencies $\nu_s$ and $\nu_c$ are significantly positively correlated related for BL Lacs having best fitting relationship $\log\nu_c=0.9\log\nu_s +10.44$ with $r_p\approx0.9$ and $p\sim 10^{-16}$, as shown in Figure \ref{fig:nu-nu}, confirming the the SSC nature of their high energy emission $\nu_c=\gamma_{3p}^2 \nu_s$ \citep{2021ApJ...915...59Z}. The FSRQs do not show any relationship between the peak frequencies. These systematic trends observed for BL Lacs only suggest that their spectra might evolve differently from FSRQs.\\

The anti-correlation between peak frequency $\nu_s$ and luminosity $\nu_sL_s$ for BL Lacs is consistent with the blazar sequence. In the blazar sequence, the higher jet power ($\propto$ accretion power) derives the inverse relationship, such that the FSRQs have higher jet power than BL Lacs due to increased ambient density $U_{tot}$, which limits the electron energy $\gamma_{3p}$ due to efficient cooling \citep{1998MNRAS.301..451G, 2008MNRAS.387.1669G}. However, the blazar sequence does not explain the internal relationship between the physical parameters, such as magnetic field $B$, particle density $n_e$, and the particle energy $\gamma_{3p}$. Moreover, the blazar sequence has been criticized as a selection effect or an artifact of Doppler boosting \citep{2008A&A...488..867N}. The systematic trends observed for BL Lacs in the synchrotron and IC $\nu_p L_p$ SED planes indicate that the jet parameters might be correlated. We, therefore, explore the internal relationship between source parameters that can mimic the blazar sequence.\\

The observed inverse relationship between $\nu_s$ and $\nu_sL_s$ for BL Lacs cannot be dominated by changes in a single jet parameter such as magnetic field $B$, peak energy $\gamma_{3p}$, or Doppler factor $\delta$; since such changes would rather lead to a positive relationship between $\nu_s$ and $\nu_sL_s$, as expected in a pure synchrotron cooling scenario. Assuming a synchrotron scenario with independent jet parameters, the peak frequency and luminosity follow a positive scaling relation $\nu_sL_s \propto \nu_s^\alpha$, where
\begin{equation}
    \nu_s\propto \gamma_{3p}^2B\delta
\end{equation}
is the synchrotron peak frequency,
\begin{equation}
  \nu_sL_s\propto n_e\gamma_{3p}^2B^2R^3\delta^4
\end{equation}
is the synchrotron peak luminosity, and $\alpha$ is the slope that indicates the parameters causing the spectral changes in the SED \citep{2008A&A...478..395M}. Individual sources usually show a positive relationship between peak frequency and luminosity and such a bluer-when-brighter trend of blazar may arise due to a single parameter, for example, the changing Doppler factor $\delta$ \citep{2022Univ....8..585F}. A negative trend in the $\nu_s-\nu_sL_s$ plane cannot be explained by independent jet parameters and suggests that some jet source parameters might be related to each other.\\ 

Figure \ref{fig:nus} shows that the synchrotron frequency $\nu_s$ is tightly correlated with electron peak energy $\gamma_{3p}$ with correlation coefficient $r_p\approx0.9$ and chance probability $p\leq10^{-18}$, however, it does not show any correlation with the parameter $B\delta$. The BL Lacs and FSRQs show the best linear relationship $\log\nu_s=1.62\log\gamma_{3p}+7.80$ and $\log\nu_s=1.56\log\gamma_{3p}+8.02$, respectively, closer to theoretical relation $\nu_s \sim \gamma_{3p}^2$. This confirms that synchrotron frequency shifts in the $\nu_s-\nu_sL_s$ plane are primarily caused by variations of particle energy $\gamma_{3p}$, whereas the parameter $B\delta$ is irrelevant. The synchrotron luminosity $L_s$ inversely correlates with Doppler factor $\delta$ and source size $R$, as shown in Figure \ref{fig:Ls}, showing that $\delta$ and $R$ are inversely related. Therefore, the luminosity in $\nu_s-\nu_sL_s$ plane should evolve mainly due to the internal relationship between the other jet parameters, namely, the magnetic field $B$, electron density $n_e$, and electron peak energy $\gamma_{3p}$. The relationship of $\gamma_{3p}$ with $n_e$ and $B$ is shown in Figure \ref{fig:gammap}. For BL Lacs, the particle energy $\gamma_{3p}$ seems to be significantly inversely relationship with particle density $n_e$ with coefficient $r_p<-0.8$, showing the best fit relationship $\log n_e=-1.35\log\gamma_{3p}+5.4$. The $\gamma_{3p}$ also shows a negative correlation with $B$, with best fit relationship $\log B=-0.53\log \gamma_{3p}+0.6$. Similarly, the magnetic field $B$ is positively correlated with density $n_e$ through a linear relation $\log n_e=1.32\log B+1.85$ with $r_p=-0.63$ and $p\simeq 10^{-6}$, and negatively correlated with $\delta$, as shown in Figure \ref{fig:BDN}. The FSRQs do not show these systematic trends, indicating that the jet parameters might be varying independently.\\

The relationship between parameters magnetic field $B$, electron energy $\gamma_{3p}$, and density $n_e$ show, for BL Lacs, that jet parameters in BL Lacs must be internally linked and explains the observed negative trend in $\nu_s-\nu_sL_s$ plan in context of equipartition \citep{2008MNRAS.387.1669G}. In the equipartition scenario, the magnetic field energy density $U_B=B^2/8\pi$ and particle energy density $U_e\approx n_e\gamma_{3p}m_ec^2$ are correlated (i.e., $U_B\approx U_e$), leading to the constraint $B^2\simeq n_e\gamma_{3p}$. Thus, the magnetic field $B$ should be positively correlated with $n_e$ and/or $\gamma_{3p}$, forcing an inverse relationship between $\gamma_{3p}$ and $n_e$, to hold the equipartition. Consequently, an increasing frequency $\nu_s \propto \gamma_{3p}^2$ manifests a decrease in the luminosity $\nu_sL_s \propto n_e$. Figure \ref{fig:gammap} shows the negative relation between $\gamma_{3p}$ and $n_e$ and confirms this scenario for BL Lacs. The BL Lacs, on average, show a Compton Dominance ($C_D\sim1$), which is another observational indicator that the physical conditions might be steady, dominated by equipartition. The FSRQs do not show such linear trends, indicating that the underlying physical conditions deviate significantly from the equipartition scenario. This difference might be related to the fact that the EC emission in FSRQs is sensitive to ambient environment, e.g. external photon fields, whereas the SSC emission in BL Lacs is unrelated to ambient environment.

\subsection{\textbf{Acceleration in Blazars}} \label{sec:acceleration}
\begin{figure*}[htb!]
    \centering
    \includegraphics[scale=0.9]{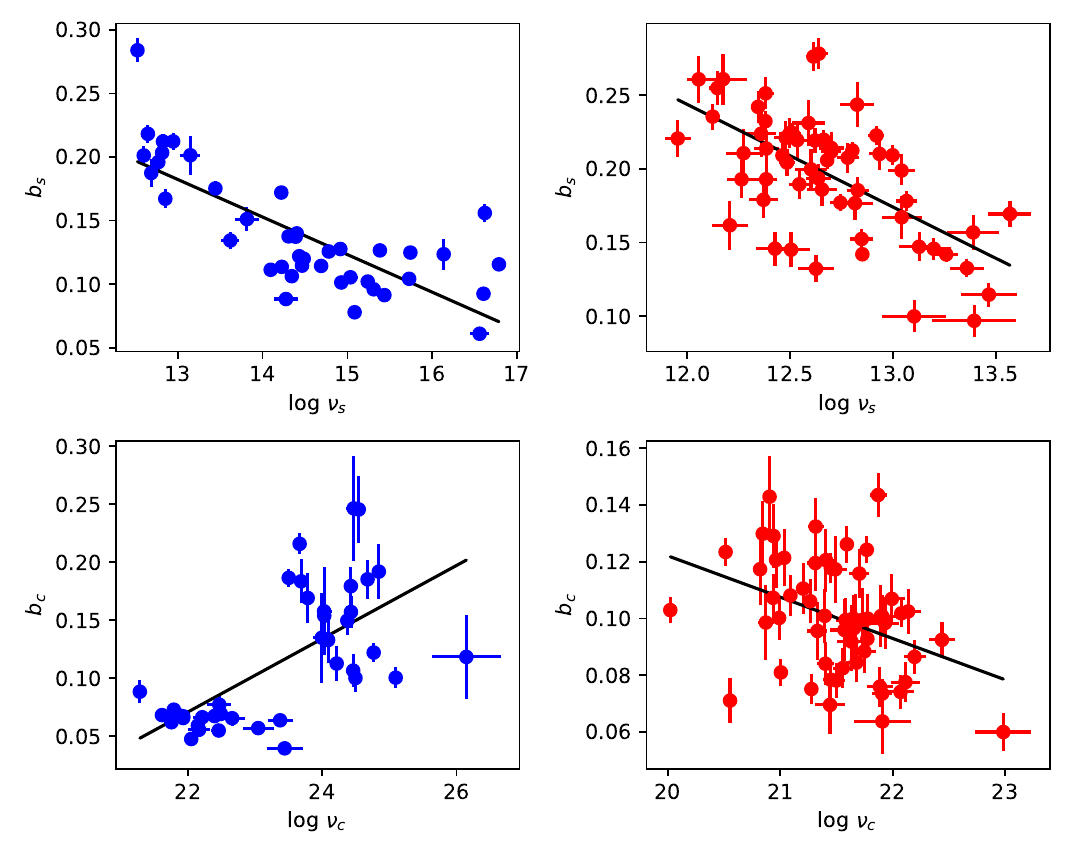}
    \caption{The relationship between the peak frequency and spectral curvature. The \textit{upper panels} show the relationship between synchrotron frequency $\nu_s$ and synchrotron curvature $b_s$, whereas the \textit{lower panels} show the relationship between IC frequency $\nu_c$ and IC curvature $b_c$. BL Lacs are shown by blue circles and FSRQs are shown by red circles. The solid lines represent the linear regression fit.}
    \label{fig:nu-b}
\end{figure*}

\begin{figure*}[htb!]
    \centering
    \includegraphics[scale=0.9]{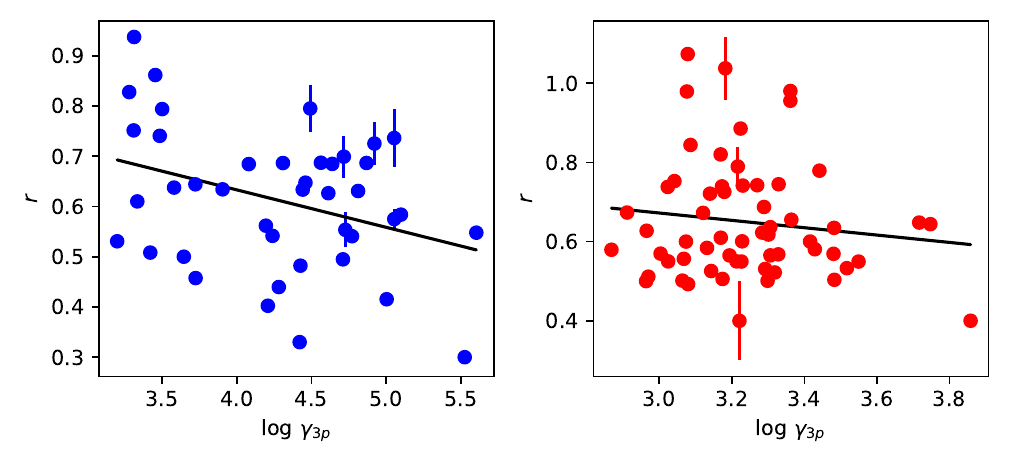}
    \caption{The intrinsic signature of stochastic acceleration, i.e., the relationship between EED peak energy $\gamma_{3p}$ and its curvature $r$. BL Lacs are shown by blue circles and FSRQs are shown by red circles. The solid lines represent the linear regression fit.}
    \label{fig:r-gammap}
\end{figure*}

\begin{figure*}[htb!]
    \centering
    \includegraphics[scale=0.9]{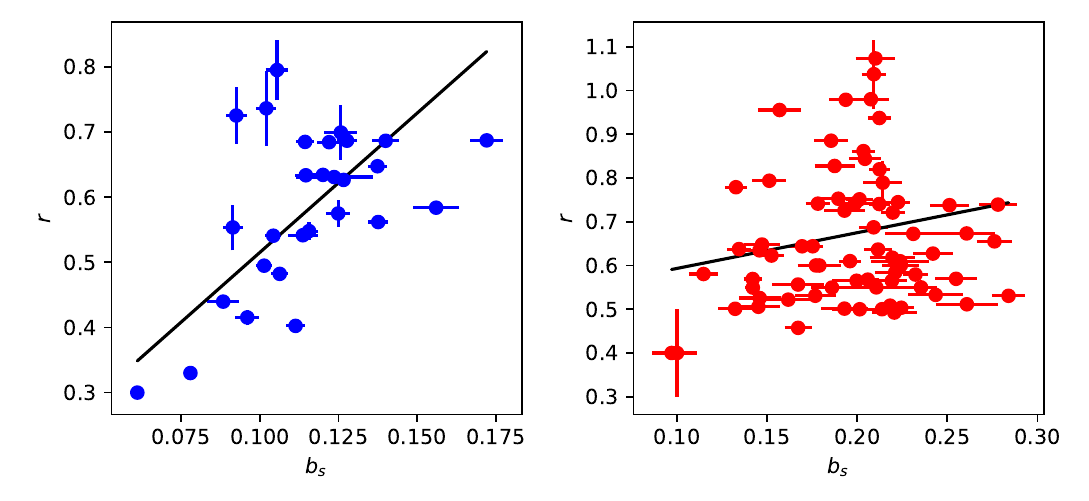}
    \caption{The relationship between synchrotron spectral curvature $b_s$ and source electron curvature $r$. The SSC blazars are shown by blue circles and the EC blazars are shown by red circles. The solid lines represent the linear regression fit.}
    \label{fig:r-b}
\end{figure*}

To investigate the statistical nature of particle acceleration in a combined acceleration+cooling scenario, we study how the synchrotron spectral curvature $b_s$ evolves with synchrotron frequency $\nu_s$. We find that $\nu_s$ significantly anti-correlates with $b_s$, for both the BL Lacs and FSRQ, as shown in Figure \ref{fig:nu-b}, suggesting that stochastic nature of the underlying acceleration is general for all blazars. Such inverse correlation has been found in previous studies for large samples \citep[see, e.g.,][]{2014ApJ...788..179C, 2016MNRAS.463.3038X}. The best regression fit for BL Lacs yields $b_s= -0.03\log\nu_s+0.56$, whereas the FSRQs show $b_s= -0.07\log\nu_s+1.08$. The slope is usually different for different blazar classes \citep{2017AcASn..58...57L}, suggesting that physical conditions or nature of acceleration mechanism might be slightly different in FSRQs and BL Lacs. \citet{2024ApJ...966...99X} showed that the blazars with low IC peak frequency follow the stochastic acceleration, whereas the high IC peak sources follow the energy-dependent acceleration mechanism, with the transition at $\nu_s \approx 10^{15}$. The IC peak frequency $\nu_c$ show a different behavior against its spectral curvature $b_c$. For BL Lacs, the $\nu_c$ and $b_c$ show a positive relationship $b_c=0.03 \log\nu_c-0.62$ with $r_p=0.66$ and $p\sim10^{-6}$, whereas, the FSRQs show mildly negative relationship $b_c=-0.015\log\nu_c+0.41$ with $r_p=-0.38$ and $p\sim10^{-3}$, as shown in Figure \ref{fig:nu-b}. This might be related to different nature of IC emission in two classes. The IC emission in FSRQs is more complicated than BL Lacs and arises as composite emission including both the SSC and EC components.\\

We investigate the evolution of source EED curvature $r$ against its peak energy $\gamma_{3p}$, as shown in Figure \ref{fig:r-gammap}. These EED parameters are intrinsic as compared to SED parameters $b_s$ and $\nu_s$ and are not affected by other physical parameters, such as $B$ or $\delta$. Furthermore, the relationship between these intrinsic parameters reveals the relevance of specific processes and nature of magnetic turbulence during the evolution of EED. The $\gamma_{3p}-r$ plane shows the BL Lacs manifest a negative relationship $r=-0.075\log\gamma_{3p}+0.93$, whereas the FSRQs show a milder relationship with less significant correlation. This less significant correlation for FSRQs might be related to a relatively poor estimates of $\gamma_{3p}$ and $r$ due to several IC components involved, such as SSC and EC. High energy resolution data at X-rays might be necessary to resolve these components. Thus, the blazars show the intrinsic signature of stochastic acceleration, i.e., an inverse relationship between peak energy $\gamma_{3p}$ and curvature $r$ \cite{2020ApJ...898...48A}. However, a small slope $<-0.1$ and a large scatter, however, shows that the cooling must be relevant and the curvature might evolve in a nearly steady state conditions, where cooling tightly competes against acceleration and the negative trend in $\gamma_{3p}-r$ plane is affected \citep{2011ApJ...739...66T}. This is especially true when the magnetic turbulence is softer than the hard-sphere ($q=2$), such as, a Kolmogorov or Kraichnan turbulence with turbulence index $q<2$. \citet{2011ApJ...739...66T} showed that a softer magnetic turbulence with index $q<2$ is not an efficient accelerator as compared to the hard-sphere turbulence with $q=2$, for which the acceleration is independent of particle energy and dominates the cooling.\\

We compare the SED curvature $b_s$ with the EED curvature $r$ by dividing our blazar sample into SSC and EC classes. In our work, the photon curvature $b_s$ is constrained by polynomial fitting of synchrotron bump, whereas the sources EED curvature $r$ is constrained by one-zone model fitting of entire SED. \citet{2006A&A...448..861M} found that an LP EED nearly yields an LP synchrotron SED and the ratio of photon and electron curvature parmaters ($b_s/r$) varies from $0.18$ to $0.22$, depending on how $b_s$ is measured. They recommend a nominal value $b_s/r= 0.2$, leading to the relationship $r=5b_s$. We find that the SSC blazars (BL Lacs) show a best linear relationship $r\approx 4.2 b_s$, closer to the expected relationship $r=5b_s$, with correlation parameters $r_p>0.5$ and $p\sim10^{-3}$, as shown in Figure \ref{fig:r-b}, indicating that the SED curvature in BL Lacs is a true indicator of source EED curvature $r$. The slight deviation from $r=5b_s$ could be related to the fact the relationship $r=5b_s$ was derived by polynomial fitting within a small frequency window around peak frequency $\nu_s$, whereas we used broadband spectral data. On the other side, the EC blazars (FSRQs) show a mild relationship $r\approx 0.8 b_s$ and a very weak correlation. The large deviation from $r=5b_s$, for FSRQs, seem to arise due to a relatively poor estimation of $b_s$ due to presence of thermal BBB in the synchrotron spectra. Moreover, when the spectral window of polynomial fitting is different than that of physical model fitting, the relationship between $b_s$ and $r$ is significantly deviates from $r=5b_s$. Removing these biases would require high quality multiband IR-to-UV spectral data.

\section{\textbf{Discussion \& Summary}} \label{sec:summary}
We systematically fit the nearly simultaneous broadband SEDs of an adequately large Fermi bright blazar sample with 4FGL-DR3 $\gamma$-ray data in the GeV-TeV band to study the spectral evolution of blazars. The blazras shows a negative trend in both the $\nu_s-\nu_sL_s$ and $\nu_s-b_s$ planes, however, these trends have been studied independently previously. We investigate the evolution of the synchrotron SED in both of these planes in a combined scenario including the nonthermal cooling and stochastic acceleration. We find that the blazar SEDs can be reproduced self-consistently by a leptonic one-zone model with LP EED. The GeV-TeV $\gamma$-ray emission of the ISP/HSP blazars is fitted satisfactorily by the SSC model component, whereas, practically all LSP blazars essentially need an EC component based on external dusty torus to explain the $\gamma$-ray emission. For a few LSP sources, a BLR component might be required to fit the high-energy tail of Fermi-LAT spectra, indicating that the energy dissipation location in blazars should be inside dusty torus photon field but closer to the edge of BLR. For some highly luminous Compton-dominated FSRQs, the dust torus component completely dominates the SSC, suggesting that the the SSC component might be irrelevant for such sources. Moreover, the location of the blob $R_b$ from the central black hole is correlated with the disk luminosity, for FSRQs, indicating that the location of energy dissipation might be an indicator of accretion power and, thus, $\gamma$-ray power.\\

The electron energy $\gamma_{3p}$ seems to be the dominant driver of synchrotron frequency $\nu_s$, whereas magnetic field $B$ and Doppler factor $\delta$ almost play no role, indicating that synchrotron frequency shifts in the blazar SED mainly arise due to the variation of particle energy$\gamma_{3p}$. The BL Lacs show a negative trend in the $\nu_s-\nu_sL_s$ SED plane, known as blazar sequence. Such a spectral behavior of the BL Lacs in seems to be dominated by equipartition scenario, corresponding to stable jet configurations with minimum jet energy budget requirement and a maximum radiative efficiency. The internal relationship between the $\gamma_{3p}$, $n_e$, and $B$ can explain the observed blazar sequence in the $\nu_s-\nu_sL_s$ plane \citep{2008MNRAS.387.1669G}. We find that the BL Lacs seem to follow the equipartition $U_B \sim U_e$, forcing the constraint $\gamma_{3p} \propto 1/n_e$. Thus, when Doppler changes are irrelevant, the negative trend in $\nu_s-\nu_sL_s$ plane can naturally arise due to equipartition constraints, corresponding to the near steady state jet conditions. Thus, in a combined scenario, while an increase in particle energy $\gamma_{3p}$ derives the synchrotron frequency, the luminosity gets lower due to a corresponding decrease in particle density $n_e$. The FSRQs do not show such equipartition constraints, indicating that jet parameters in these sources might be changing more independently, most likely due to variable ambient conditions.\\

We find that the synchrotron frequency $\nu_s$ and the synchrotron spectral curvature $b_s$ of blazars are negatively related, however, the slope is slightly different for BL Lacs and FSRQ. This confirms the stochastic nature of particle acceleration in the blazar jets. The hallmark of stochastic acceleration is the negative relationship between intrinsic EED parameters, namely, the electron peak energy $\gamma_{3p}$ and electron curvature parameter $r$. The $\gamma_{3p}-r$ plane shows an indication of anti-correlation, thus, the negative trend observed in the $\nu_s-b_s$ plane seems to manifested in $\gamma_{3p}-r$ plane. However, the negative relationship is mild, especially for FSRQs. In the hard-sphere acceleration, the $\gamma_{3p}$ and $r$ are tightly inversely related. Thus, a deviation from strong hard-sphere acceleration suggests that the physical conditions in the blazar jet should be near to steady state, consistent with equipartition, and the magnetic turbulence might be soft, either Kraichnan or kolmogorov type with turbulence index $q<2$. The soft turbulence is not an efficient accelerator and the electron curvature $r$ only mildly correlates with peak energy $\gamma_{3p}$ \citep{2011ApJ...739...66T}. The synchrotron photon curvature $b_s$ is generally an indicator of source electron curvature $r$ and, within a small energy window, the two curvature parameters are found to be related as $r= 5b_s$. We find the relationship $r\approx 4 b_s$ for BL Lacs, closer to the theoretical relationship $r=5b_s$, as compared to FSRQs which show a mild relationship $r\approx b_s$. The significantly large deviation from $r=5b_s$ for FSRQs might be related to the presence of thermal BBB in the synchrotron SED and slightly different energy windows in which $b_s$ and $r$ are measured. High energy resolution broadband spectra data of FSRQs might be required to tightly constrain the spectral curvature.

\begin{acknowledgments}
We thank the editor and the anonymous referee for providing important insights and useful suggestions to improve this work. This work is supported by the Natural Science Foundation of Henan Province (242300420261), Key scientific research projects of colleges and universities in Henan Province (24a160001), and National Natural Science Foundation of China (grant No. U1938116). LC is supported by the National Science Foundation of China (grant 12173066). MFG is supported by the National Science Foundation of China (grant 12473019). Both LC and MFG are supported by the National SKA Program of China (grant No. 2022SKA0120102), the Shanghai Pilot Program for Basic Research-Chinese Academy of Science, Shanghai Branch (JCYJ-SHFY-2021-013), and the China Manned Space Project with No. CMS-CSST-2025-A07.
\end{acknowledgments}

\newpage

\startlongtable
\begin{deluxetable}{ccccccccccccc}
\tabletypesize{\footnotesize}
\tablecaption{Parameters of Log-parabolic Polynomial Fitting for BL Lacs.}  
\label{tab:LP-BL}
\tablewidth{0pt}
\tablehead{
\colhead{Name} & 
\colhead{Association} & 
\colhead{SED} & 
\colhead{$z$} & 
\colhead{$b_s$} & 
\colhead{$\nu_s$} & 
\colhead{$\nu_sF_s$} & 
\colhead{$\nu_sL_s$} & 
\colhead{$b_c$} & 
\colhead{$\nu_c$} & 
\colhead{$\nu_cF_c$} & 
\colhead{$\nu_cL_c$} & 
\colhead{$C_D$} \\
\colhead{(1)} & 
\colhead{(2)} & 
\colhead{(3)} & 
\colhead{(4)} & 
\colhead{(5)} & 
\colhead{(6)} & 
\colhead{(7)} & 
\colhead{(8)} & 
\colhead{(9)} & 
\colhead{(10)} & 
\colhead{(11)} & 
\colhead{(12)} & 
\colhead{(13)}
}
\startdata
0033.5-1921 & KUV00311-1938    & HSP & 0.61  & 0.12 & 15.74 & -11.09 & 46.11 & 0.25 & 24.54 & -11.18 & 46.03 & 0.82 \\
0050.7-0929 & PKS0048-09       & HSP & 0.634 & 0.08 & 15.09 & -11.06 & 46.18 & 0.18 & 23.69 & -11.16 & 46.09 & 0.8  \\
0120.4-2701 & 1Jy0118-272      & ISP & 0.56  & 0.11 & 14.34 & -11.05 & 46.06 & 0.13 & 24.09 & -11.21 & 45.9  & 0.69 \\
0136.5+3906 & B30133+388       & HSP & 0.75  & 0.12 & 16.14 & -10.74 & 46.69 & 0.25 & 24.47 & -10.94 & 46.48 & 0.63 \\
0222.6+4302 & 3C66A            & ISP & 0.444 & 0.11 & 14.46 & -10.56 & 46.31 & 0.04 & 23.44 & -10.68 & 46.19 & 0.76 \\
0238.6+1637 & PKS0235+164      & LSP & 0.94  & 0.19 & 12.69 & -11.31 & 46.36 & 0.07 & 21.79 & -10.57 & 47.1  & 5.51 \\
0303.4-2407 & PKS0301-243      & ISP & 0.26  & 0.13 & 14.78 & -10.96 & 45.37 & 0.11 & 24.22 & -11.05 & 45.27 & 0.8  \\
0334.2-4008 & PKS0332-403      & LSP & 1.445 & 0.2  & 12.81 & -11.16 & 46.97 & 0.07 & 21.93 & -11.11 & 47.02 & 1.12 \\
0428.6-3756 & PKS0426-380      & LSP & 1.11  & 0.21 & 12.94 & -10.97 & 46.88 & 0.08 & 22.47 & -10.43 & 47.42 & 3.45 \\
0449.4-4350 & PKS0447-439      & HSP & 0.205 & 0.13 & 15.38 & -10.36 & 45.73 & 0.15 & 24.37 & -10.61 & 45.49 & 0.57 \\
0507.9+6737 & 1ES0502+675      & HSP & 0.416 & 0.12 & 16.79 & -10.73 & 46.08 & 0.12 & 26.15 & -10.94 & 45.86 & 0.61 \\
0516.7-6207 & PKS0516-621      & LSP & 1.3   & 0.22 & 12.64 & -11.54 & 46.47 & 0.06 & 22.16 & -11.36 & 46.65 & 1.52 \\
0538.8-4405 & PKS0537-441      & LSP & 0.894 & 0.18 & 13.44 & -10.21 & 47.4  & 0.07 & 22.21 & -10.48 & 47.13 & 0.54 \\
0700.5-6610 & PKS0700-661      & ISP & 0.26  & 0.14 & 14.4  & -10.9  & 45.43 & 0.19 & 23.5  & -11.11 & 45.22 & 0.61 \\
0712.7+5033 & GB6J0712+5033    & LSP & 0.502 & 0.2  & 13.15 & -11.41 & 45.59 & 0.06 & 22.15 & -11.55 & 45.45 & 0.73 \\
0721.9+7120 & S50716+714       & ISP & 0.31  & 0.14 & 14.31 & -10.14 & 46.36 & 0.07 & 22.49 & -10.38 & 46.13 & 0.58 \\
0738.1+1742 & PKS0735+17       & ISP & 0.45  & 0.09 & 14.27 & -11.16 & 45.73 & 0.06 & 23.05 & -11.02 & 45.86 & 1.36 \\
0818.2+4222 & S40814+425       & LSP & 0.53  & 0.17 & 12.85 & -11.44 & 45.62 & 0.07 & 22.66 & -11.01 & 46.05 & 2.7  \\
1015.0+4926 & 1H1013+498       & HSP & 0.212 & 0.09 & 15.44 & -10.87 & 45.25 & 0.11 & 24.46 & -10.81 & 45.32 & 1.17 \\
1054.5+2211 & SDSSJ105430.62+2 & ISP & 2.05  & 0.12 & 14.48 & -11.86 & 46.65 & 0.05 & 22.46 & -11.58 & 46.92 & 1.88 \\
1058.4+0133 & 4C01.28          & LSP & 0.89  & 0.2  & 12.77 & -10.9  & 46.71 & 0.07 & 21.93 & -10.84 & 46.77 & 1.16 \\
1058.6+5627 & RXJ10586+5628    & HSP & 0.143 & 0.11 & 15.04 & -10.98 & 44.76 & 0.13 & 24.0  & -11.24 & 44.51 & 0.56 \\
1058.6-8003 & PKS1057-79       & LSP & 0.581 & 0.21 & 12.82 & -11.05 & 46.1  & 0.09 & 21.28 & -11.07 & 46.08 & 0.96 \\
1104.4+3812 & MKN421           & HSP & 0.03  & 0.16 & 16.62 & -9.34  & 44.99 & 0.19 & 24.68 & -9.9   & 44.43 & 0.28 \\
1147.0-3812 & PKS1144-379      & LSP & 1.048 & 0.2  & 12.6  & -11.34 & 46.44 & 0.06 & 21.75 & -11.2  & 46.58 & 1.37 \\
1217.9+3007 & ON325            & HSP & 0.13  & 0.1  & 14.93 & -10.95 & 44.71 & 0.15 & 24.03 & -10.73 & 44.92 & 1.64 \\
1221.5+2814 & ON231            & ISP & 0.102 & 0.12 & 14.43 & -10.85 & 44.58 & 0.07 & 22.4  & -11.13 & 44.3  & 0.52 \\
1248.3+5820 & PG1246+586       & ISP & 0.847 & 0.11 & 14.69 & -10.99 & 46.56 & 0.16 & 24.03 & -10.99 & 46.56 & 1.0  \\
1253.2+5301 & S41250+53        & ISP & 0.445 & 0.11 & 14.23 & -11.38 & 45.49 & 0.22 & 23.66 & -11.18 & 45.7  & 1.61 \\
1427.0+2348 & PG1424+240       & HSP & 0.16  & 0.13 & 14.92 & -10.5  & 45.35 & 0.16 & 24.43 & -10.65 & 45.21 & 0.72 \\
1517.7-2422 & APLIB            & ISP & 0.048 & 0.11 & 14.09 & -10.73 & 44.01 & 0.05 & 22.05 & -10.84 & 43.9  & 0.78 \\
1543.0+6130 & 1RXSJ154256.6+6  & ISP & 0.117 & 0.14 & 14.39 & -11.16 & 44.4  & 0.17 & 23.78 & -11.0  & 44.55 & 1.44 \\
1555.7+1111 & PG1553+113       & HSP & 0.36  & 0.1  & 15.24 & -10.59 & 46.07 & 0.19 & 24.84 & -10.37 & 46.29 & 1.66 \\
1653.8+3945 & MKN501           & HSP & 0.034 & 0.06 & 16.56 & -10.3  & 44.13 & 0.1  & 25.09 & -10.57 & 43.86 & 0.54 \\
1751.5+0938 & OT081            & LSP & 0.322 & 0.28 & 12.52 & -10.84 & 45.7  & 0.07 & 21.61 & -10.82 & 45.73 & 1.06 \\
1800.6+7828 & S51803+784       & LSP & 0.68  & 0.13 & 13.62 & -10.86 & 46.46 & 0.07 & 21.89 & -10.82 & 46.5  & 1.09 \\
2000.0+6508 & 1ES1959+650      & HSP & 0.047 & 0.09 & 16.61 & -10.23 & 44.5  & 0.12 & 24.77 & -10.65 & 44.08 & 0.38 \\
2009.4-4849 & 1Jy2005-489      & HSP & 0.071 & 0.1  & 15.31 & -10.42 & 44.68 & 0.18 & 24.43 & -11.18 & 43.92 & 0.18 \\
2139.4-4235 & MH2136-428       & ISP & 0.28  & 0.17 & 14.22 & -10.9  & 45.51 & 0.06 & 23.37 & -11.03 & 45.38 & 0.74 \\
2158.8-3013 & PKS2155-304      & HSP & 0.116 & 0.1  & 15.73 & -9.97  & 45.58 & 0.1  & 24.5  & -10.41 & 45.14 & 0.36 \\
2202.7+4216 & BLLAC            & ISP & 0.069 & 0.15 & 13.81 & -10.16 & 44.92 & 0.06 & 21.75 & -10.16 & 44.91 & 0.99 \\
\enddata
\tablecomments{
Column (1) gives the source name (4FGL J).
Column (2) gives the source association.
Column (3) gives the SED type of source based on spectral fitting.
Column (4) gives the redshift of the source.
Column (5-8) gives the curvature, peak frequency, peak flux, and peak luminosity of the synchrotron bump, respectively.
Column (9-12) gives the curvature, peak frequency, peak flux, and peak luminosity of the IC bump, respectively.
Column (13) shows Compton dominance (ratio of IC to synchrotron peak flux).
}
\end{deluxetable}


\startlongtable
\begin{deluxetable}{ccccccccccccc}
\tabletypesize{\footnotesize}
\tablecaption{Parameters of Log-parabolic Polynomial Fitting for FSRQs.}  
\label{tab:LP-FQ}
\tablewidth{0pt}
\tablehead{
\colhead{Name} & 
\colhead{Association} & 
\colhead{SED} & 
\colhead{$z$} & 
\colhead{$b_s$} & 
\colhead{$\nu_s$} & 
\colhead{$\nu_sF_s$} & 
\colhead{$\nu_sL_s$} & 
\colhead{$b_c$} & 
\colhead{$\nu_c$} & 
\colhead{$\nu_cF_c$} & 
\colhead{$\nu_cL_c$} & 
\colhead{$C_D$} \\
\colhead{(1)} & 
\colhead{(2)} & 
\colhead{(3)} & 
\colhead{(4)} & 
\colhead{(5)} & 
\colhead{(6)} & 
\colhead{(7)} & 
\colhead{(8)} & 
\colhead{(9)} & 
\colhead{(10)} & 
\colhead{(11)} & 
\colhead{(12)} & 
\colhead{(13)}
}
\startdata
0017.5-0514 & PMNJ0017-0512 & LSP & 0.227 & 0.2  & 13.04 & -11.67 & 44.53 & 0.1  & 20.87 & -11.12 & 45.07 & 3.48  \\
0051.1-0648 & PKS0048-071   & LSP & 1.975 & 0.26 & 12.17 & -11.69 & 46.77 & 0.08 & 21.4  & -11.42 & 47.05 & 1.89  \\
0118.9-2141 & PKS0116-219   & LSP & 1.165 & 0.22 & 12.66 & -11.52 & 46.38 & 0.1  & 21.91 & -11.01 & 46.88 & 3.18  \\
0137.0+4751 & OC457         & LSP & 0.859 & 0.22 & 12.62 & -11.03 & 46.54 & 0.08 & 21.45 & -10.95 & 46.62 & 1.22  \\
0145.0-2732 & PKS0142-278   & LSP & 1.148 & 0.22 & 12.48 & -11.61 & 46.27 & 0.12 & 20.82 & -11.09 & 46.79 & 3.3   \\
0205.0-1700 & PKS0202-17    & LSP & 1.74  & 0.18 & 12.37 & -11.86 & 46.47 & 0.13 & 20.94 & -10.83 & 47.5  & 10.72 \\
0210.7-5101 & PKS0208-512   & LSP & 1.003 & 0.15 & 12.85 & -11.24 & 46.5  & 0.1  & 21.65 & -10.44 & 47.3  & 6.35  \\
0217.8+0144 & PKS0215+015   & LSP & 1.715 & 0.21 & 12.8  & -11.2  & 47.12 & 0.07 & 22.07 & -11.09 & 47.23 & 1.29  \\
0221.1+3556 & B20218+357    & LSP & 0.944 & 0.18 & 12.82 & -11.56 & 46.11 & 0.1  & 22.08 & -10.64 & 47.03 & 8.29  \\
0229.5-3644 & PKS0227-369   & LSP & 2.115 & 0.17 & 13.04 & -11.98 & 46.55 & 0.12 & 20.96 & -11.05 & 47.48 & 8.51  \\
0237.8+2848 & 4C28.07       & LSP & 1.213 & 0.22 & 12.5  & -11.22 & 46.72 & 0.1  & 21.78 & -10.39 & 47.55 & 6.69  \\
0245.9-4650 & PKS0244-470   & LSP & 1.385 & 0.23 & 12.59 & -11.5  & 46.59 & 0.11 & 21.27 & -10.85 & 47.24 & 4.49  \\
0349.8-2103 & PKS0347-211   & LSP & 2.944 & 0.26 & 12.06 & -12.16 & 46.73 & 0.13 & 21.31 & -11.05 & 47.83 & 12.76 \\
0407.0-3826 & PKS0405-385   & LSP & 1.285 & 0.21 & 12.38 & -11.67 & 46.33 & 0.1  & 21.58 & -10.99 & 47.01 & 4.78  \\
0423.3-0120 & PKS0420-01    & LSP & 0.916 & 0.28 & 12.64 & -10.55 & 47.09 & 0.08 & 21.28 & -10.88 & 46.76 & 0.46  \\
0457.0-2324 & PKS0454-234   & LSP & 1.003 & 0.18 & 12.74 & -11.28 & 46.46 & 0.1  & 22.14 & -10.32 & 47.42 & 9.19  \\
0530.9+1332 & PKS0528+134   & LSP & 2.06  & 0.22 & 11.95 & -11.77 & 46.74 & 0.11 & 20.94 & -10.78 & 47.73 & 9.79  \\
0654.4+4514 & S40650+45     & LSP & 0.933 & 0.21 & 12.7  & -11.68 & 45.97 & 0.07 & 21.91 & -11.44 & 46.21 & 1.74  \\
0730.3-1141 & PKS0727-11    & LSP & 1.591 & 0.24 & 12.34 & -10.96 & 47.27 & 0.1  & 21.89 & -10.43 & 47.8  & 3.39  \\
0824.7+5552 & OJ535         & LSP & 1.417 & 0.15 & 12.43 & -12.09 & 46.02 & 0.1  & 20.99 & -11.28 & 46.83 & 6.48  \\
0920.9+4441 & S40917+44     & LSP & 2.19  & 0.21 & 12.46 & -11.59 & 46.98 & 0.1  & 21.33 & -10.71 & 47.86 & 7.61  \\
0948.9+0022 & PMNJ0948+0022 & LSP & 0.585 & 0.28 & 12.61 & -11.52 & 45.64 & 0.14 & 20.9  & -10.35 & 46.81 & 14.86 \\
0957.6+5523 & 4C55.17       & LSP & 0.896 & 0.1  & 13.39 & -11.69 & 45.93 & 0.06 & 22.98 & -10.84 & 46.78 & 7.04  \\
1012.7+2439 & B21011+25     & LSP & 1.634 & 0.15 & 13.13 & -11.7  & 46.56 & 0.06 & 21.91 & -10.93 & 47.33 & 5.93  \\
1016.0+0512 & TXS1013+054   & LSP & 1.713 & 0.22 & 12.36 & -12.12 & 46.19 & 0.08 & 22.11 & -11.5  & 46.81 & 4.13  \\
1033.9+6050 & S41030+61     & LSP & 1.401 & 0.17 & 13.57 & -11.01 & 47.09 & 0.08 & 21.89 & -10.89 & 47.2  & 1.3   \\
1129.8-1447 & PKS1127-145   & LSP & 1.184 & 0.15 & 12.5  & -11.49 & 46.43 & 0.12 & 20.51 & -10.68 & 47.24 & 6.5   \\
1159.5+2914 & 4C29.45       & LSP & 0.729 & 0.18 & 13.07 & -11.06 & 46.33 & 0.09 & 22.19 & -10.58 & 46.81 & 3.02  \\
1229.0+0202 & 3C273         & LSP & 0.158 & 0.11 & 13.47 & -10.07 & 45.77 & 0.1  & 20.02 & -9.7   & 46.14 & 2.35  \\
1246.7-2548 & PKS1244-255   & LSP & 0.635 & 0.21 & 12.7  & -11.22 & 46.03 & 0.1  & 21.63 & -10.59 & 46.66 & 4.24  \\
1256.1-0547 & 3C279         & LSP & 0.536 & 0.23 & 12.38 & -10.48 & 46.59 & 0.09 & 21.75 & -9.95  & 47.12 & 3.35  \\
1310.5+3221 & 1Jy1308+326   & LSP & 0.997 & 0.22 & 12.51 & -11.24 & 46.49 & 0.08 & 21.5  & -10.9  & 46.83 & 2.18  \\
1332.0-0509 & PKS1329-049   & LSP & 2.15  & 0.21 & 12.93 & -11.28 & 47.27 & 0.12 & 21.03 & -10.6  & 47.95 & 4.76  \\
1354.8-1041 & PKS1352-104   & LSP & 0.332 & 0.2  & 12.6  & -11.69 & 44.88 & 0.07 & 20.55 & -11.21 & 45.36 & 3.03  \\
1457.4-3539 & PKS1454-354   & LSP & 1.424 & 0.22 & 12.53 & -11.57 & 46.54 & 0.1  & 21.57 & -10.92 & 47.2  & 4.51  \\
1504.4+1029 & PKS1502+106   & LSP & 1.84  & 0.14 & 12.85 & -11.61 & 46.78 & 0.11 & 21.99 & -10.27 & 48.12 & 22.21 \\
1510.8-0542 & PKS1508-05    & LSP & 1.191 & 0.1  & 13.1  & -11.79 & 46.14 & 0.1  & 21.4  & -11.11 & 46.81 & 4.75  \\
1512.8-0906 & PKS1510-08    & LSP & 0.36  & 0.15 & 13.2  & -11.05 & 45.61 & 0.09 & 21.77 & -9.96  & 46.7  & 12.3  \\
1522.1+3144 & B21520+31     & LSP & 1.487 & 0.19 & 12.38 & -12.24 & 45.92 & 0.14 & 21.87 & -10.34 & 47.82 & 79.39 \\
1553.6+1257 & PKS1551+130   & LSP & 1.29  & 0.21 & 12.27 & -12.02 & 45.98 & 0.09 & 21.63 & -11.26 & 46.75 & 5.87  \\
1625.7-2527 & PKS1622-253   & LSP & 0.786 & 0.19 & 12.54 & -11.34 & 46.13 & 0.12 & 21.77 & -10.71 & 46.77 & 4.33  \\
1635.2+3808 & 4C38.41       & LSP & 1.814 & 0.2  & 12.48 & -11.13 & 47.24 & 0.12 & 21.49 & -10.19 & 48.18 & 8.79  \\
1640.4+3945 & NRAO512       & LSP & 1.666 & 0.16 & 12.21 & -12.22 & 46.07 & 0.1  & 21.67 & -11.1  & 47.18 & 13.0  \\
1848.4+3217 & B21846+32A    & LSP & 0.789 & 0.21 & 12.68 & -11.64 & 45.84 & 0.1  & 21.65 & -11.12 & 46.36 & 3.33  \\
1849.2+6705 & 4C66.20       & LSP & 0.657 & 0.24 & 12.83 & -11.04 & 46.25 & 0.08 & 21.56 & -10.88 & 46.41 & 1.45  \\
1911.2-2006 & 1908-201      & LSP & 1.119 & 0.19 & 12.83 & -11.02 & 46.83 & 0.12 & 21.41 & -10.58 & 47.27 & 2.78  \\
1923.5-2104 & PMNJ1923-2104 & LSP & 0.874 & 0.14 & 13.26 & -11.0  & 46.59 & 0.07 & 21.44 & -11.18 & 46.4  & 0.66  \\
2025.6-0735 & PKS2023-07    & LSP & 1.388 & 0.19 & 12.65 & -11.5  & 46.59 & 0.1  & 21.93 & -10.57 & 47.51 & 8.48  \\
2056.2-4714 & PKS2052-47    & LSP & 1.489 & 0.19 & 12.64 & -11.25 & 46.92 & 0.12 & 21.31 & -10.41 & 47.76 & 6.92  \\
2143.5+1743 & S32141+17     & LSP & 0.213 & 0.16 & 13.39 & -11.31 & 44.82 & 0.12 & 21.45 & -10.53 & 45.6  & 6.0   \\
2147.1+0931 & 1Jy2144+092   & LSP & 1.113 & 0.24 & 12.12 & -11.94 & 45.91 & 0.11 & 21.2  & -10.9  & 46.95 & 10.94 \\
2157.5+3127 & B22155+31     & LSP & 1.486 & 0.21 & 12.78 & -11.63 & 46.53 & 0.1  & 21.73 & -11.05 & 47.11 & 3.78  \\
2203.4+1725 & PKS2201+171   & LSP & 1.076 & 0.21 & 13.0  & -11.07 & 46.74 & 0.08 & 21.68 & -11.08 & 46.73 & 0.97  \\
2207.5-5346 & PKS2204-54    & LSP & 1.206 & 0.13 & 12.63 & -11.94 & 45.99 & 0.08 & 21.01 & -11.33 & 46.61 & 4.13  \\
2229.7-0832 & PKS2227-08    & LSP & 1.56  & 0.25 & 12.38 & -11.26 & 46.95 & 0.11 & 21.09 & -10.68 & 47.53 & 3.8   \\
2232.6+1143 & CTA102        & LSP & 1.037 & 0.19 & 12.26 & -11.4  & 46.37 & 0.12 & 21.7  & -9.81  & 47.97 & 39.43 \\
2253.9+1609 & 3C454.3       & LSP & 0.859 & 0.13 & 13.36 & -10.46 & 47.11 & 0.13 & 21.59 & -9.5   & 48.07 & 9.03  \\
2327.5+0939 & PKS2325+093   & LSP & 1.843 & 0.25 & 12.15 & -11.85 & 46.55 & 0.13 & 20.84 & -10.73 & 47.66 & 12.99 \\
2345.2-1555 & PMNJ2345-1555 & LSP & 0.621 & 0.22 & 12.92 & -11.26 & 45.96 & 0.09 & 22.44 & -10.76 & 46.46 & 3.16 \\
\enddata
\tablecomments{
Column (1-13) are same as in \ref{tab:LP-BL}.
}
\end{deluxetable}


\startlongtable
\begin{deluxetable}{cccccccccccc}
\tabletypesize{\footnotesize}
\tablecaption{One-zone IC Model Jet Parameters of BL Lacs.}  
\label{tab:FIT-BL}
\tablewidth{0pt}
\tablehead{
\colhead{Name} & 
\colhead{SED} & 
\colhead{$z$} & 
\colhead{$R_{\rm 17}$} & 
\colhead{$B$} &  
\colhead{$\delta$} & 
\colhead{$n_e$} & 
\colhead{$\gamma_{\rm min,2}$} & 
\colhead{$\gamma_0$} & 
\colhead{$\gamma_{\rm max,6}$} & 
\colhead{$s$} & 
\colhead{$r$} \\
\colhead{(1)} & 
\colhead{(2)} & 
\colhead{(3)} & 
\colhead{(4)} & 
\colhead{(5)} & 
\colhead{(6)} & 
\colhead{(7)} & 
\colhead{(8)} & 
\colhead{(9)} & 
\colhead{(10)} &
\colhead{(11)} &
\colhead{(12)} 
}
\startdata
0033.5-1921 & HSP & 0.61  & 8.0   & 0.018 & 15.04 & 0.01  & 4.5   & 19977.0 & 1.3   & 2.13 & 0.57 \\
0050.7-0929 & HSP & 0.634 & 2.74  & 0.033 & 22.44 & 1.37  & 1.7   & 6002.0  & 9.4   & 2.58 & 0.33 \\
0120.4-2701 & ISP & 0.56  & 4.8   & 0.005 & 44.09 & 0.16  & 3.5   & 427.0   & 1.1   & 1.27 & 0.48 \\
0136.5+3906 & HSP & 0.75  & 2.17  & 0.087 & 20.0  & 0.07  & 1.0   & 17108.0 & 4.6   & 2.27 & 0.63 \\
0222.6+4302 & ISP & 0.444 & 5.29  & 0.006 & 39.13 & 0.12  & 2.0   & 5126.0  & 2.4   & 2.07 & 0.63 \\
0238.6+1637 & LSP & 0.94  & 3.64  & 0.072 & 13.74 & 13.68 & 0.4   & 370.0   & 0.0   & 1.82 & 0.83 \\
0303.4-2407 & ISP & 0.26  & 4.09  & 0.004 & 31.86 & 0.03  & 5.0   & 8252.0  & 4.4   & 1.88 & 0.7  \\
0334.2-4008 & LSP & 1.445 & 8.38  & 0.063 & 17.09 & 1.15  & 0.4   & 448.0   & 0.0   & 1.62 & 0.86 \\
0428.6-3756 & LSP & 1.11  & 5.45  & 0.091 & 17.05 & 2.04  & 0.1   & 1493.0  & 0.0   & 2.74 & 0.94 \\
0449.4-4350 & HSP & 0.205 & 1.52  & 0.024 & 27.24 & 0.18  & 1.6   & 20851.0 & 1.6   & 2.63 & 0.63 \\
0507.9+6737 & HSP & 0.416 & 5.0   & 0.007 & 26.43 & 0.0   & 3.6   & 14347.0 & 3.9   & 1.42 & 0.55 \\
0516.7-6207 & LSP & 1.3   & 8.12  & 0.078 & 10.65 & 4.91  & 0.4   & 1142.0  & 0.0   & 2.63 & 0.51 \\
0538.8-4405 & LSP & 0.894 & 14.75 & 0.049 & 19.19 & 0.4   & 0.2   & 905.0   & 0.0   & 2.01 & 0.64 \\
0700.5-6610 & ISP & 0.26  & 2.22  & 0.012 & 26.17 & 0.29  & 2.8   & 58.0    & 1.0   & -0.5 & 0.69 \\
0712.7+5033 & LSP & 0.502 & 2.73  & 0.139 & 9.28  & 9.18  & 0.1   & 1181.0  & 0.0   & 2.43 & 0.5  \\
0721.9+7120 & ISP & 0.31  & 7.43  & 0.014 & 23.2  & 0.24  & 2.0   & 266.0   & 5.7   & 1.01 & 0.56 \\
0738.1+1742 & ISP & 0.45  & 2.92  & 0.008 & 29.25 & 0.83  & 5.0   & 1468.0  & 0.4   & 2.02 & 0.44 \\
0818.2+4222 & LSP & 0.53  & 7.5   & 0.013 & 18.14 & 1.64  & 0.5   & 971.0   & 0.0   & 2.33 & 0.46 \\
1015.0+4926 & HSP & 0.212 & 0.61  & 0.019 & 33.35 & 0.76  & 1.8   & 16606.0 & 0.6   & 2.44 & 0.55 \\
1054.5+2211 & ISP & 2.05  & 0.39  & 0.233 & 20.75 & 60.63 & 0.5   & 140.0   & 0.8   & 0.77 & 0.63 \\
1058.4+0133 & LSP & 0.89  & 8.74  & 0.049 & 17.8  & 3.72  & 0.0   & 600.0   & 0.0   & 2.32 & 0.61 \\
1058.6+5627 & HSP & 0.143 & 1.11  & 0.035 & 16.77 & 0.15  & 3.0   & 10880.0 & 1.9   & 2.28 & 0.8  \\
1058.6-8003 & LSP & 0.581 & 8.5   & 0.038 & 12.35 & 1.58  & 0.3   & 1166.0  & 0.0   & 2.38 & 0.74 \\
1104.4+3812 & HSP & 0.03  & 0.38  & 0.036 & 25.15 & 0.13  & 1.4   & 15494.0 & 2.4   & 1.94 & 0.58 \\
1147.0-3812 & LSP & 1.048 & 6.64  & 0.032 & 18.3  & 4.7   & 0.1   & 864.0   & 0.0   & 2.44 & 0.75 \\
1217.9+3007 & HSP & 0.13  & 2.8   & 0.008 & 16.03 & 0.24  & 4.2   & 315.0   & 6.7   & 0.81 & 0.49 \\
1221.5+2814 & ISP & 0.102 & 0.42  & 0.03  & 26.78 & 3.61  & 0.8   & 183.0   & 3.7   & 0.51 & 0.68 \\
1248.3+5820 & ISP & 0.847 & 10.0  & 0.012 & 22.21 & 0.02  & 3.6   & 1782.0  & 2.2   & 1.1  & 0.68 \\
1253.2+5301 & ISP & 0.445 & 2.01  & 0.015 & 22.17 & 1.93  & 1.6   & 800.0   & 9.3   & 1.55 & 0.54 \\
1427.0+2348 & HSP & 0.16  & 9.0   & 0.002 & 28.83 & 0.01  & 4.1   & 1580.0  & 2.9   & 0.71 & 0.69 \\
1517.7-2422 & ISP & 0.048 & 1.99  & 0.011 & 12.13 & 1.56  & 3.2   & 501.0   & 4.3   & 1.79 & 0.4  \\
1543.0+6130 & ISP & 0.117 & 1.78  & 0.006 & 19.64 & 0.43  & 4.8   & 614.0   & 6.9   & 0.84 & 0.65 \\
1555.7+1111 & HSP & 0.36  & 9.0   & 0.002 & 38.11 & 0.01  & 4.2   & 34583.0 & 2.1   & 2.24 & 0.74 \\
1653.8+3945 & HSP & 0.034 & 1.72  & 0.003 & 26.42 & 0.12  & 2.2   & 17928.0 & 19.4  & 2.24 & 0.3  \\
1751.5+0938 & LSP & 0.322 & 3.92  & 0.084 & 9.51  & 23.86 & 0.2   & 722.0   & 0.0   & 2.64 & 0.53 \\
1800.6+7828 & LSP & 0.68  & 4.09  & 0.143 & 11.15 & 4.83  & 0.2   & 646.0   & 0.0   & 2.02 & 0.64 \\
2000.0+6508 & HSP & 0.047 & 0.14  & 0.066 & 29.0  & 0.34  & 1.4   & 15211.0 & 2.8   & 1.93 & 0.73 \\
2009.4-4849 & HSP & 0.071 & 8.99  & 0.007 & 15.89 & 0.01  & 3.1   & 25260.0 & 1.2   & 2.5  & 0.42 \\
2139.4-4235 & ISP & 0.28  & 4.47  & 0.005 & 28.63 & 0.06  & 5.3   & 1797.0  & 0.3   & 1.2  & 0.69 \\
2158.8-3013 & HSP & 0.116 & 2.77  & 0.026 & 20.0  & 0.06  & 1.0   & 21131.0 & 1.3   & 2.52 & 0.54 \\
2202.7+4216 & ISP & 0.069 & 0.66  & 0.146 & 10.77 & 25.13 & 0.3   & 1384.0  & 0.0   & 2.43 & 0.79 \\
\enddata
\tablecomments{
Column (1) gives the source name (4FGL J).
Column (2) gives the SED type of source.
Column (2) gives the redshift of the source.
Column (4) gives the source size in $10^{17}$ cm.
Column (8) gives the minimum energy of EED in $10^2$.
Column (9) gives the reference energy of EED peak energy.
Column (10) gives the maximum energy of EED in $10^6$.
}
\end{deluxetable}


\startlongtable
\begin{deluxetable}{ccccccccccccccc}
\tabletypesize{\footnotesize}
\tablecaption{One-zone IC Model Jet Parameters of FSRQs.}  
\label{tab:FIT-FQ}
\tablewidth{0pt}
\tablehead{
\colhead{Name} & 
\colhead{$z$} & 
\colhead{$R_{\rm 17}$} & 
\colhead{$B$} & 
\colhead{$n_e$} & 
\colhead{$\Gamma$} & 
\colhead{$\delta$} & 
\colhead{$\gamma_{\rm min}$} & 
\colhead{$\gamma_{\rm 0,2 }$} & 
\colhead{$\gamma_{\rm max, 4}$} & 
\colhead{$s$} & 
\colhead{$r$} &
\colhead{$R_{\rm b,18}$} &
\colhead{$R_{\rm T,19}$} &
\colhead{$L_{\rm D,45}$} \\
\colhead{(1)} & 
\colhead{(2)} & 
\colhead{(3)} & 
\colhead{(4)} & 
\colhead{(5)} & 
\colhead{(6)} & 
\colhead{(7)} & 
\colhead{(8)} & 
\colhead{(9)} & 
\colhead{(10)} &
\colhead{(11)} &
\colhead{(12)} &
\colhead{(13)} &
\colhead{(14)} &
\colhead{(15)}
}
\startdata
0017.5-0514 & 0.227 & 0.4  & 0.1  & 242.4 & 7.0   & 12.3 & 2.8  & 8.1     & 5.0   & 2.46 & 0.74 & 0.4  & 0.4  & 0.33   \\
0051.1-0648 & 1.975 & 7.2  & 0.08 & 15.6  & 8.84  & 14.5 & 24.9 & 3.8     & 9.0   & 2.6  & 0.51 & 6.8  & 3.3  & 27.97  \\
0118.9-2141 & 1.165 & 3.8  & 0.11 & 4.2   & 28.0  & 17.8 & 25.2 & 7.8     & 5.0   & 2.65 & 0.72 & 4.5  & 1.8  & 7.69   \\
0137.0+4751 & 0.859 & 8.5  & 0.1  & 4.2   & 6.0   & 10.9 & 39.0 & 5.2     & 7.0   & 2.46 & 0.57 & 8.4  & 1.7  & 7.04   \\
0145.0-2732 & 1.148 & 4.8  & 0.09 & 39.3  & 6.52  & 11.6 & 2.2  & 5.0     & 2.0   & 2.49 & 0.58 & 4.6  & 2.8  & 19.32  \\
0205.0-1700 & 1.74  & 4.0  & 0.07 & 47.2  & 30.0  & 17.3 & 3.3  & 3.5     & 3.0   & 2.36 & 0.6  & 3.8  & 6.5  & 106.02 \\
0210.7-5101 & 1.003 & 6.6  & 0.07 & 9.9   & 8.06  & 13.7 & 3.5  & 7.0     & 6.0   & 2.45 & 0.62 & 6.3  & 3.1  & 23.87  \\
0217.8+0144 & 1.715 & 8.7  & 0.13 & 1.9   & 9.24  & 15.0 & 13.5 & 6.8     & 4.0   & 2.45 & 0.82 & 8.9  & 14.8 & 545.24 \\
0221.1+3556 & 0.944 & 4.5  & 0.06 & 11.5  & 27.79 & 15.8 & 12.4 & 5.5     & 8.0   & 2.42 & 0.53 & 4.3  & 6.2  & 96.07  \\
0229.5-3644 & 2.115 & 3.0  & 0.08 & 70.7  & 15.0  & 18.6 & 9.4  & 3.5     & 6.0   & 2.42 & 0.56 & 3.9  & 3.2  & 26.14  \\
0237.8+2848 & 1.213 & 7.4  & 0.07 & 4.8   & 36.26 & 15.8 & 30.1 & 5.4     & 6.0   & 2.4  & 0.6  & 7.6  & 2.3  & 13.46  \\
0245.9-4650 & 1.385 & 3.8  & 0.1  & 28.5  & 9.19  & 14.9 & 10.6 & 4.5     & 10.0  & 2.38 & 0.67 & 3.6  & 3.9  & 37.77  \\
0349.8-2103 & 2.944 & 2.3  & 0.08 & 35.9  & 45.0  & 26.0 & 33.2 & 4.2     & 5.0   & 2.61 & 0.67 & 2.3  & 5.1  & 63.88  \\
0407.0-3826 & 1.285 & 5.4  & 0.08 & 18.9  & 17.11 & 19.0 & 2.6  & 2.9     & 6.0   & 2.5  & 0.5  & 5.1  & 1.8  & 8.5    \\
0423.3-0120 & 0.916 & 13.9 & 0.15 & 1.5   & 5.59  & 10.2 & 22.8 & 7.5     & 4.0   & 2.56 & 0.74 & 13.3 & 2.4  & 15.0   \\
0457.0-2324 & 1.003 & 11.0 & 0.05 & 2.1   & 7.31  & 14.0 & 2.3  & 3.3     & 3.0   & 1.92 & 0.6  & 13.4 & 1.3  & 4.5    \\
0530.9+1332 & 2.06  & 16.8 & 0.04 & 23.5  & 7.0   & 12.3 & 2.5  & 2.7     & 7.0   & 2.37 & 0.49 & 16.0 & 5.2  & 67.24  \\
0654.4+4514 & 0.933 & 3.8  & 0.09 & 6.9   & 6.89  & 12.2 & 4.6  & 4.3     & 5.0   & 2.07 & 0.79 & 3.8  & 1.2  & 3.62   \\
0730.3-1141 & 1.591 & 13.6 & 0.08 & 1.6   & 10.0  & 17.8 & 47.3 & 3.5     & 7.0   & 2.46 & 0.63 & 16.4 & 10.1 & 253.4  \\
0824.7+5552 & 1.417 & 3.8  & 0.07 & 73.0  & 45.0  & 13.7 & 2.4  & 1.3     & 10.0  & 1.9  & 0.53 & 4.4  & 2.8  & 19.4   \\
0920.9+4441 & 2.19  & 9.2  & 0.06 & 6.6   & 8.77  & 14.5 & 14.8 & 4.7     & 5.0   & 2.15 & 0.69 & 8.9  & 5.0  & 62.98  \\
0948.9+0022 & 0.585 & 1.9  & 0.05 & 87.4  & 24.15 & 11.5 & 20.5 & 6.0     & 3.0   & 2.23 & 0.66 & 1.8  & 1.4  & 5.0    \\
0957.6+5523 & 0.896 & 8.5  & 0.04 & 9.0   & 5.99  & 10.0 & 2.1  & 11.9    & 10.0  & 2.37 & 0.4  & 8.2  & 1.9  & 8.7    \\
1012.7+2439 & 1.634 & 9.1  & 0.05 & 5.1   & 5.0   & 9.3  & 22.1 & 7.5     & 5.0   & 1.91 & 0.65 & 8.9  & 7.5  & 140.09 \\
1016.0+0512 & 1.713 & 9.0  & 0.06 & 2.4   & 10.0  & 15.7 & 3.5  & 1.6     & 10.0  & 1.81 & 0.61 & 8.8  & 2.8  & 20.0   \\
1033.9+6050 & 1.401 & 7.5  & 0.13 & 1.9   & 5.84  & 10.6 & 47.8 & 7.4     & 10.0  & 1.87 & 0.64 & 7.3  & 5.8  & 85.19  \\
1129.8-1447 & 1.184 & 6.1  & 0.1  & 28.3  & 6.0   & 10.9 & 24.4 & 1.1     & 4.0   & 1.85 & 0.51 & 5.8  & 4.5  & 51.32  \\
1159.5+2914 & 0.729 & 3.0  & 0.12 & 5.5   & 25.0  & 18.4 & 4.1  & 6.4     & 6.0   & 2.37 & 0.74 & 3.3  & 0.3  & 0.27   \\
1229.0+0202 & 0.158 & 4.8  & 0.2  & 55.3  & 7.78  & 3.7  & 19.7 & 4.0     & 3.0   & 2.04 & 0.58 & 4.8  & 3.4  & 28.86  \\
1246.7-2548 & 0.635 & 4.1  & 0.07 & 13.5  & 16.59 & 12.2 & 9.1  & 8.0     & 3.0   & 2.49 & 0.64 & 4.0  & 2.0  & 9.79   \\
1256.1-0547 & 0.536 & 4.7  & 0.08 & 38.6  & 13.38 & 17.4 & 11.1 & 2.7     & 2.0   & 2.49 & 0.58 & 4.5  & 0.4  & 0.5    \\
1310.5+3221 & 0.997 & 17.8 & 0.06 & 2.2   & 4.13  & 7.8  & 47.7 & 6.2     & 10.0  & 2.3  & 0.5  & 17.0 & 2.8  & 20.03  \\
1332.0-0509 & 2.15  & 4.6  & 0.12 & 8.4   & 12.77 & 17.7 & 15.6 & 4.3     & 6.0   & 2.04 & 1.07 & 4.1  & 5.0  & 61.67  \\
1354.8-1041 & 0.332 & 2.3  & 0.05 & 144.6 & 4.0   & 7.6  & 8.4  & 6.1     & 10.0  & 2.41 & 0.57 & 2.3  & 1.1  & 2.78   \\
1457.4-3539 & 1.424 & 6.0  & 0.08 & 6.8   & 8.13  & 13.7 & 21.8 & 5.8     & 5.0   & 2.34 & 0.62 & 5.9  & 5.2  & 68.87  \\
1504.4+1029 & 1.84  & 9.6  & 0.04 & 5.9   & 10.04 & 15.7 & 9.5  & 5.3     & 5.0   & 2.1  & 0.55 & 13.2 & 1.3  & 4.05   \\
1510.8-0542 & 1.191 & 8.4  & 0.13 & 1.1   & 17.0  & 19.0 & 6.5  & 3.9     & 42.0  & 2.5  & 0.4  & 9.6  & 4.5  & 49.73  \\
1512.8-0906 & 0.36  & 2.2  & 0.04 & 37.4  & 45.0  & 13.7 & 11.1 & 3.6     & 10.0  & 1.82 & 0.63 & 2.6  & 1.1  & 2.82   \\
1522.1+3144 & 1.487 & 1.7  & 0.03 & 22.9  & 23.85 & 27.9 & 23.8 & 2.3     & 6.0   & 1.81 & 0.73 & 1.7  & 1.3  & 4.26   \\
1553.6+1257 & 1.29  & 6.3  & 0.05 & 8.8   & 23.7  & 12.4 & 33.5 & 3.6     & 7.0   & 2.28 & 0.55 & 7.5  & 4.0  & 39.46  \\
1625.7-2527 & 0.786 & 6.6  & 0.09 & 1.0   & 9.98  & 15.7 & 50.0 & 2.1     & 4.0   & 1.93 & 0.75 & 6.3  & 0.6  & 1.0    \\
1635.2+3808 & 1.814 & 14.0 & 0.06 & 2.9   & 35.0  & 16.1 & 4.5  & 5.1     & 4.0   & 2.36 & 0.84 & 13.3 & 4.4  & 48.54  \\
1640.4+3945 & 1.666 & 4.4  & 0.04 & 30.1  & 19.49 & 14.2 & 15.0 & 4.9     & 8.0   & 2.35 & 0.52 & 4.3  & 3.2  & 25.3   \\
1848.4+3217 & 0.789 & 3.8  & 0.08 & 15.7  & 6.92  & 11.6 & 5.5  & 6.0     & 3.0   & 2.37 & 0.57 & 4.0  & 1.0  & 2.31   \\
1849.2+6705 & 0.657 & 5.2  & 0.11 & 6.9   & 5.38  & 9.9  & 24.2 & 5.2     & 10.0  & 2.14 & 0.53 & 7.4  & 2.0  & 10.34  \\
1911.2-2006 & 1.119 & 8.6  & 0.13 & 1.2   & 11.0  & 13.7 & 25.7 & 6.1     & 1.0   & 2.22 & 0.89 & 10.2 & 3.4  & 29.11  \\
1923.5-2104 & 0.874 & 10.0 & 0.15 & 2.6   & 4.58  & 8.6  & 12.6 & 6.4     & 4.0   & 2.23 & 0.57 & 9.6  & 3.7  & 33.38  \\
2025.6-0735 & 1.388 & 5.0  & 0.07 & 11.0  & 27.47 & 17.9 & 24.3 & 4.7     & 7.0   & 2.39 & 0.55 & 4.7  & 1.9  & 9.42   \\
2056.2-4714 & 1.489 & 5.5  & 0.1  & 4.3   & 14.0  & 18.2 & 18.4 & 5.9     & 2.0   & 2.4  & 0.98 & 7.8  & 2.4  & 13.89  \\
2143.5+1743 & 0.213 & 0.4  & 0.11 & 59.4  & 14.13 & 16.0 & 28.7 & 5.0     & 8.0   & 1.73 & 0.96 & 0.4  & 1.2  & 3.62   \\
2147.1+0931 & 1.113 & 4.2  & 0.05 & 68.0  & 8.8   & 14.5 & 4.1  & 3.1     & 9.0   & 2.41 & 0.55 & 4.1  & 2.5  & 15.02  \\
2157.5+3127 & 1.486 & 4.6  & 0.1  & 2.3   & 8.71  & 14.4 & 18.2 & 6.3     & 1.0   & 1.9  & 0.98 & 4.4  & 0.4  & 0.49   \\
2203.4+1725 & 1.076 & 4.1  & 0.18 & 2.2   & 9.4   & 15.1 & 8.9  & 8.5     & 10.0  & 2.47 & 1.04 & 3.9  & 0.4  & 0.38   \\
2207.5-5346 & 1.206 & 5.1  & 0.07 & 37.8  & 16.71 & 10.4 & 11.6 & 1.6     & 9.0   & 1.91 & 0.5  & 5.0  & 3.2  & 25.28  \\
2229.7-0832 & 1.56  & 5.4  & 0.08 & 12.0  & 15.0  & 18.6 & 24.4 & 4.0     & 6.0   & 2.38 & 0.74 & 5.3  & 3.5  & 31.42  \\
2232.6+1143 & 1.037 & 5.1  & 0.03 & 61.6  & 28.56 & 17.7 & 19.1 & 2.6     & 4.0   & 2.35 & 0.5  & 5.1  & 3.9  & 38.89  \\
2253.9+1609 & 0.859 & 11.2 & 0.07 & 1.6   & 38.0  & 15.3 & 14.6 & 2.1     & 8.0   & 1.24 & 0.78 & 13.0 & 4.1  & 41.67  \\
2327.5+0939 & 1.843 & 5.1  & 0.05 & 46.5  & 17.48 & 15.3 & 31.6 & 2.3     & 9.0   & 2.27 & 0.57 & 4.8  & 3.1  & 24.31  \\
2345.2-1555 & 0.621 & 3.4  & 0.09 & 2.2   & 35.0  & 16.1 & 17.9 & 5.0     & 2.0   & 2.06 & 0.74 & 3.6  & 1.5  & 5.75  \\
\enddata
\tablecomments{
Column (3) gives the source size $R$ in $10^{17}$ cm.
Column (8) gives the minimum energy of EED $\gamma_{\rm min}$ in $10^2$.
Column (10) gives the maximum energy $\gamma_{\rm max}$ of EED in $10^4$.
Column (13) gives the blob location $R_b$ in $10^{18}$ cm.
Column (14) gives the DT radius $R_T$ in $10^{19}$ cm.
Column (15) gives the disk luminosity in $10^{45}$ erg s$^{-1}$.
}
\end{deluxetable}


\bibliography{ms}{}
\bibliographystyle{ms}


\begin{figure}
    \centering
    \includegraphics[width=0.48\linewidth]{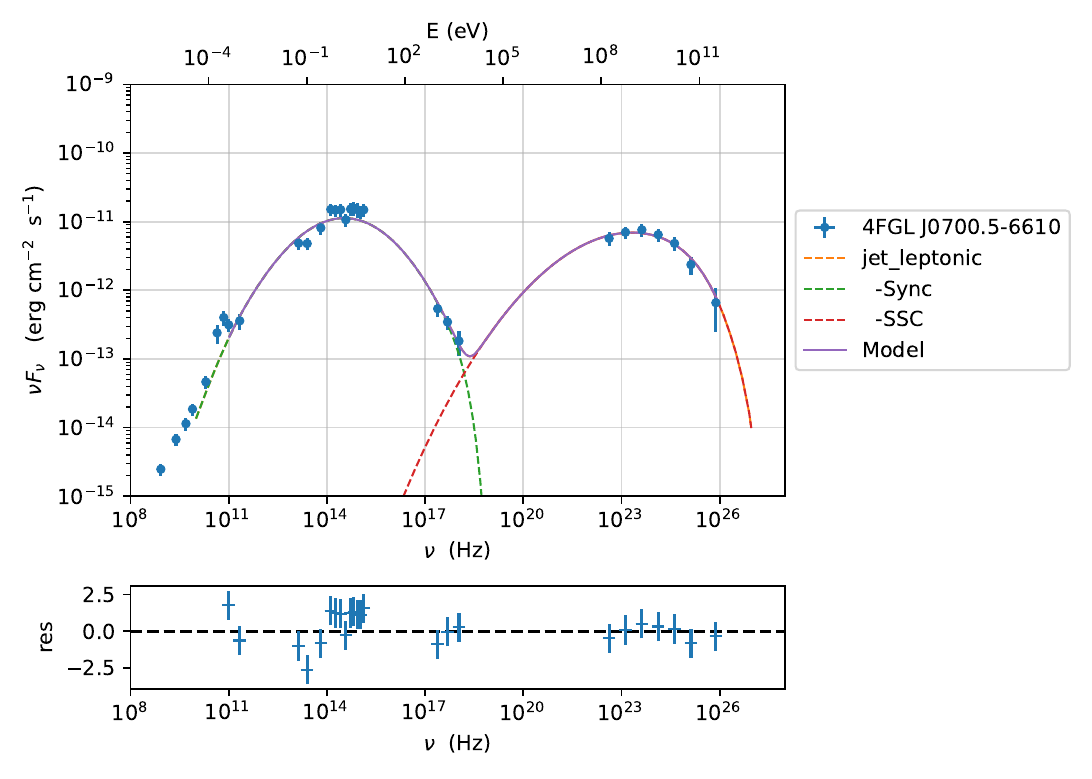}
    \includegraphics[width=0.48\linewidth]{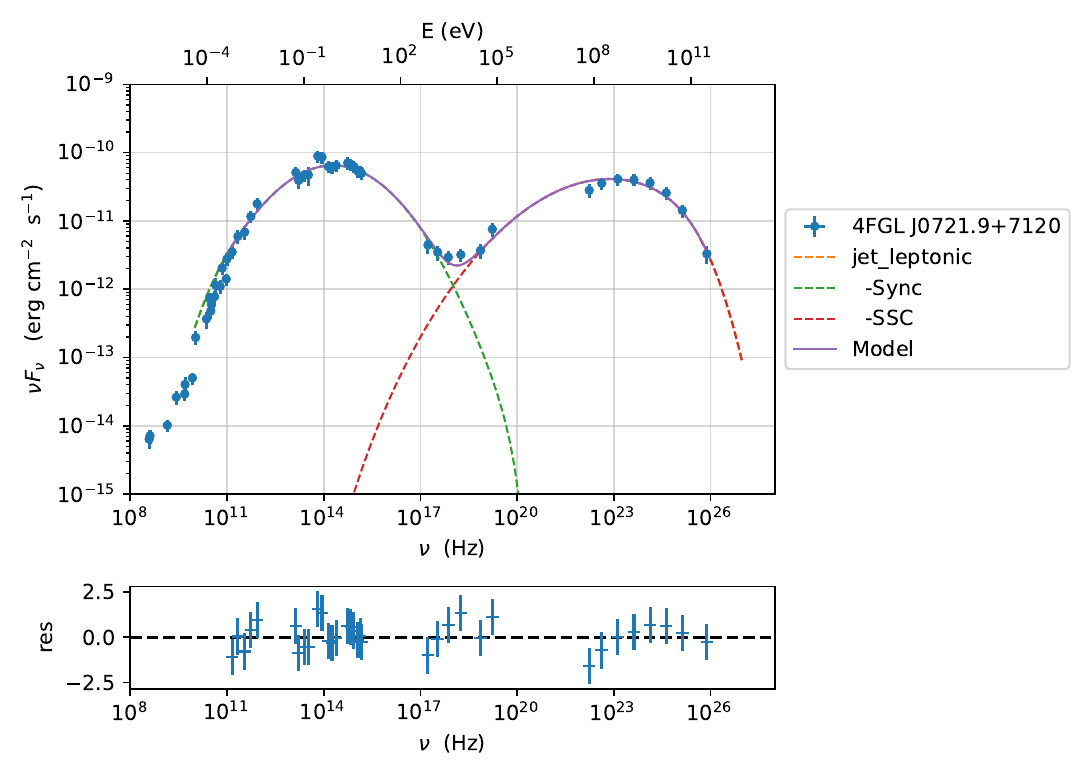}
    \includegraphics[width=0.48\linewidth]{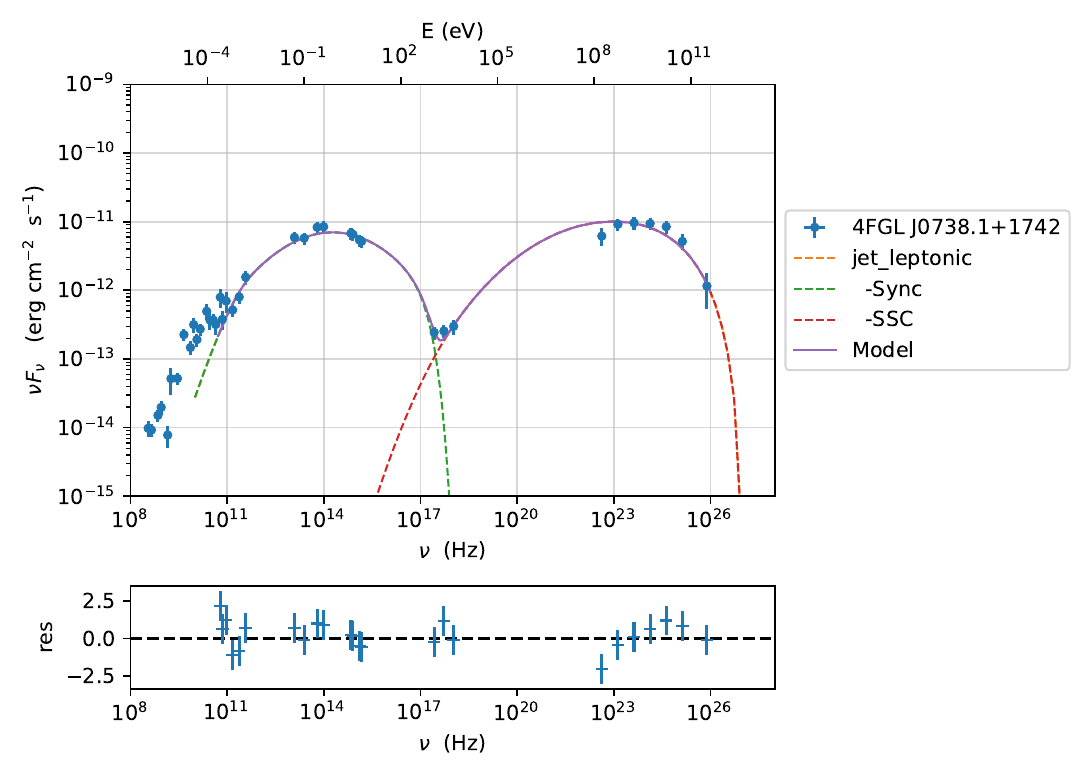}
    \includegraphics[width=0.48\linewidth]{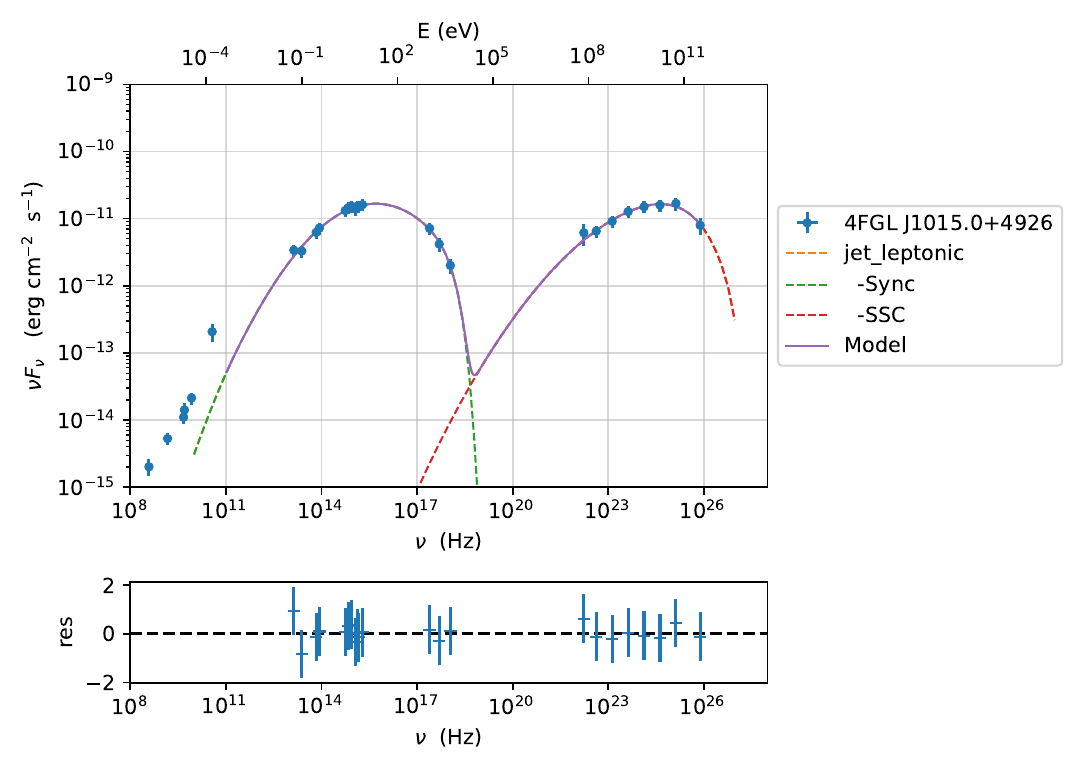}
    \includegraphics[width=0.48\linewidth]{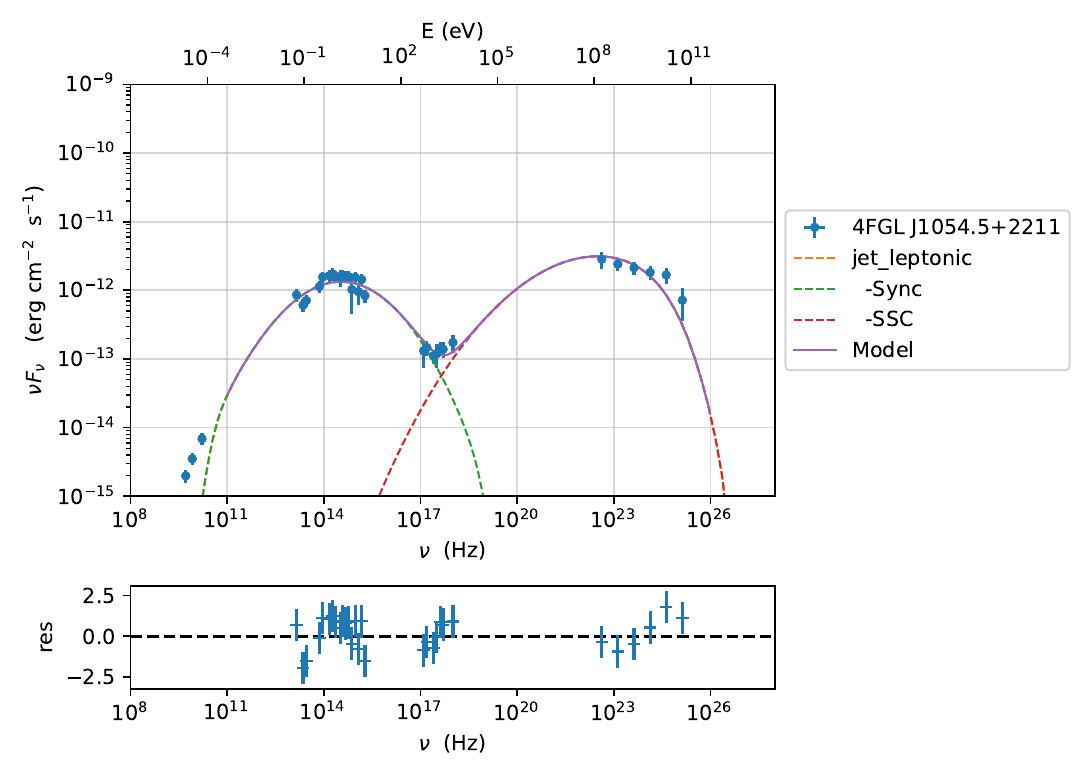}
    \includegraphics[width=0.48\linewidth]{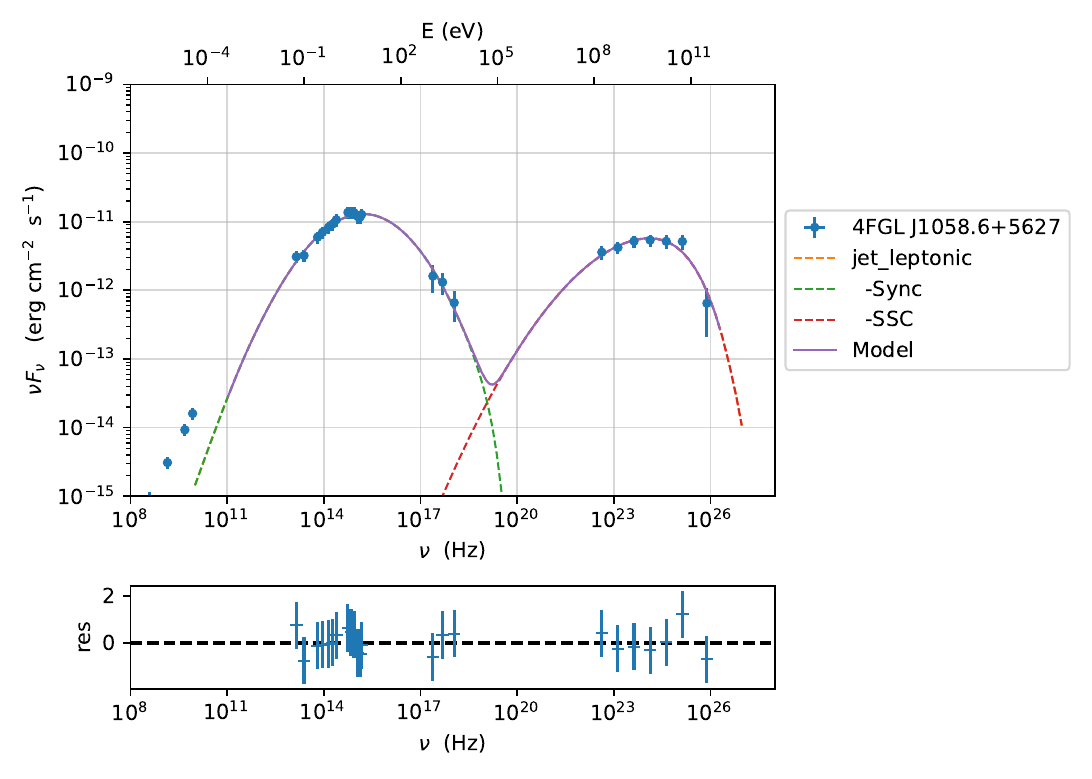}
    \includegraphics[width=0.48\linewidth]{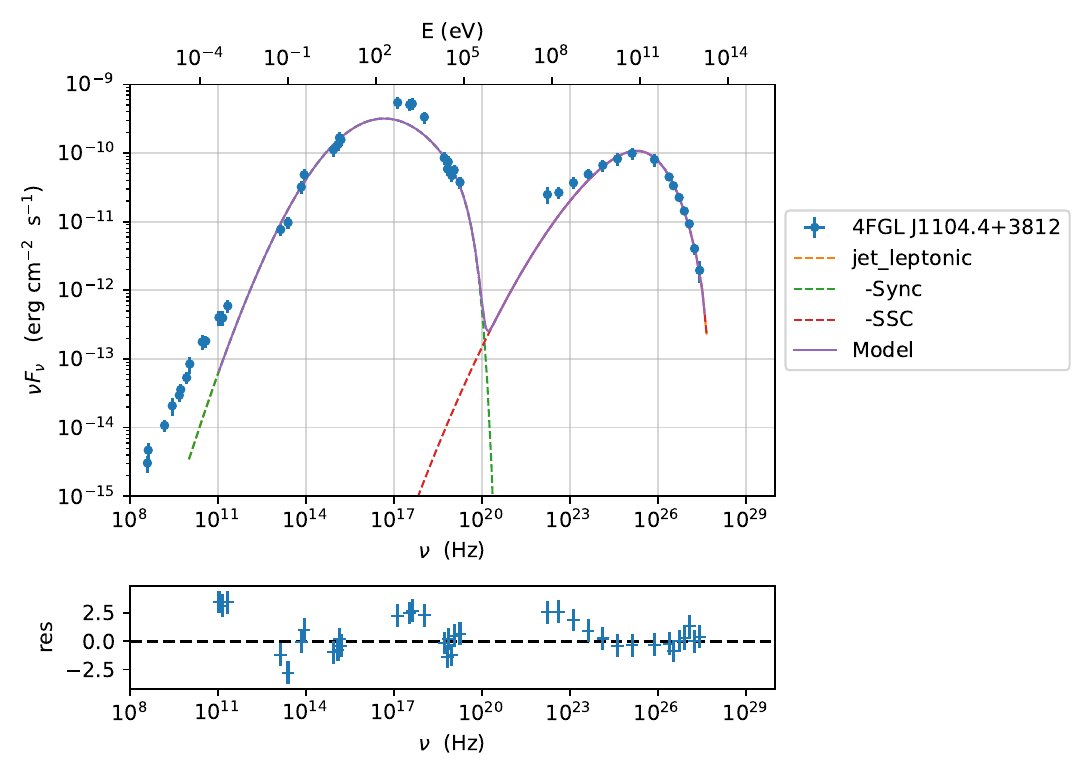}
    \includegraphics[width=0.48\linewidth]{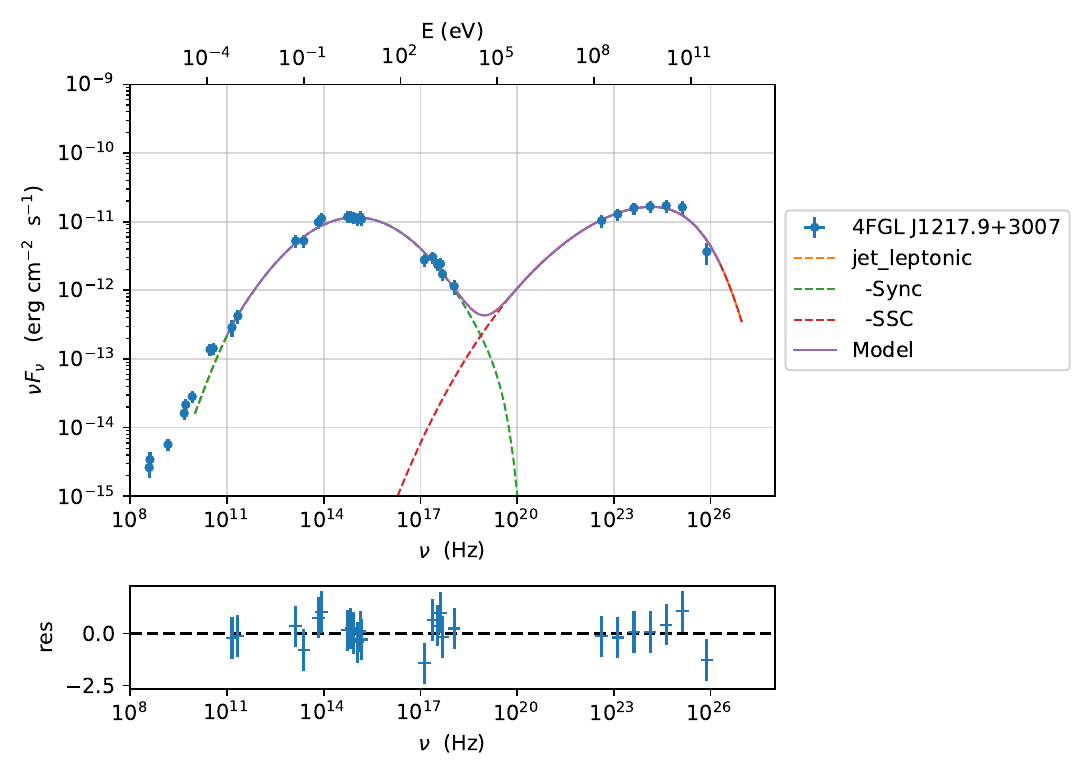}
    \center{Figure \ref{fig:SED-SSC} --- continued.}
\end{figure}

\begin{figure*}
    \centering
    \includegraphics[width=0.48\linewidth]{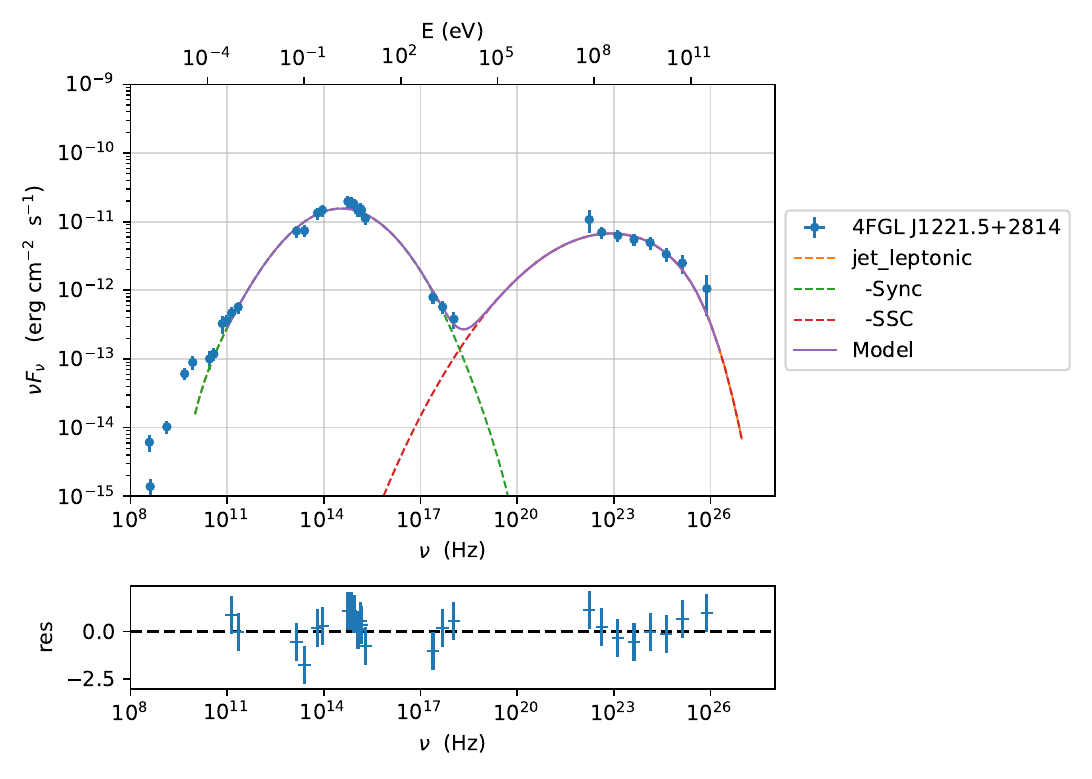}
    \includegraphics[width=0.48\linewidth]{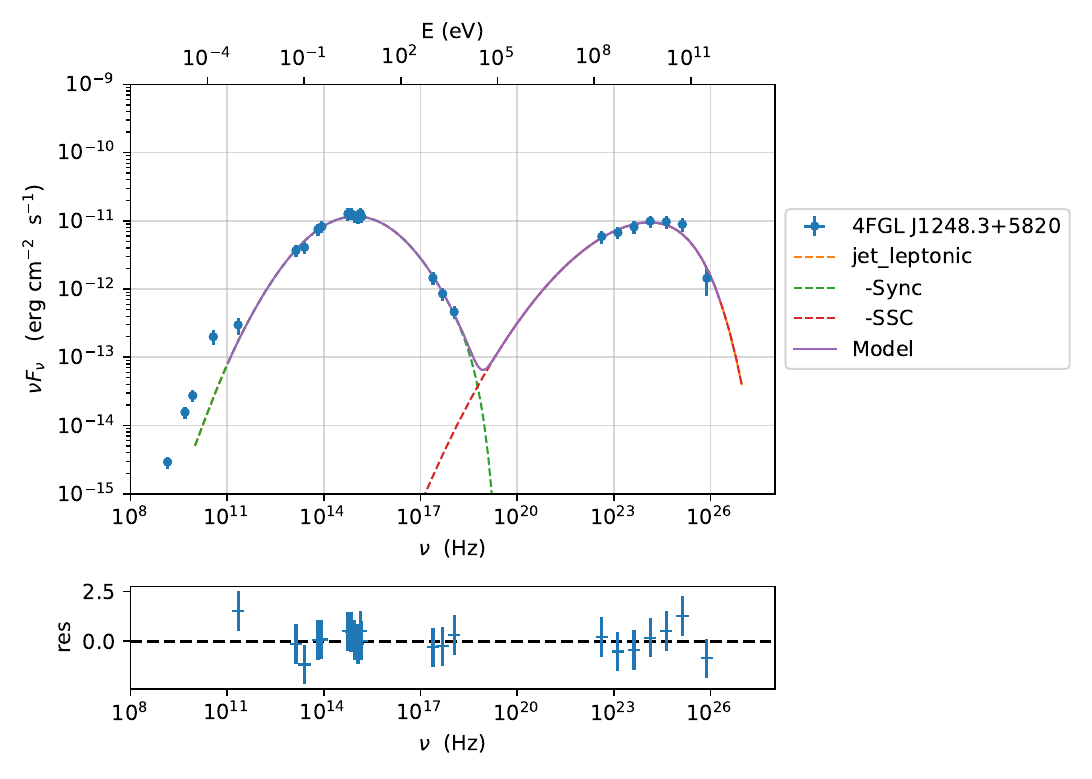}
    \includegraphics[width=0.48\linewidth]{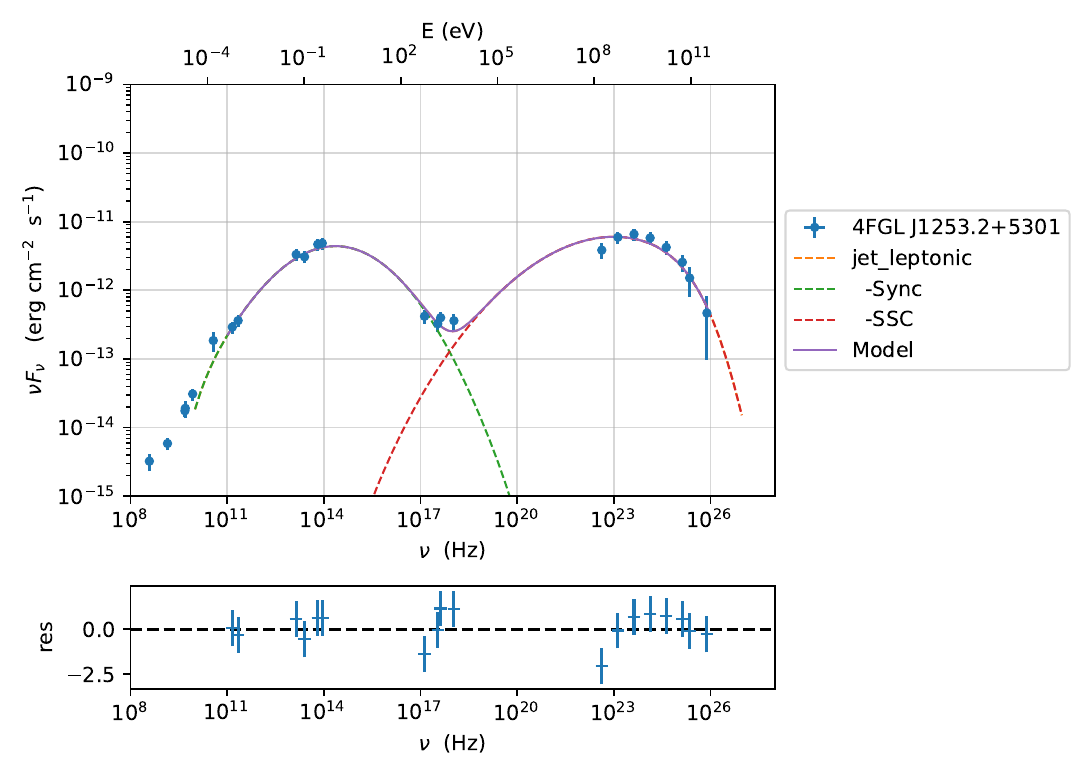}
    \includegraphics[width=0.48\linewidth]{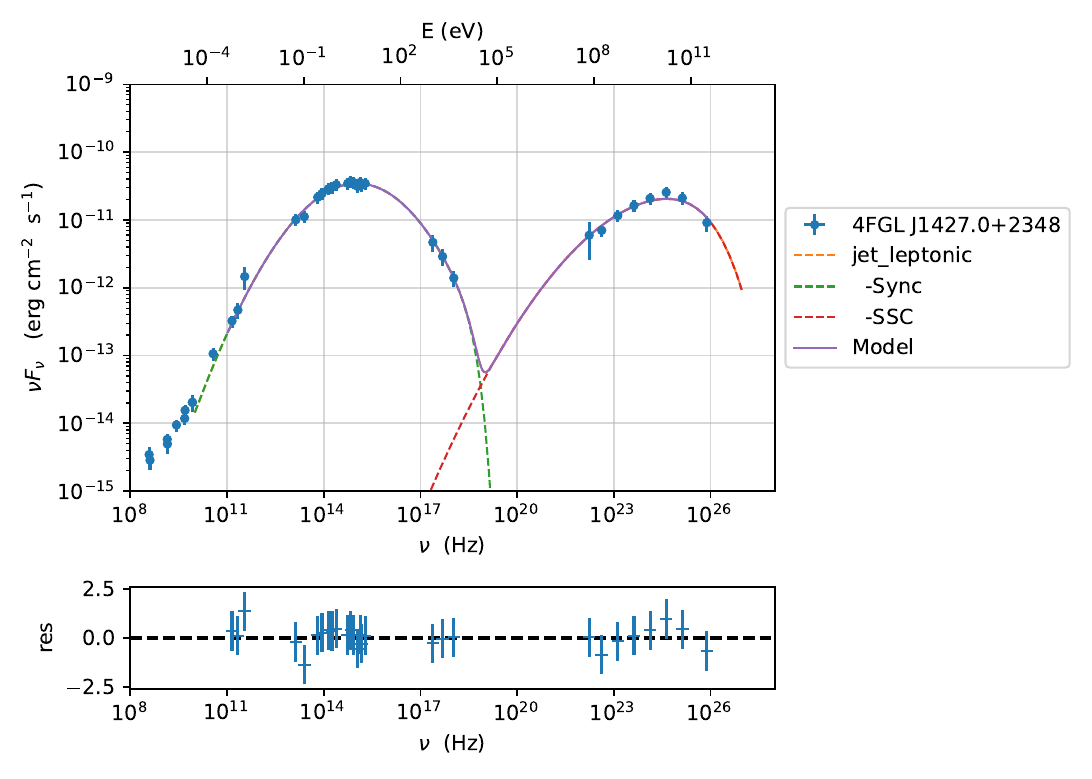}
    \includegraphics[width=0.48\linewidth]{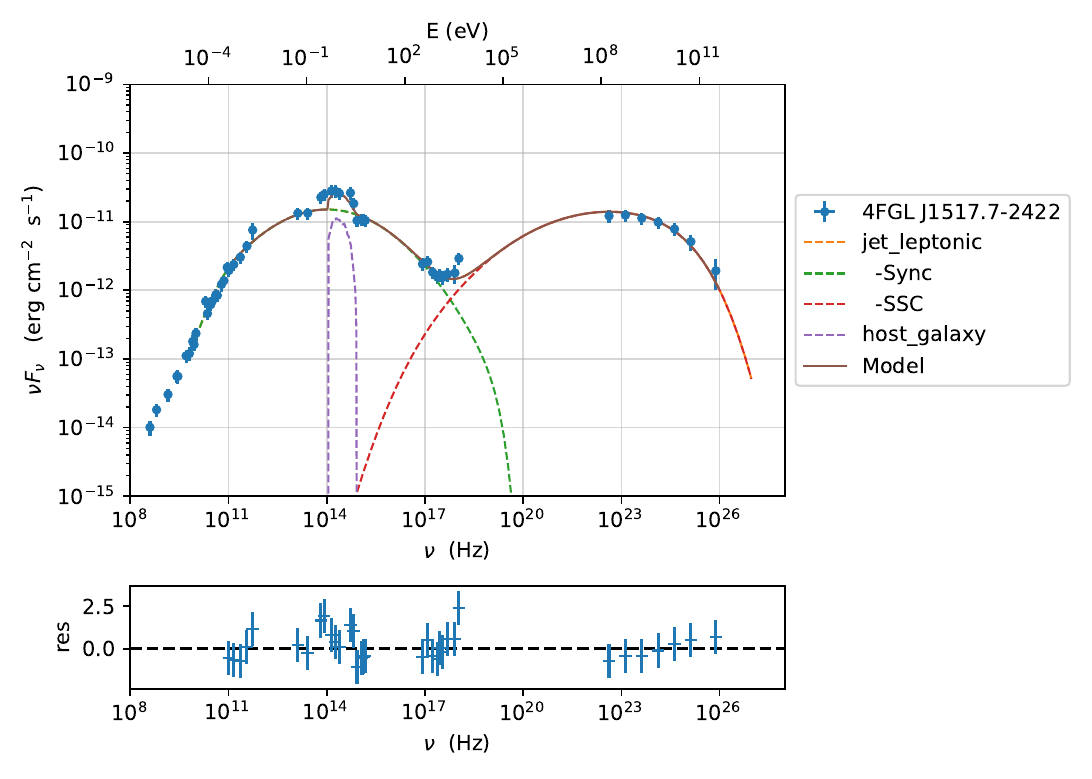}
    \includegraphics[width=0.48\linewidth]{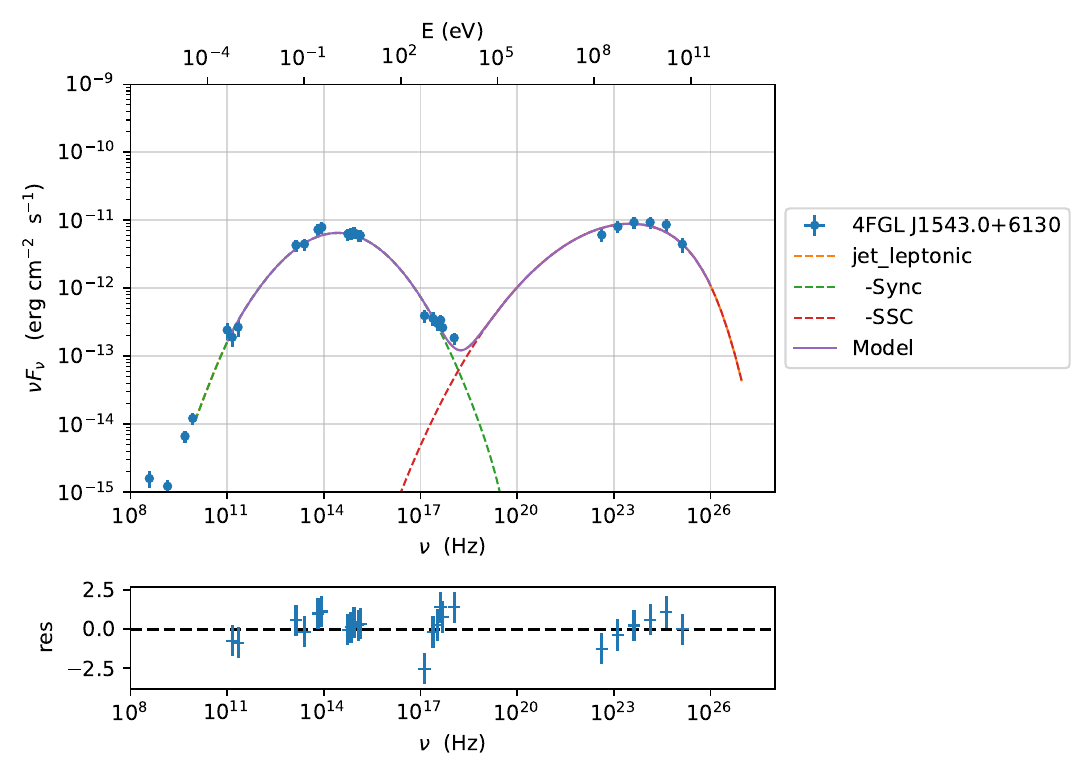}
    \includegraphics[width=0.48\linewidth]{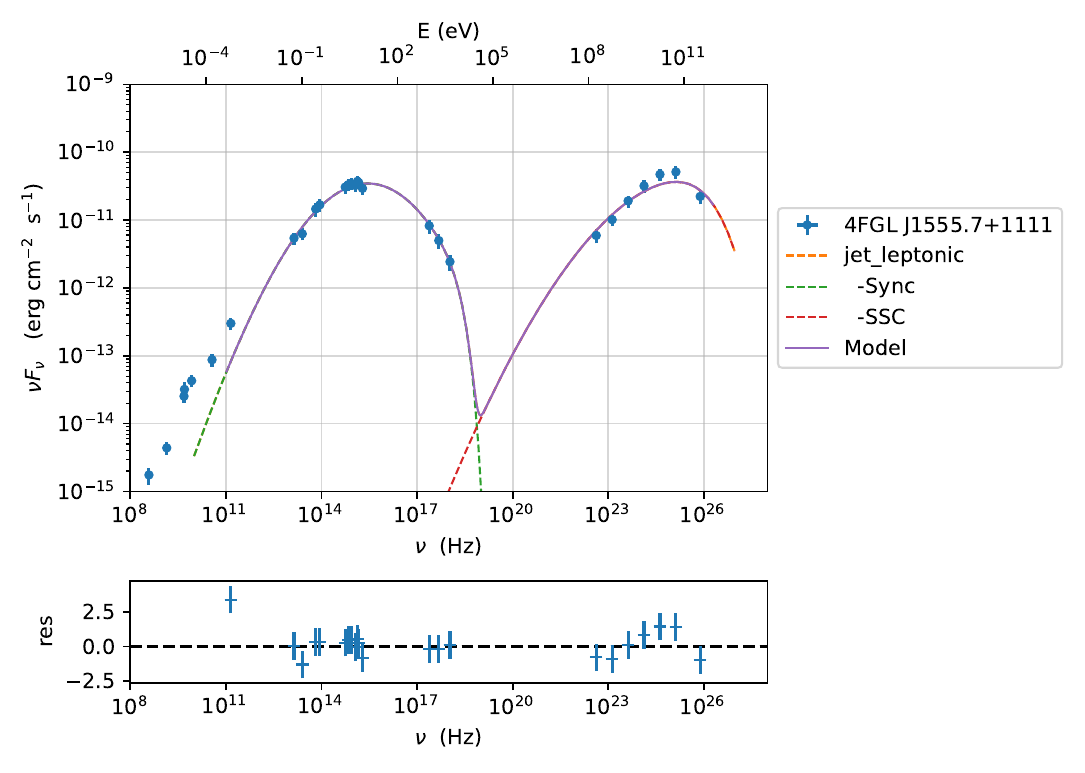}
    \includegraphics[width=0.48\linewidth]{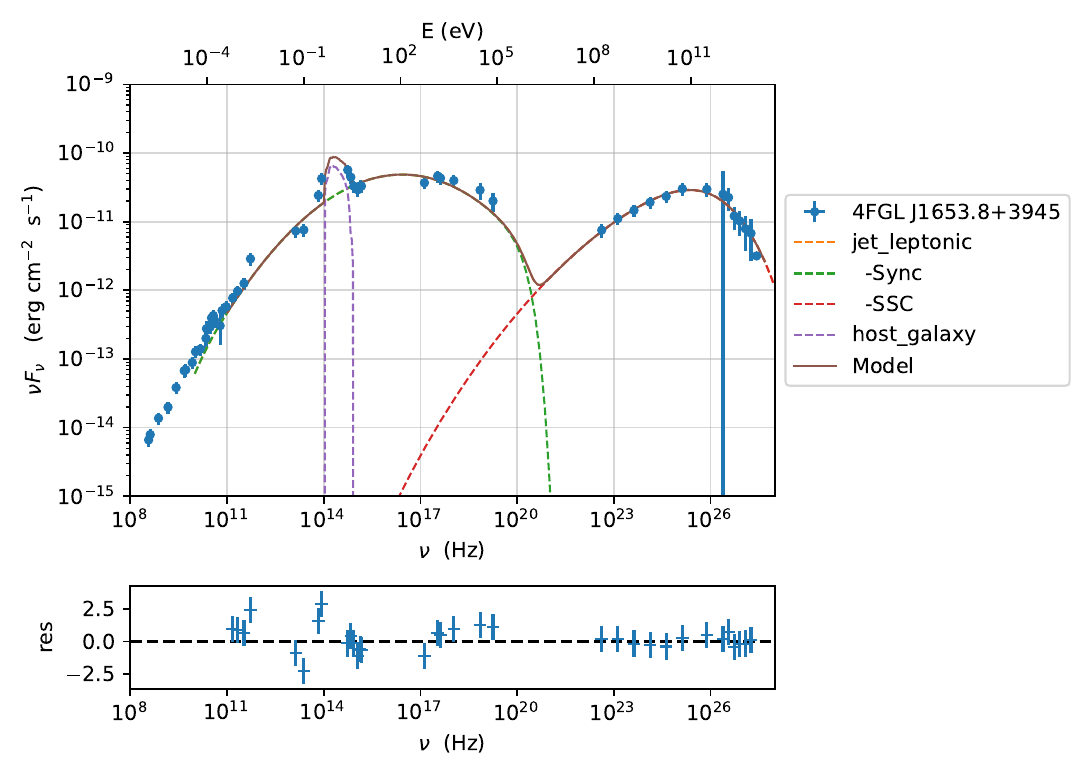}
    \center{Figure \ref{fig:SED-SSC} --- continued.}
\end{figure*}

\begin{figure*}
    \centering
    \includegraphics[width=0.48\linewidth]{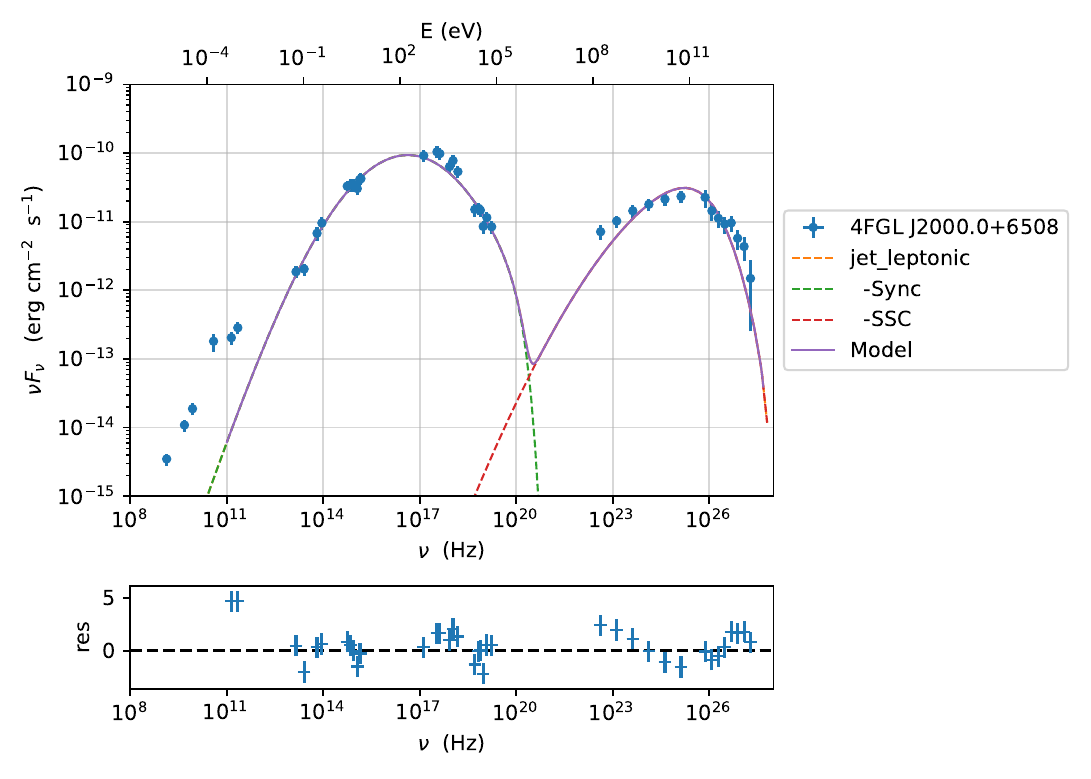}
    \includegraphics[width=0.48\linewidth]{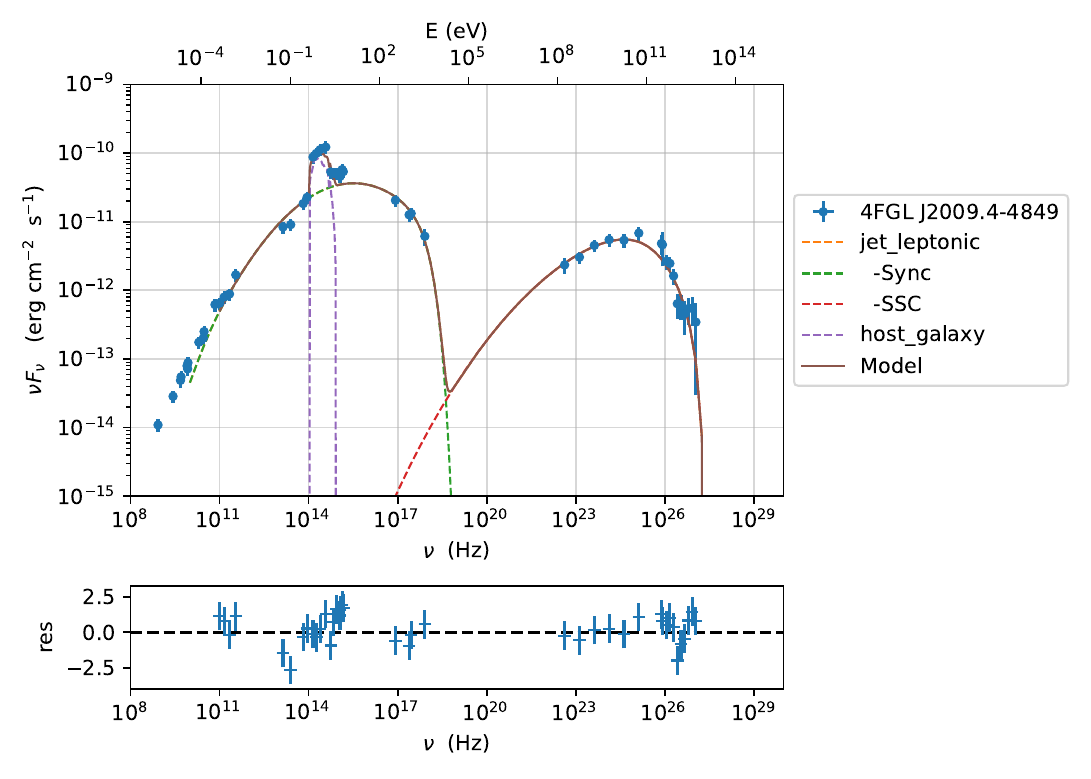}
    \includegraphics[width=0.48\linewidth]{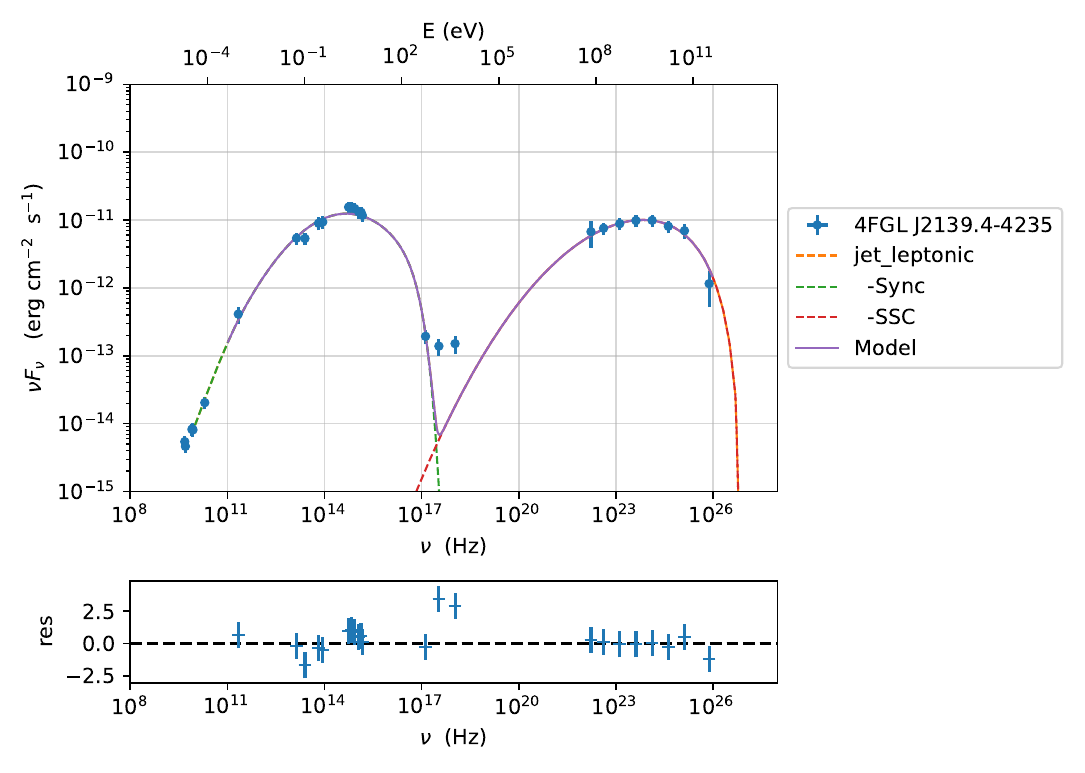}
    \includegraphics[width=0.48\linewidth]{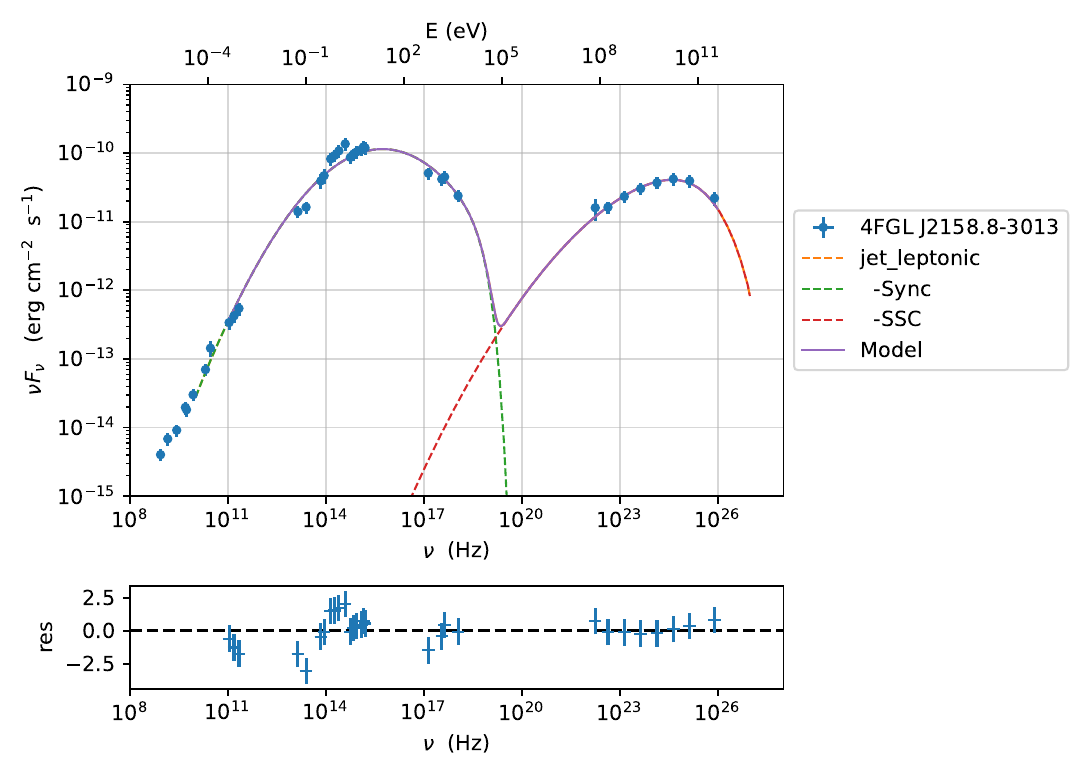}
    \center{Figure \ref{fig:SED-SSC} --- continued.}
\end{figure*}


\begin{figure*}
\centering
\includegraphics[width=0.48\linewidth]{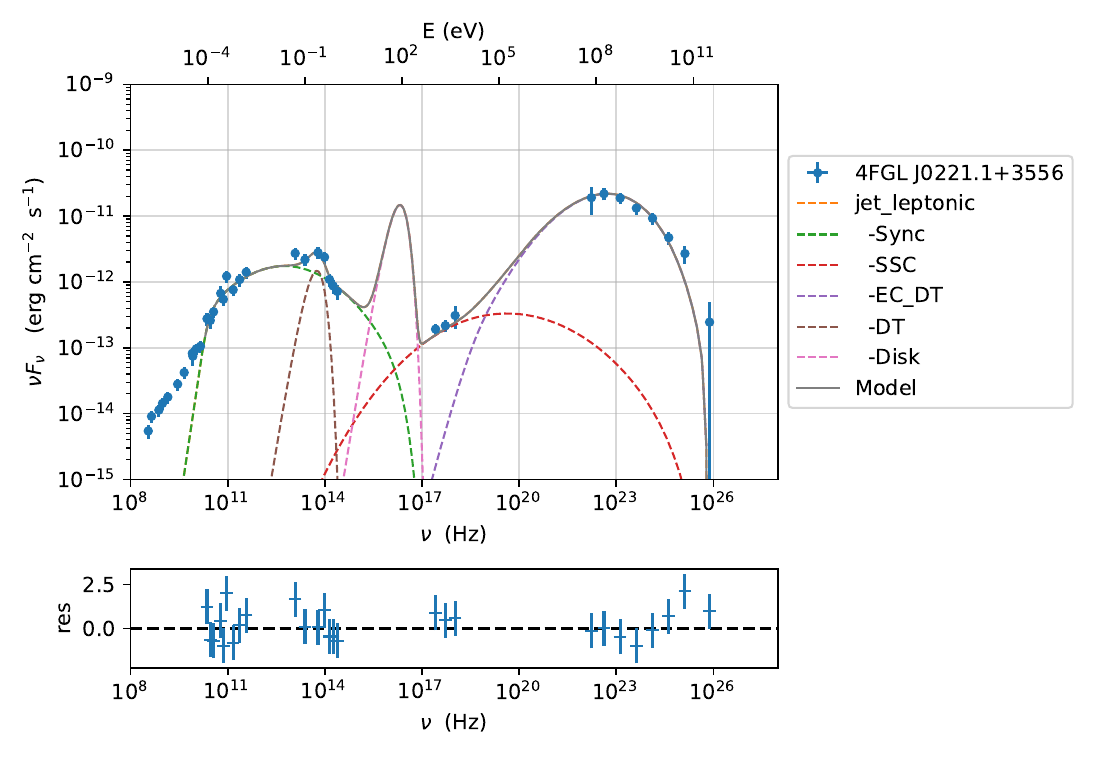}
\includegraphics[width=0.48\linewidth]{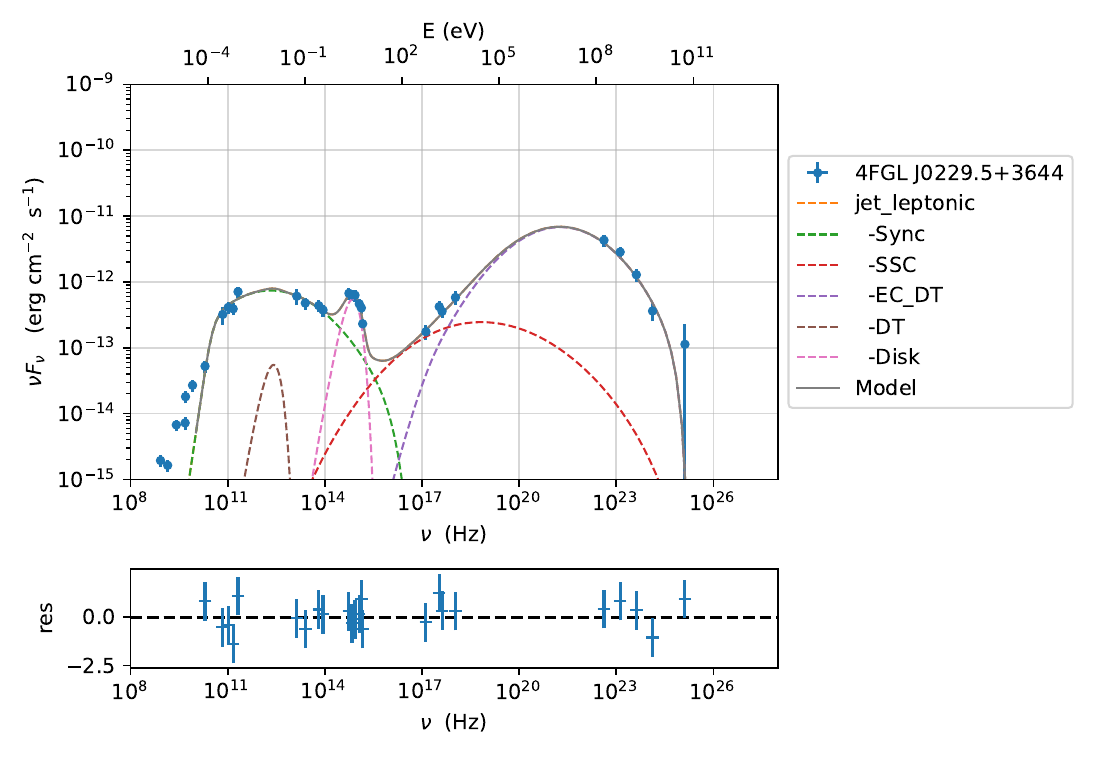}
\includegraphics[width=0.48\linewidth]{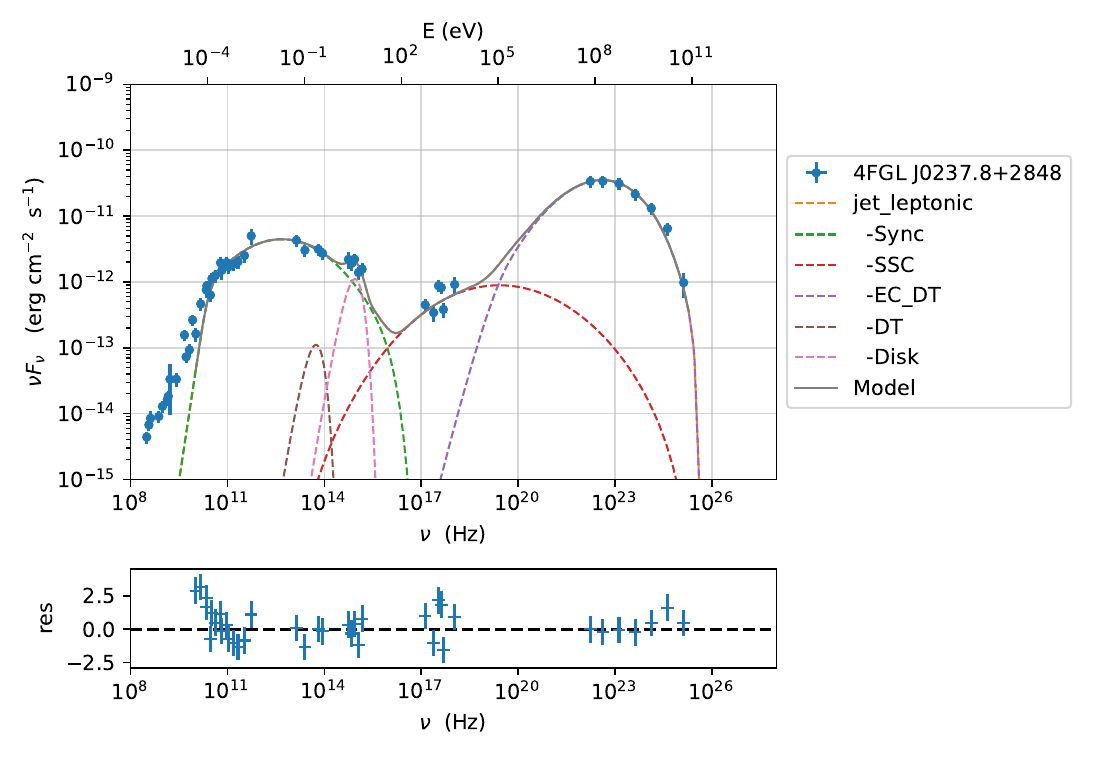}
\includegraphics[width=0.48\linewidth]{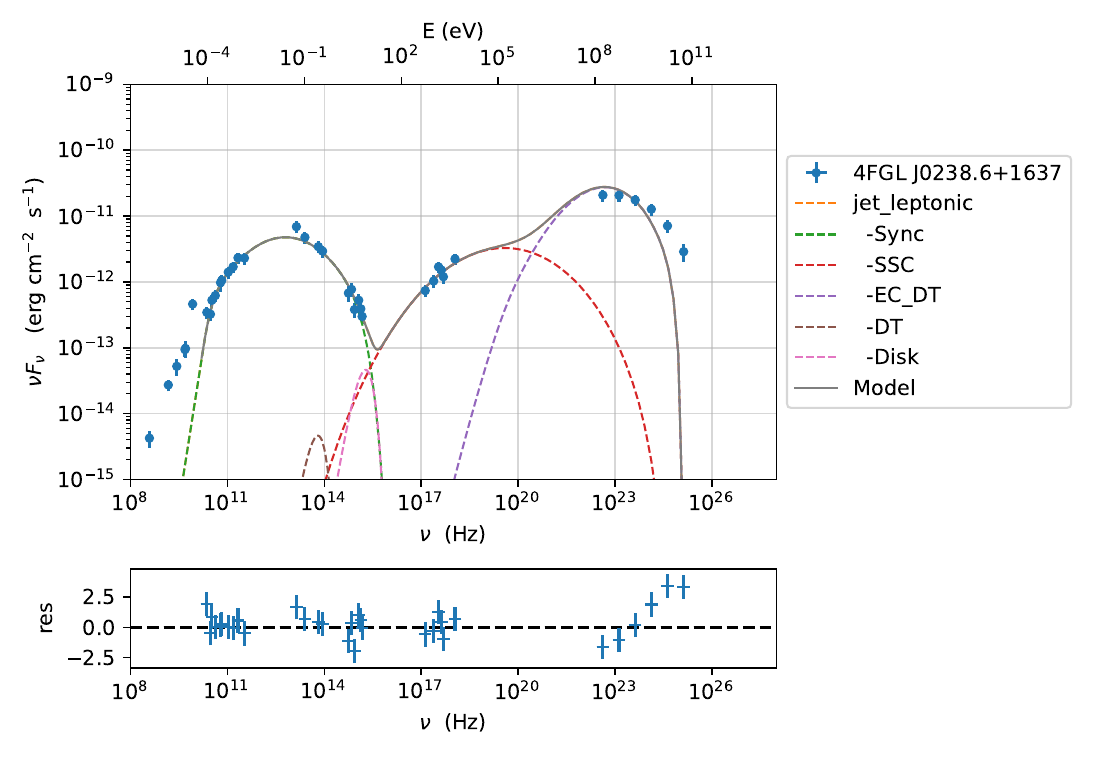}
\includegraphics[width=0.48\linewidth]{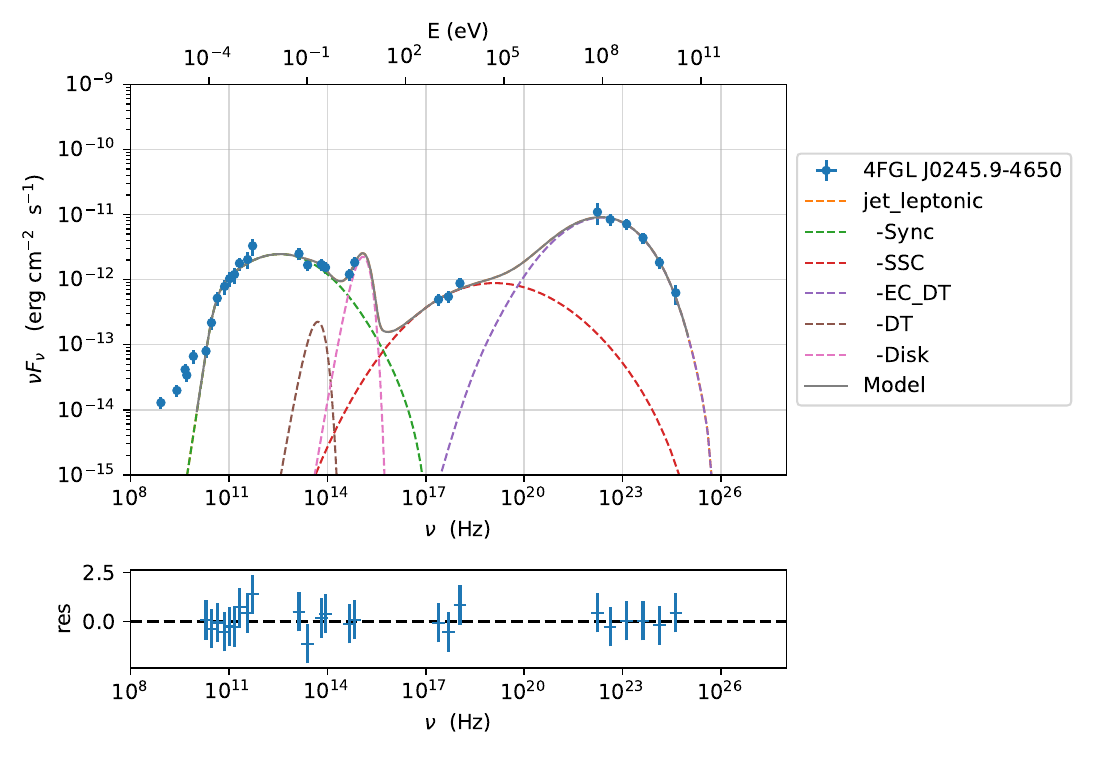}
\includegraphics[width=0.48\linewidth]{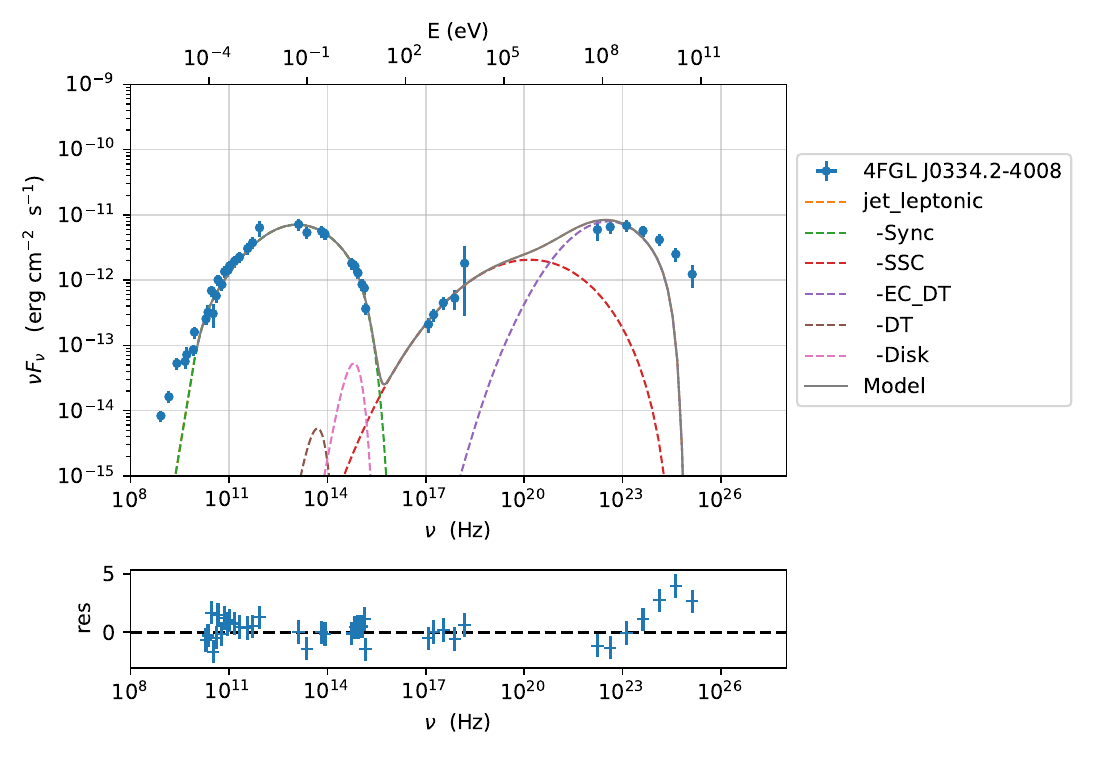}
\includegraphics[width=0.48\linewidth]{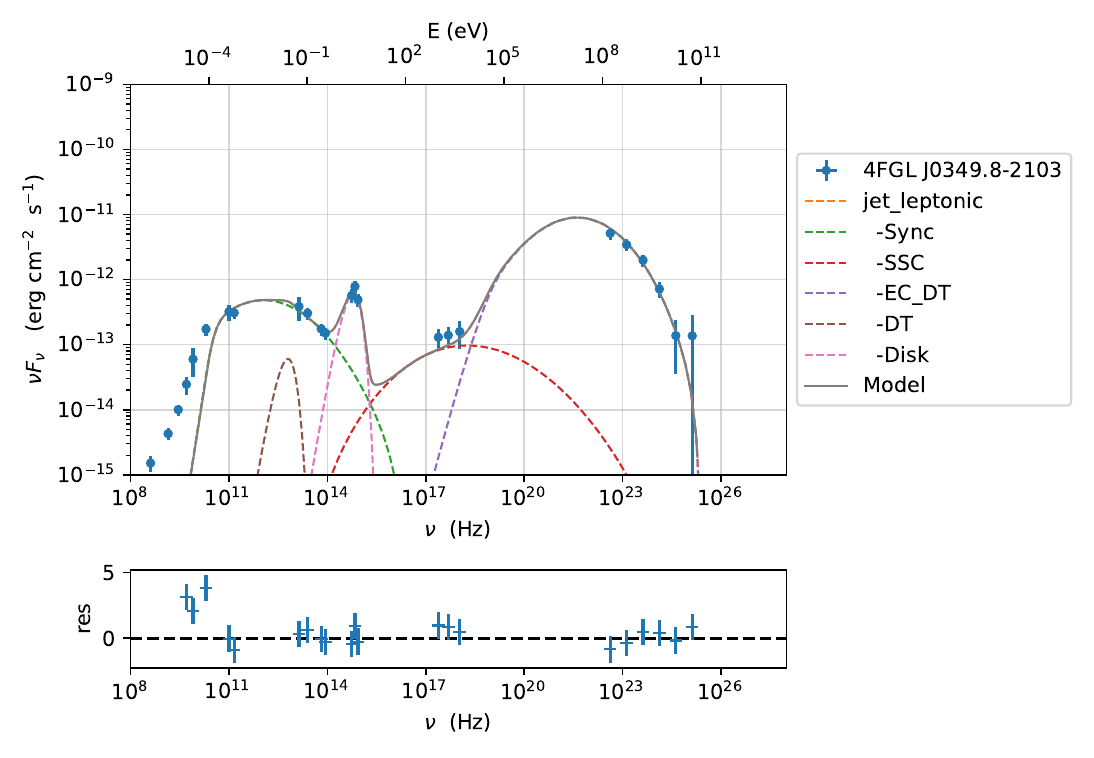}
\includegraphics[width=0.48\linewidth]{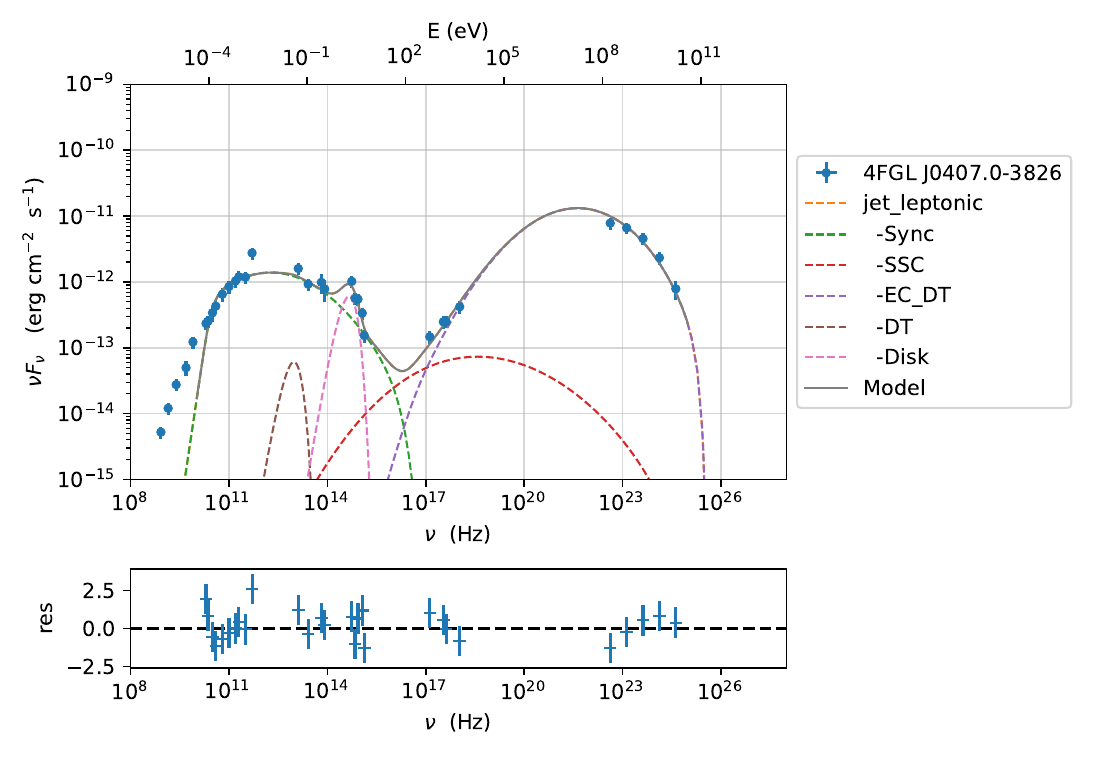}
\center{Figure \ref{fig:SED-EC} --- continued.}
\end{figure*}

\begin{figure*}
\centering
\includegraphics[width=0.48\linewidth]{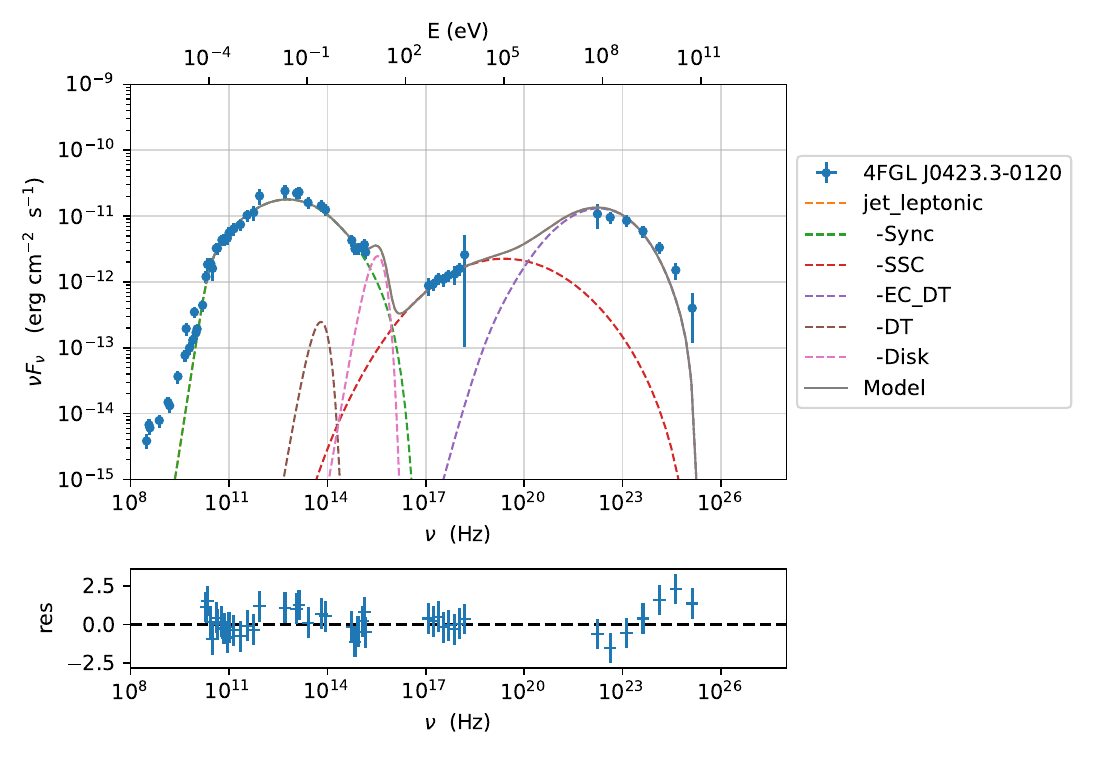}
\includegraphics[width=0.48\linewidth]{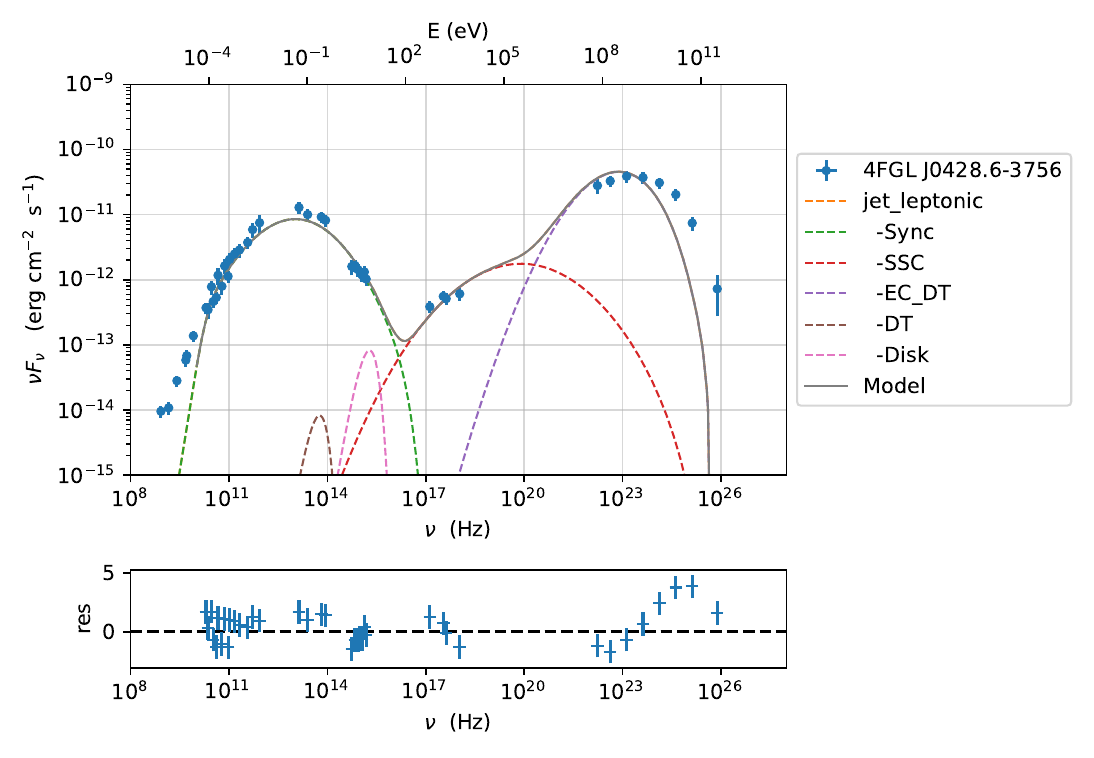}
\includegraphics[width=0.48\linewidth]{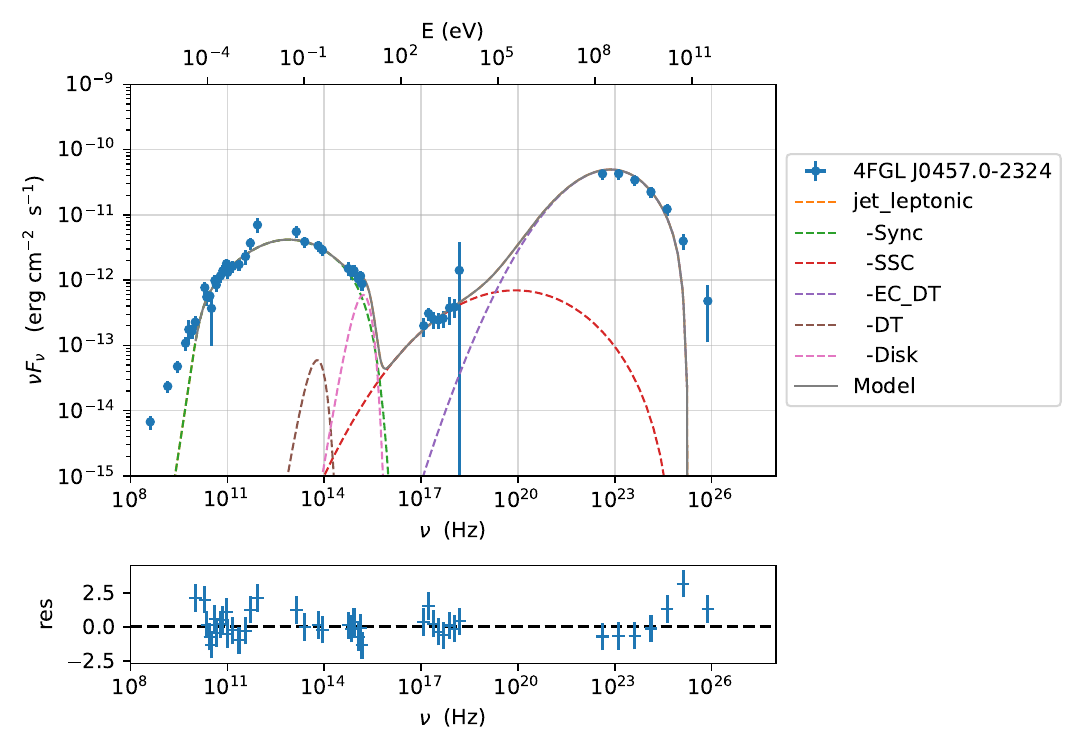}
\includegraphics[width=0.48\linewidth]{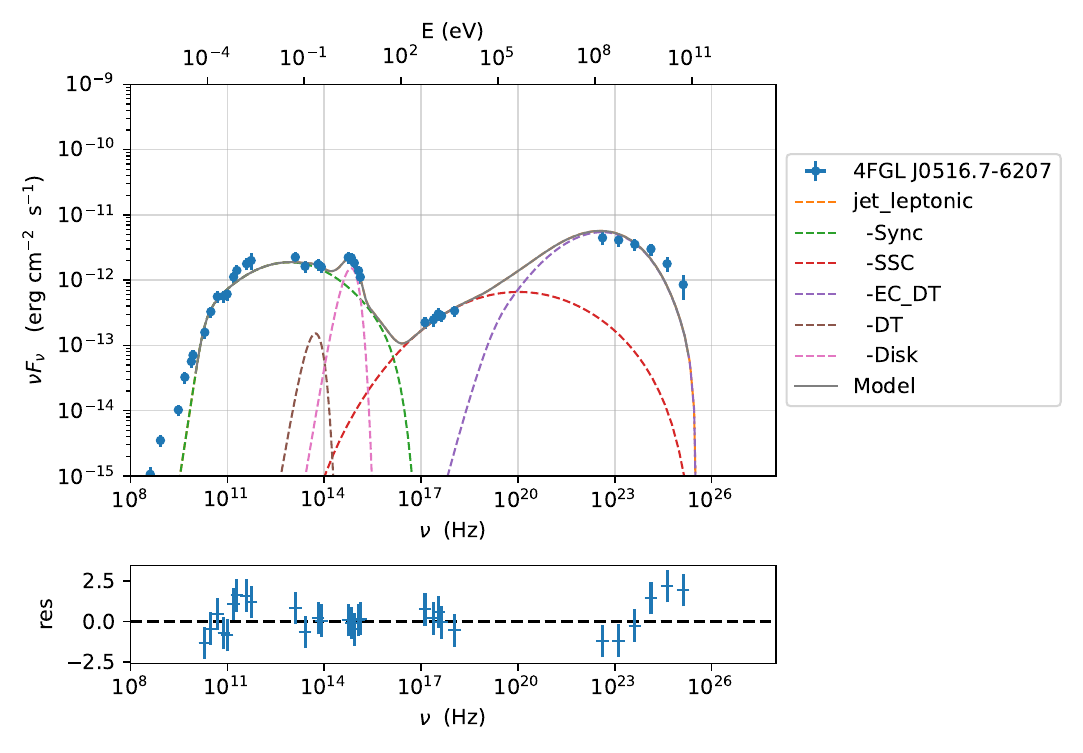}
\includegraphics[width=0.48\linewidth]{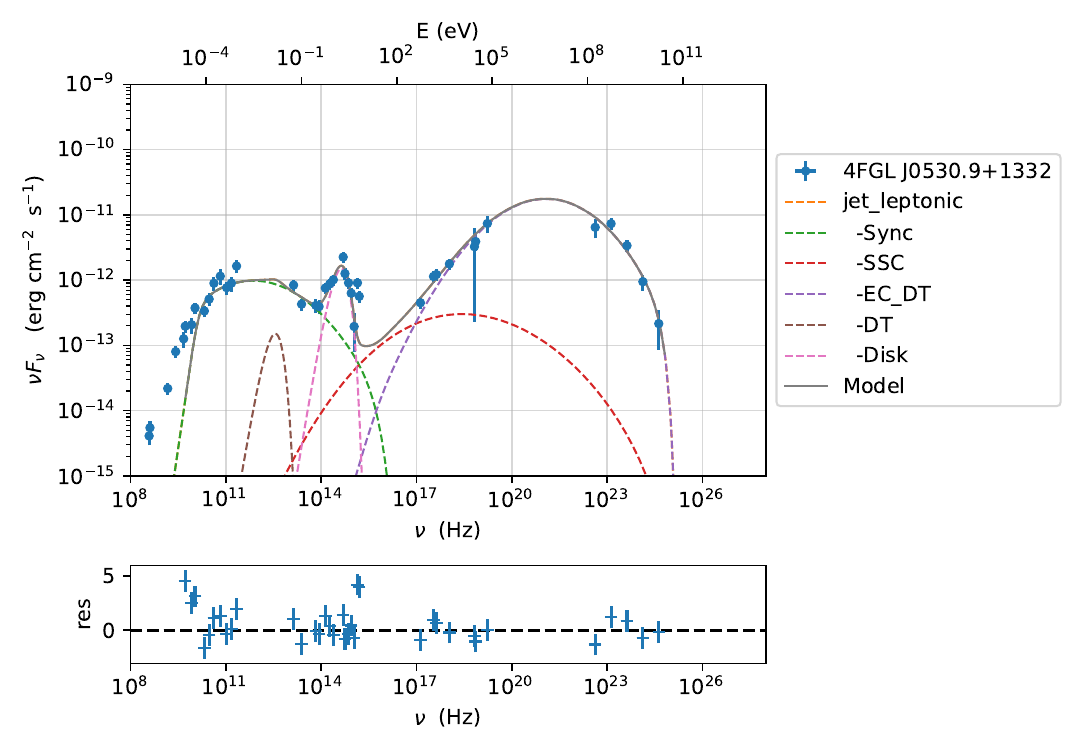}    
\includegraphics[width=0.48\linewidth]{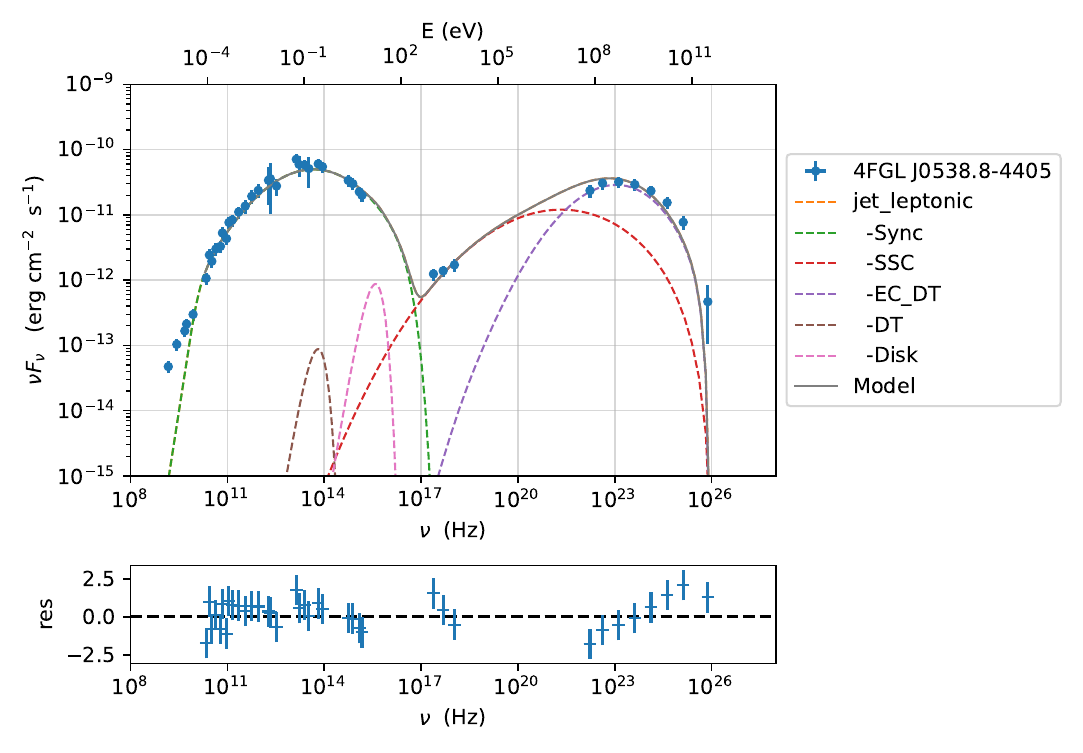}
\includegraphics[width=0.48\linewidth]{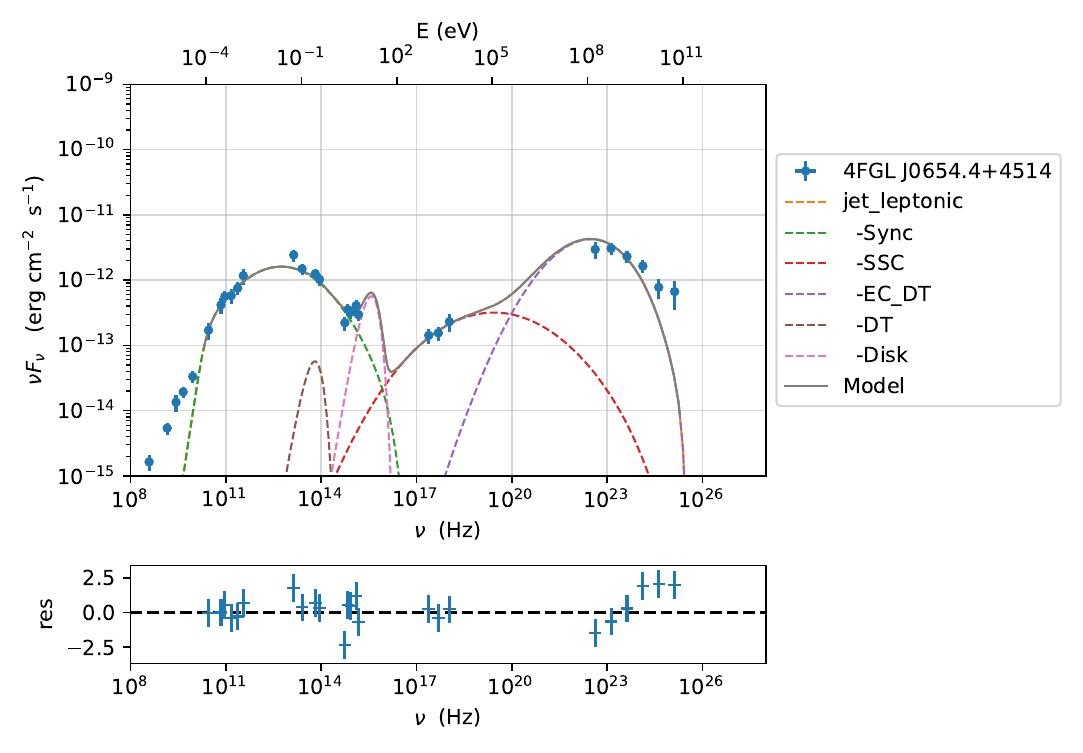}
\includegraphics[width=0.48\linewidth]{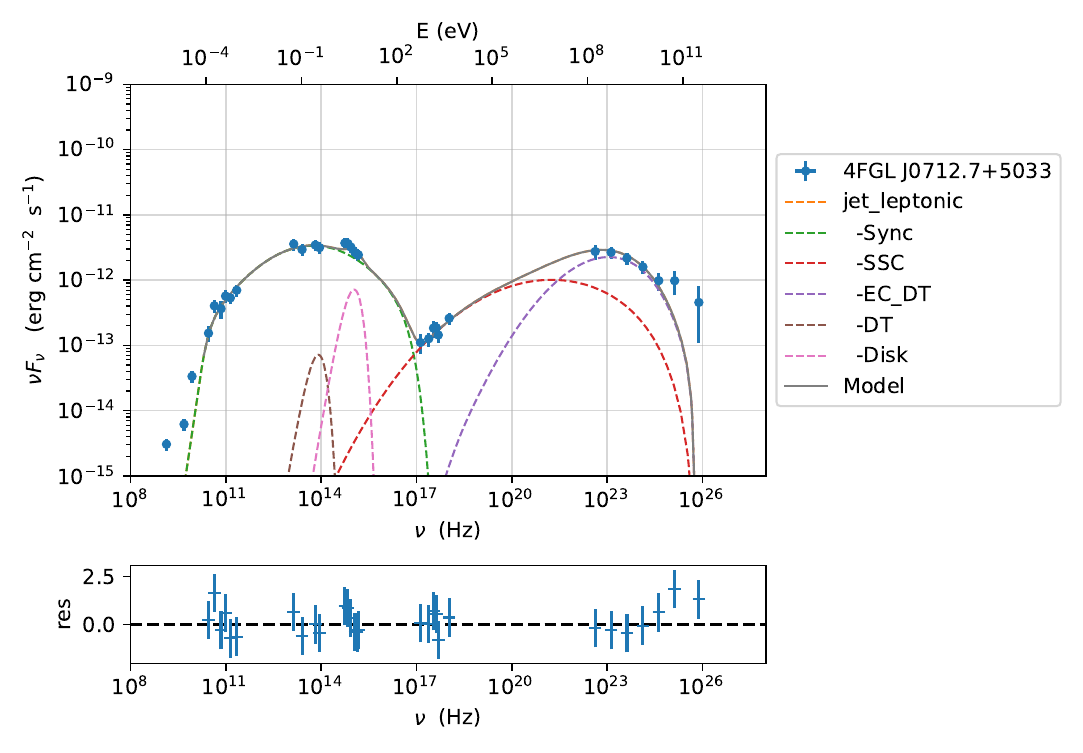}
\center{Figure \ref{fig:SED-EC} --- continued.}
\end{figure*}

\begin{figure*}
\centering
\includegraphics[width=0.48\linewidth]{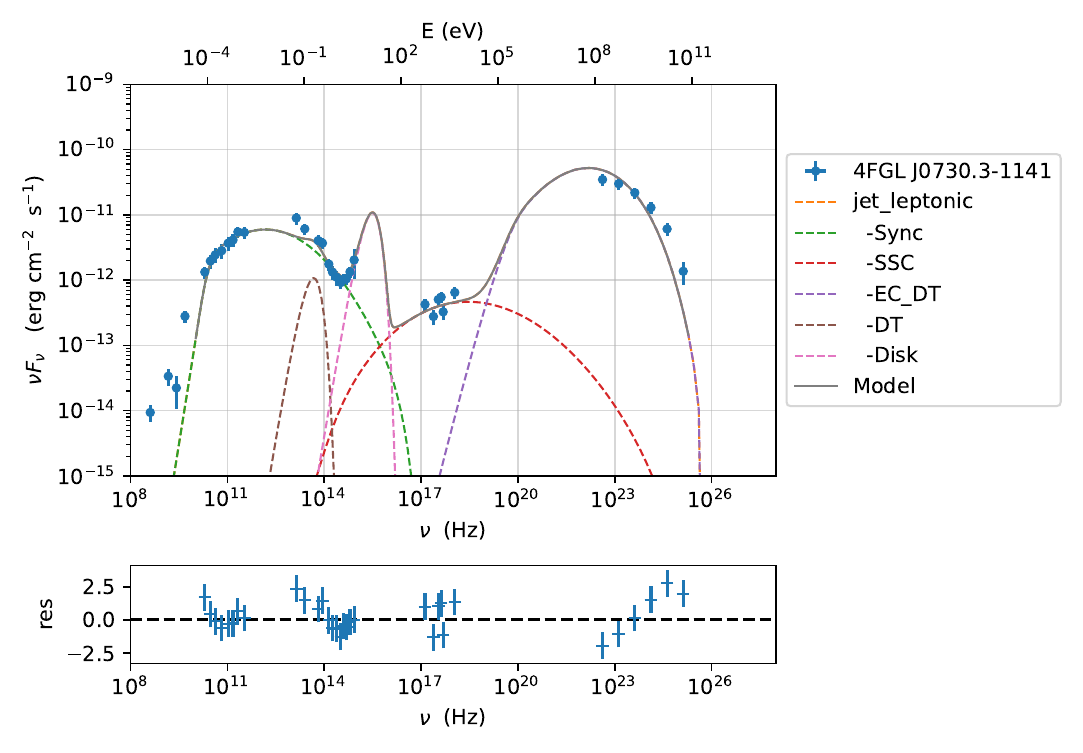}
\includegraphics[width=0.48\linewidth]{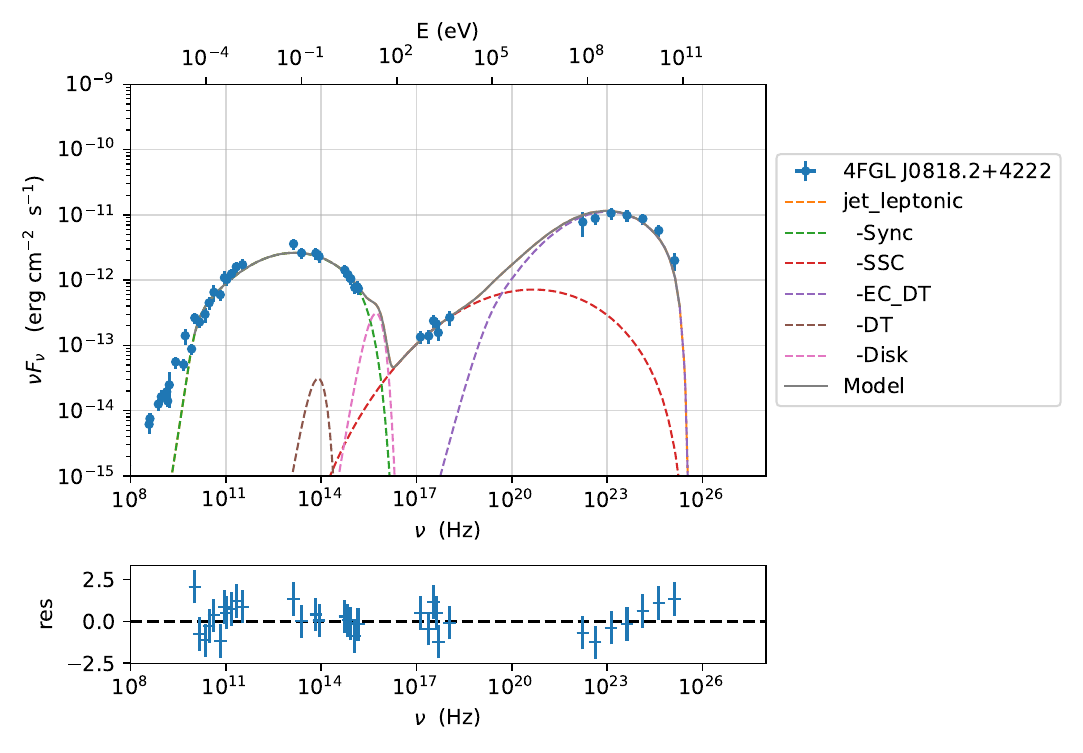}
\includegraphics[width=0.48\linewidth]{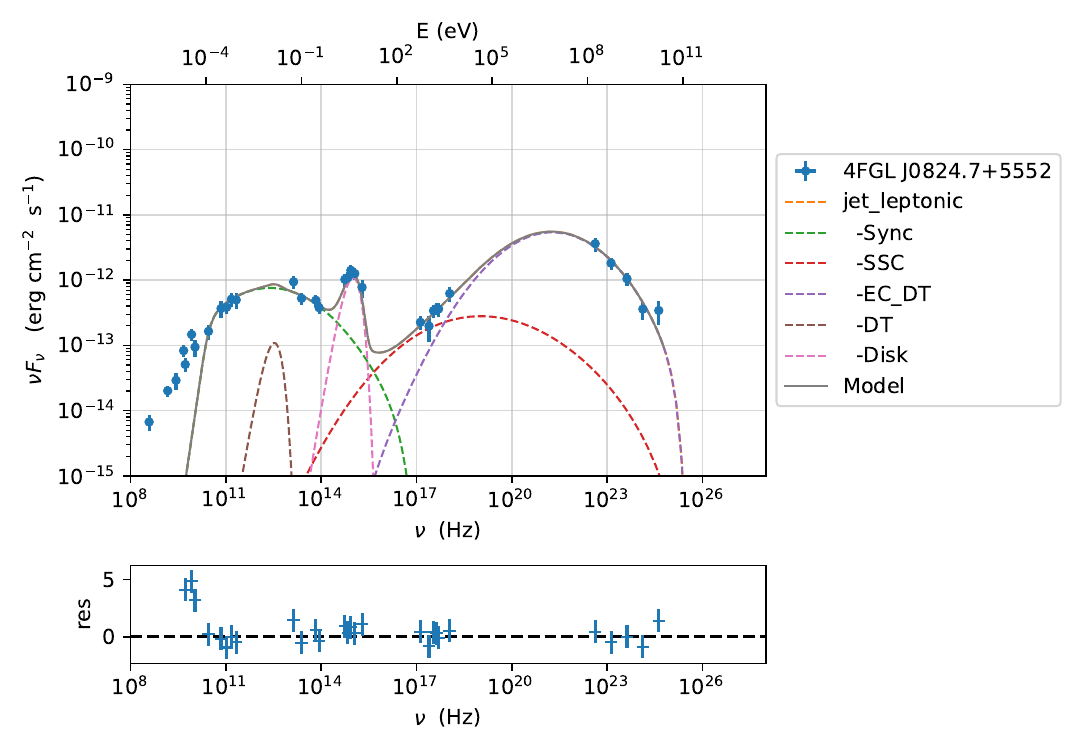}
\includegraphics[width=0.48\linewidth]{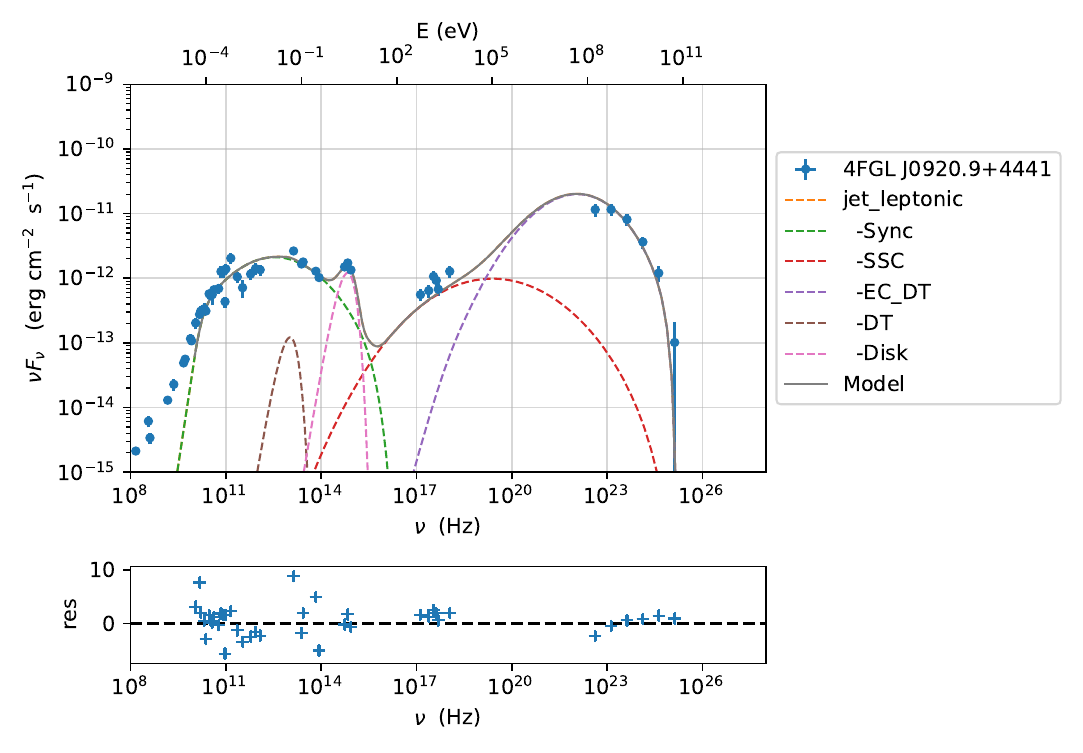}
\includegraphics[width=0.48\linewidth]{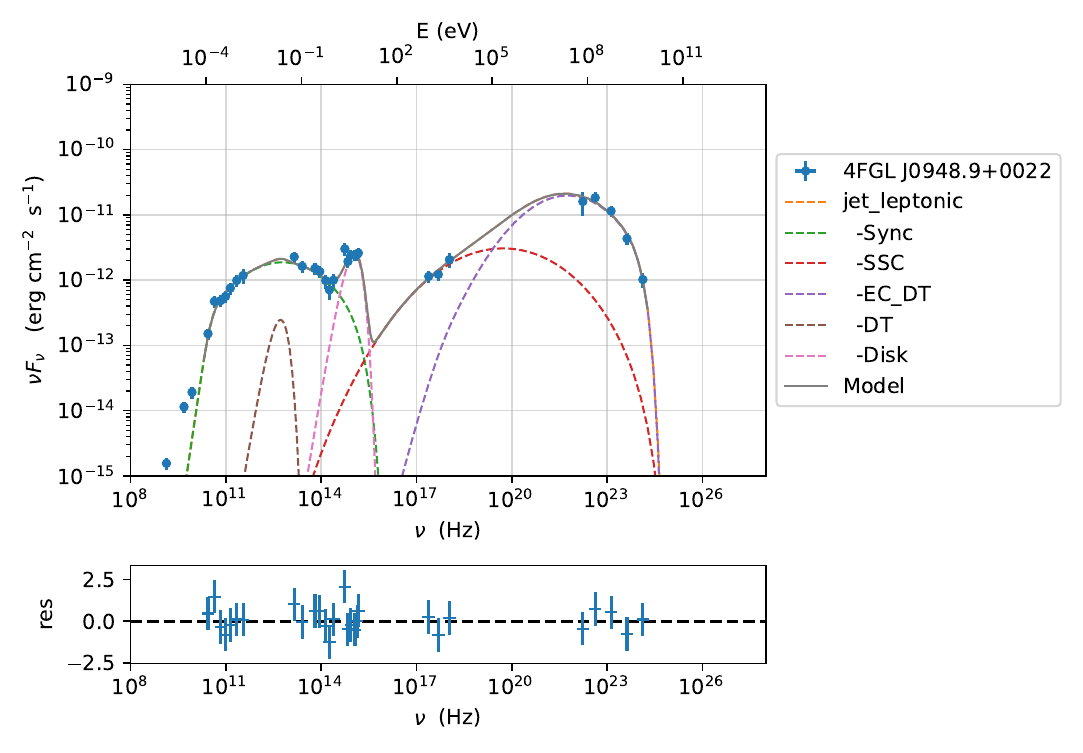}
\includegraphics[width=0.48\linewidth]{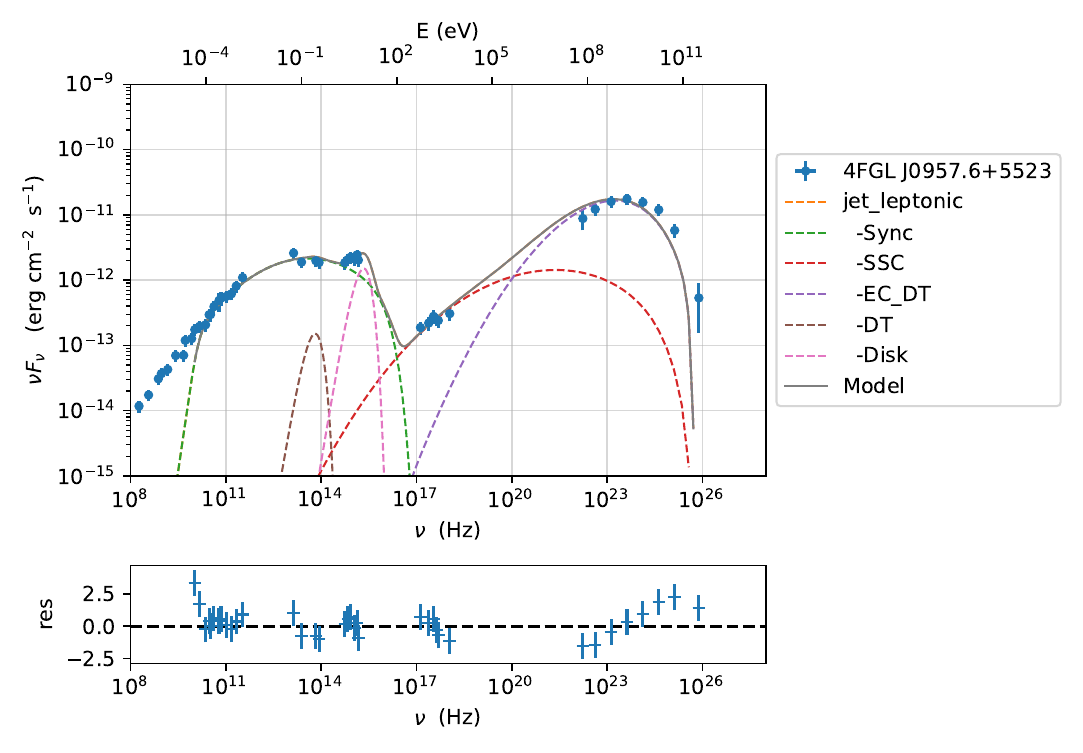}
\includegraphics[width=0.48\linewidth]{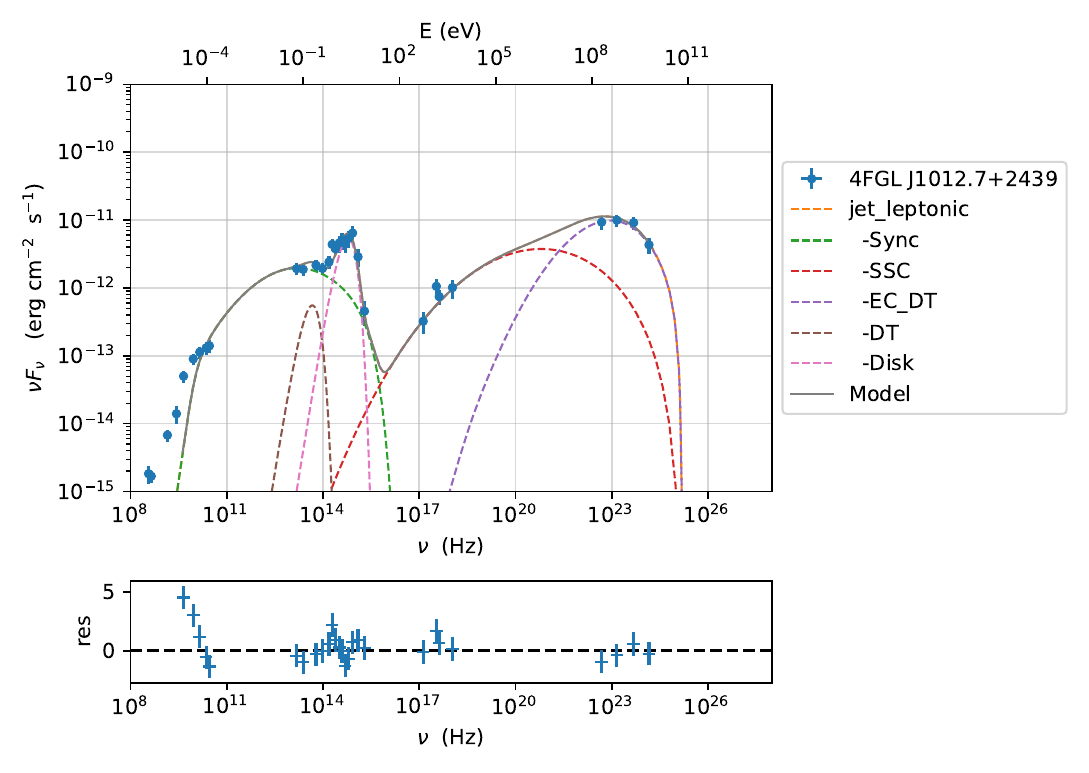}
\includegraphics[width=0.48\linewidth]{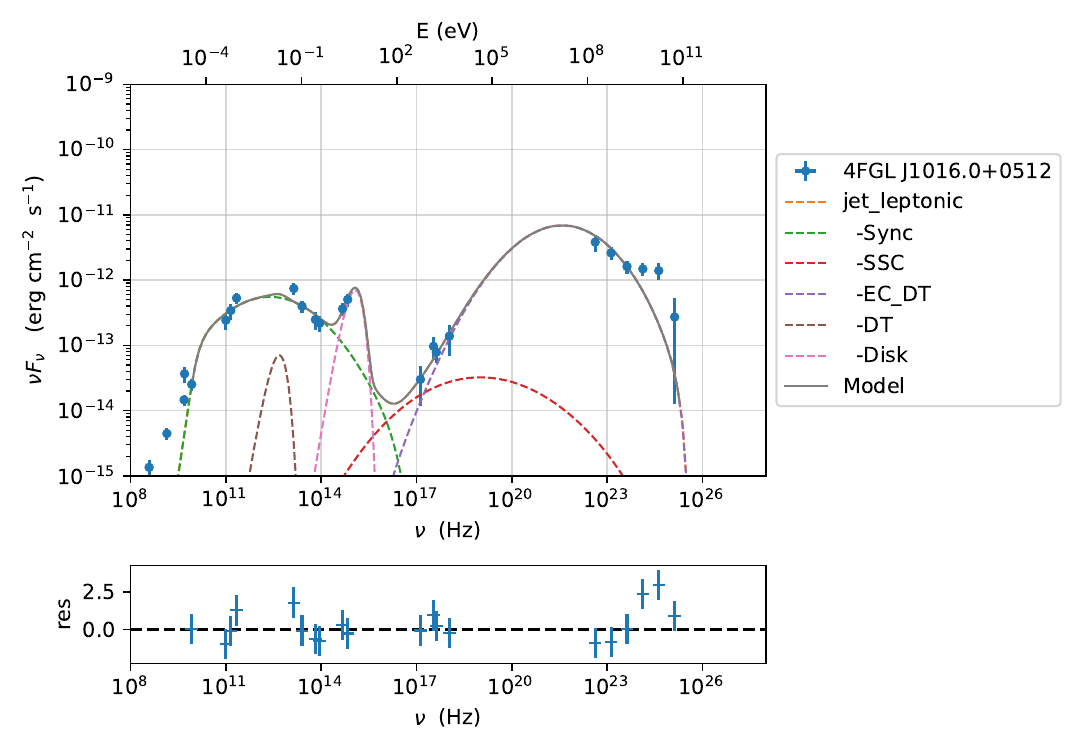}
\center{Figure \ref{fig:SED-EC} --- continued.}
\end{figure*}

\begin{figure*}
\centering
\includegraphics[width=0.48\linewidth]{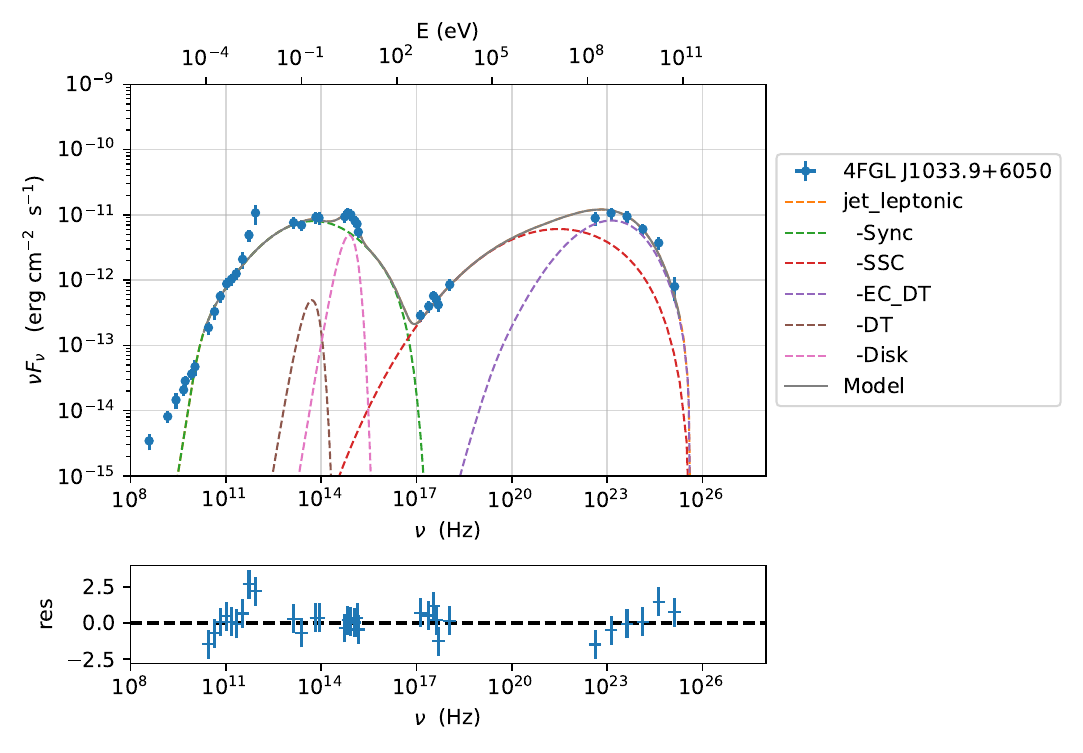}
\includegraphics[width=0.48\linewidth]{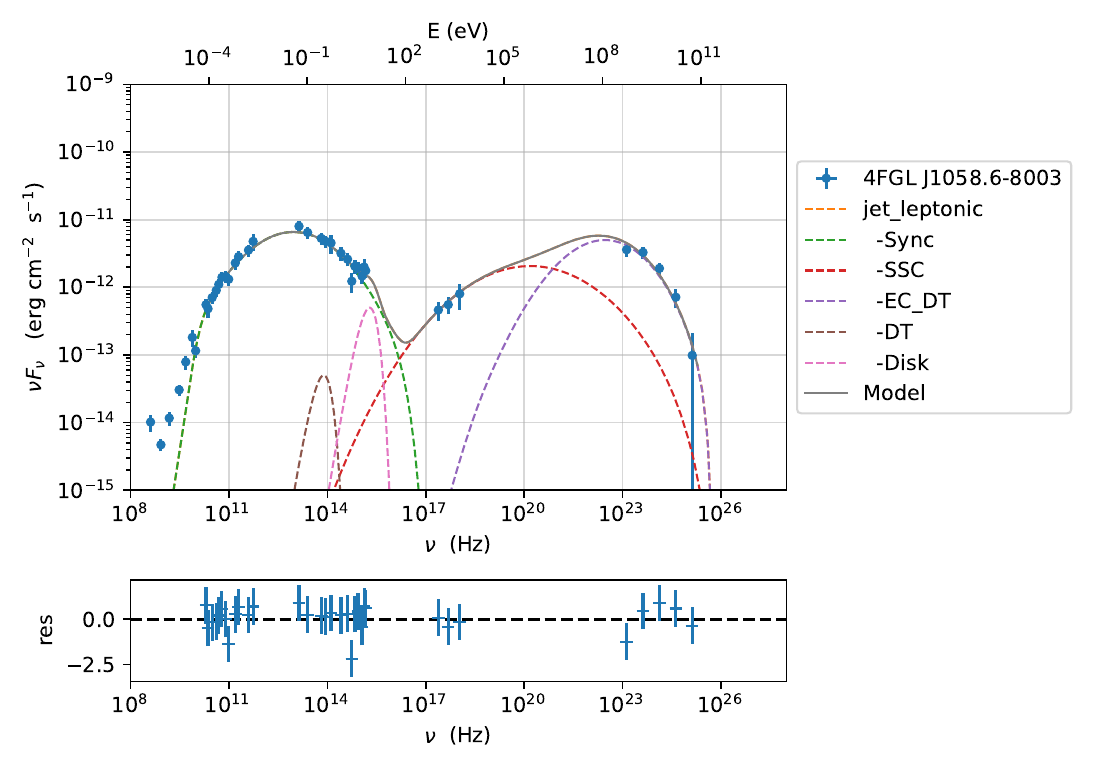}
\includegraphics[width=0.48\linewidth]{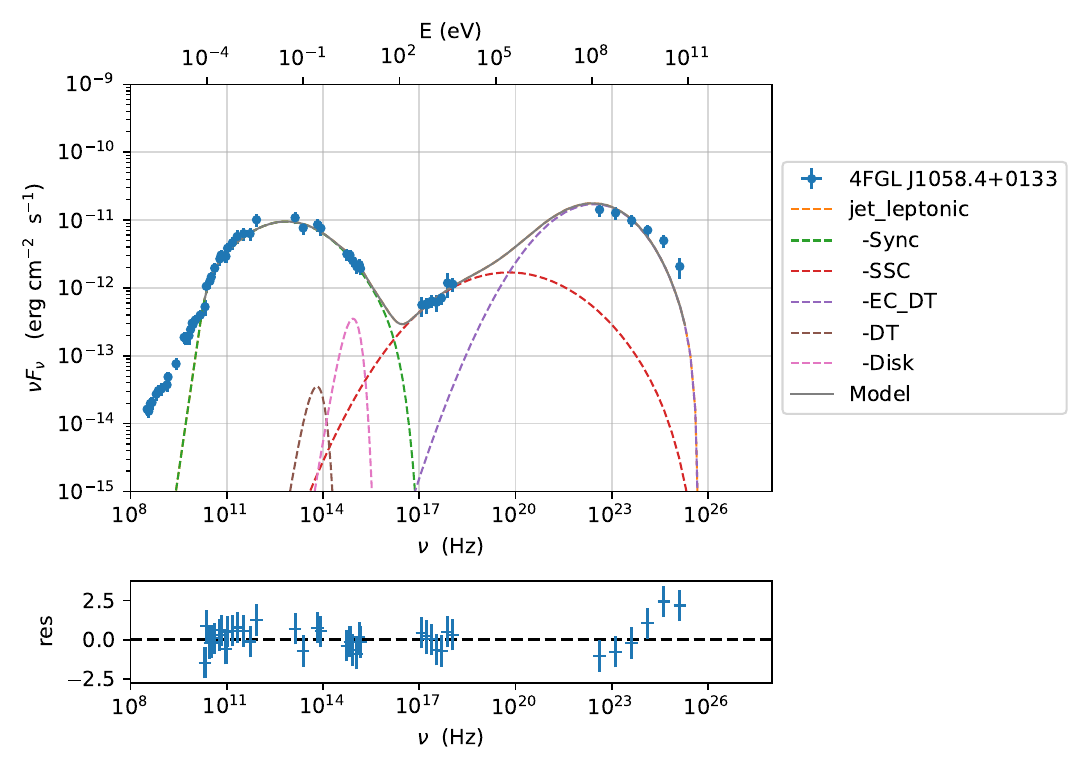}
\includegraphics[width=0.48\linewidth]{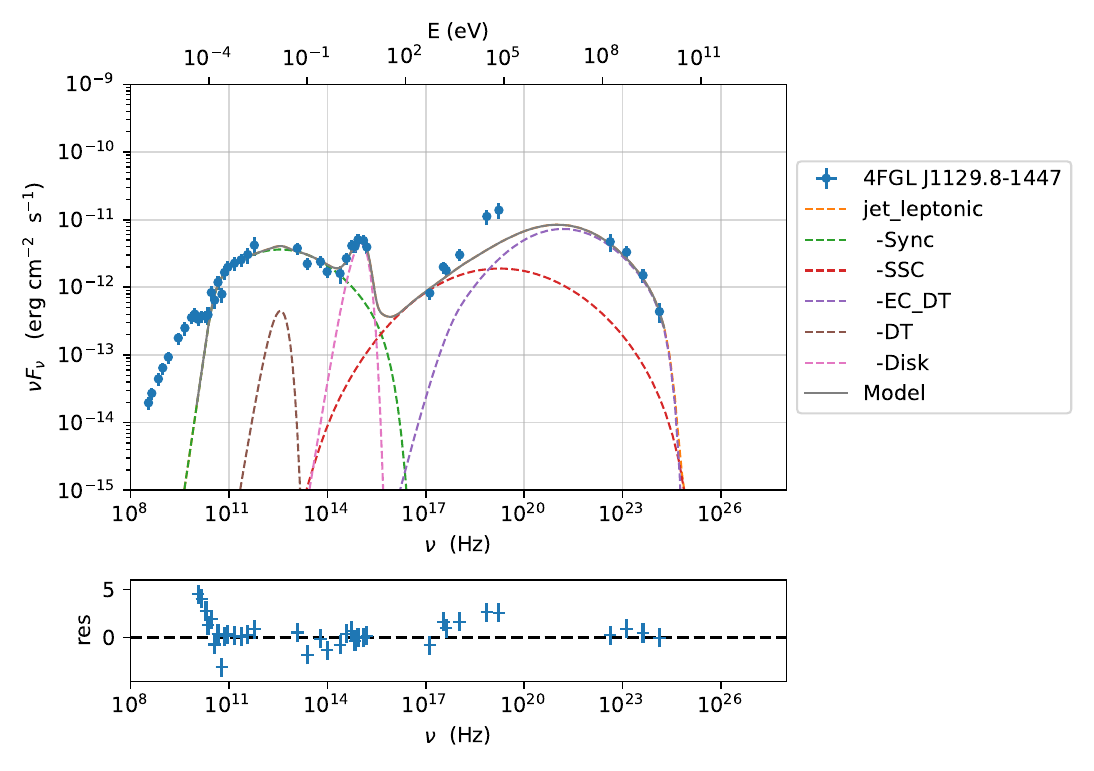}
\includegraphics[width=0.48\linewidth]{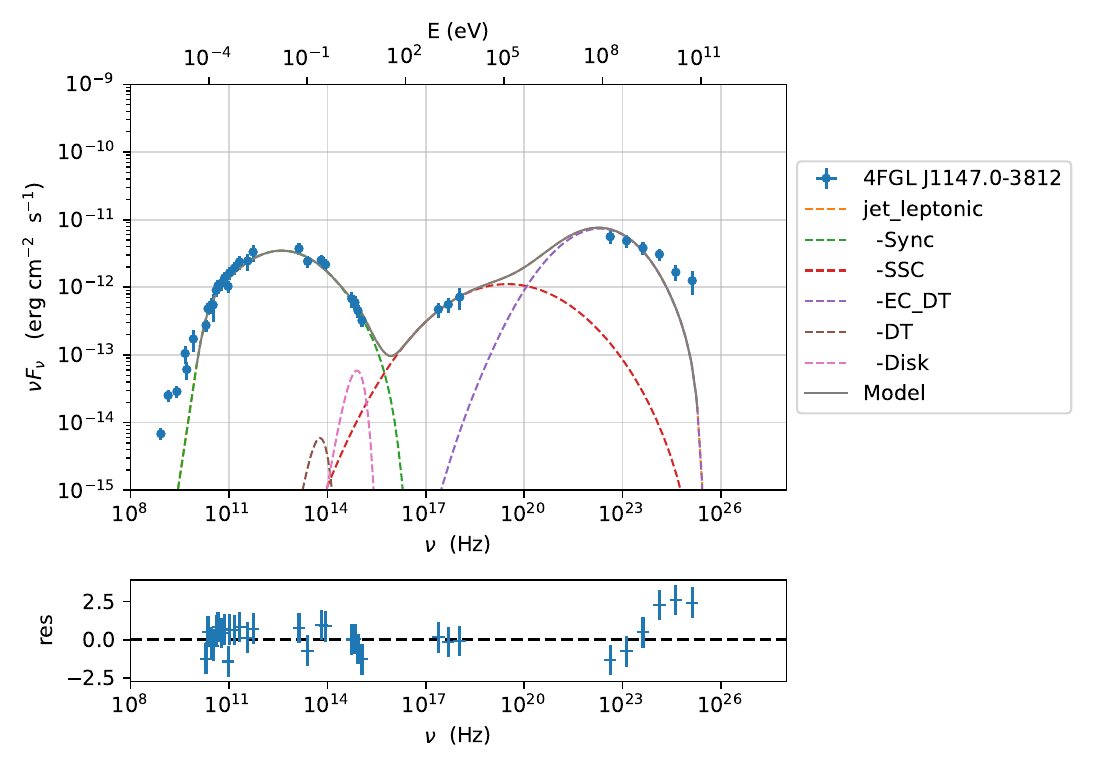}
\includegraphics[width=0.48\linewidth]{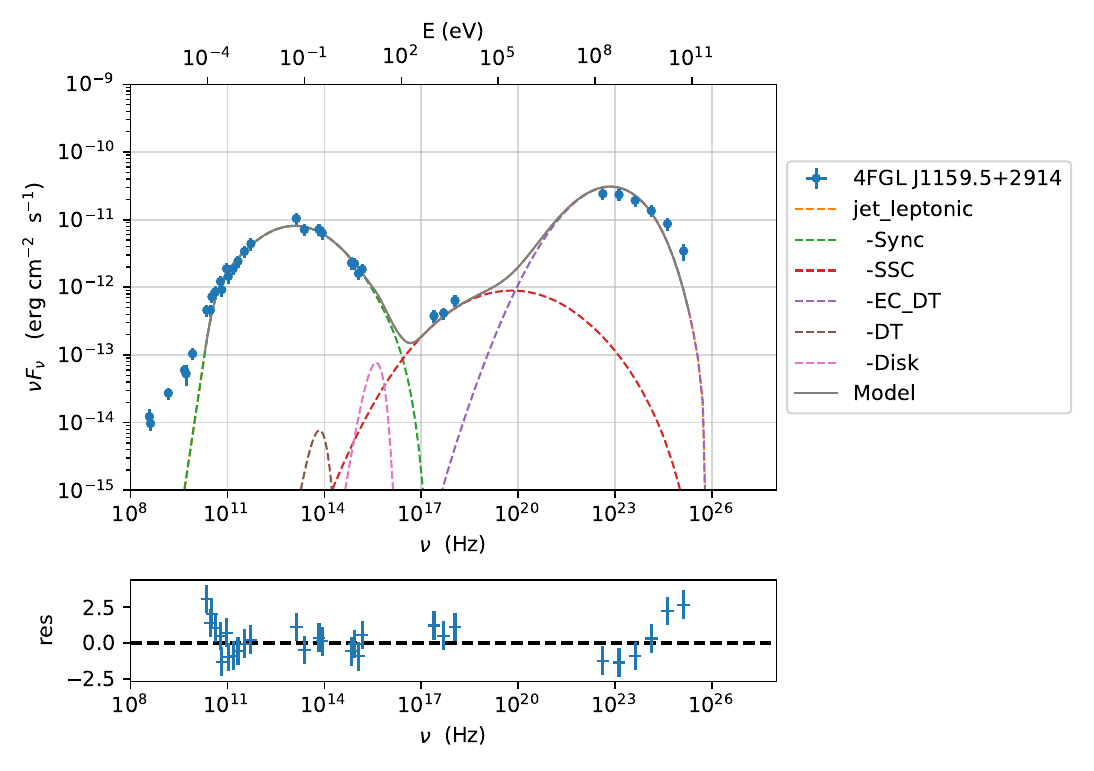}
\includegraphics[width=0.48\linewidth]{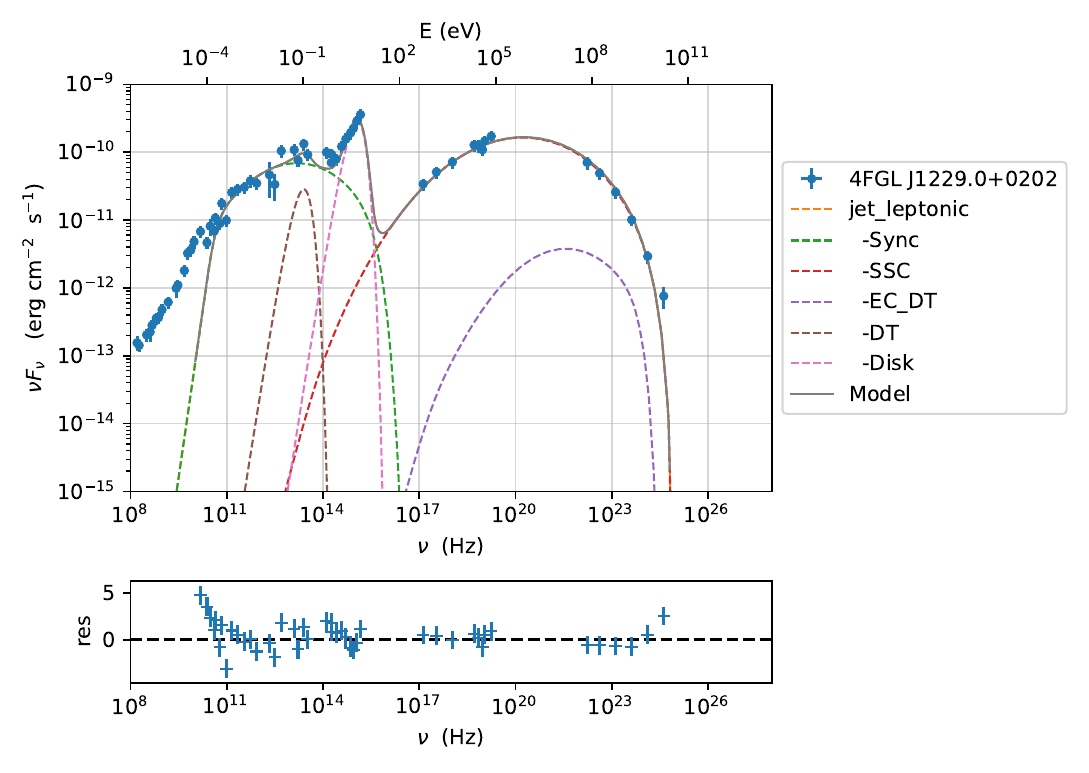}
\includegraphics[width=0.48\linewidth]{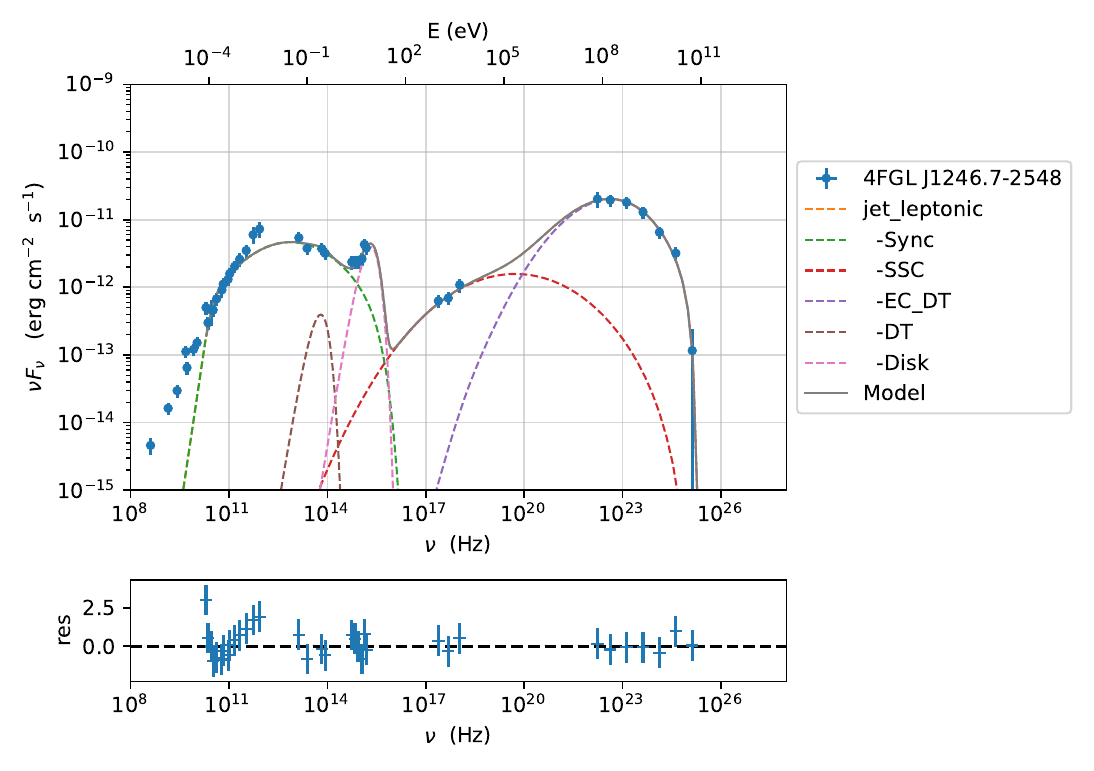}
\center{Figure \ref{fig:SED-EC} --- continued.}
\end{figure*}

\begin{figure*}
\centering
\includegraphics[width=0.48\linewidth]{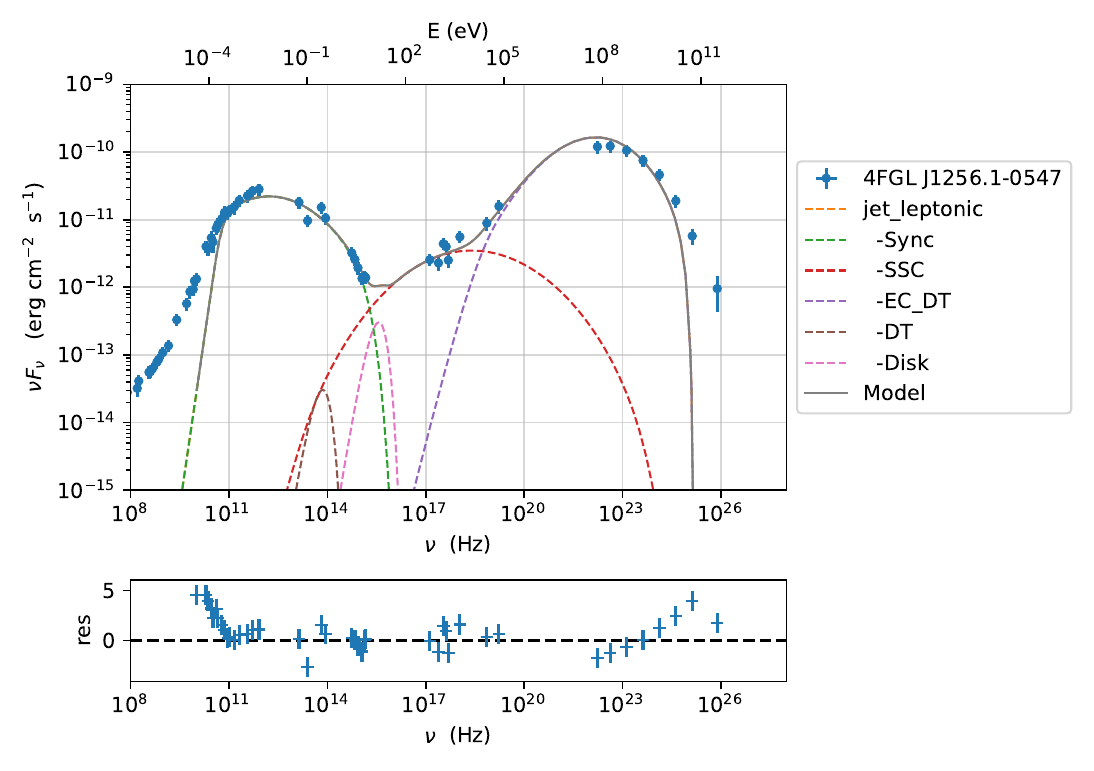}
\includegraphics[width=0.48\linewidth]{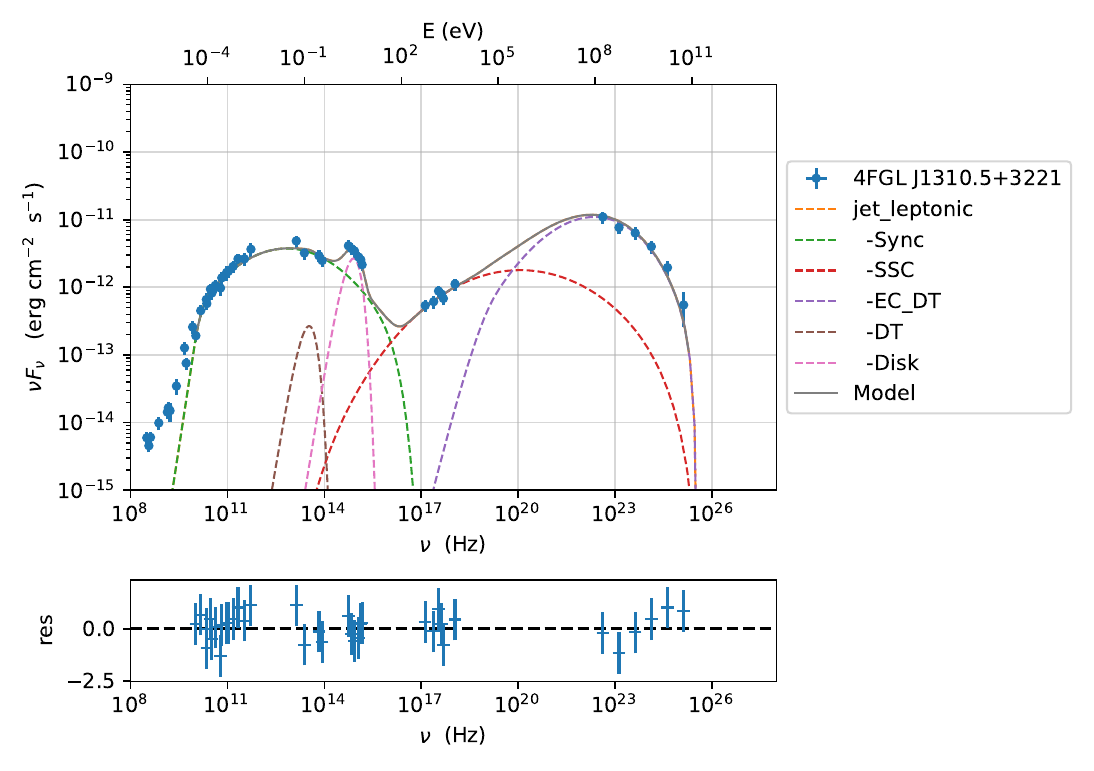}
\includegraphics[width=0.48\linewidth]{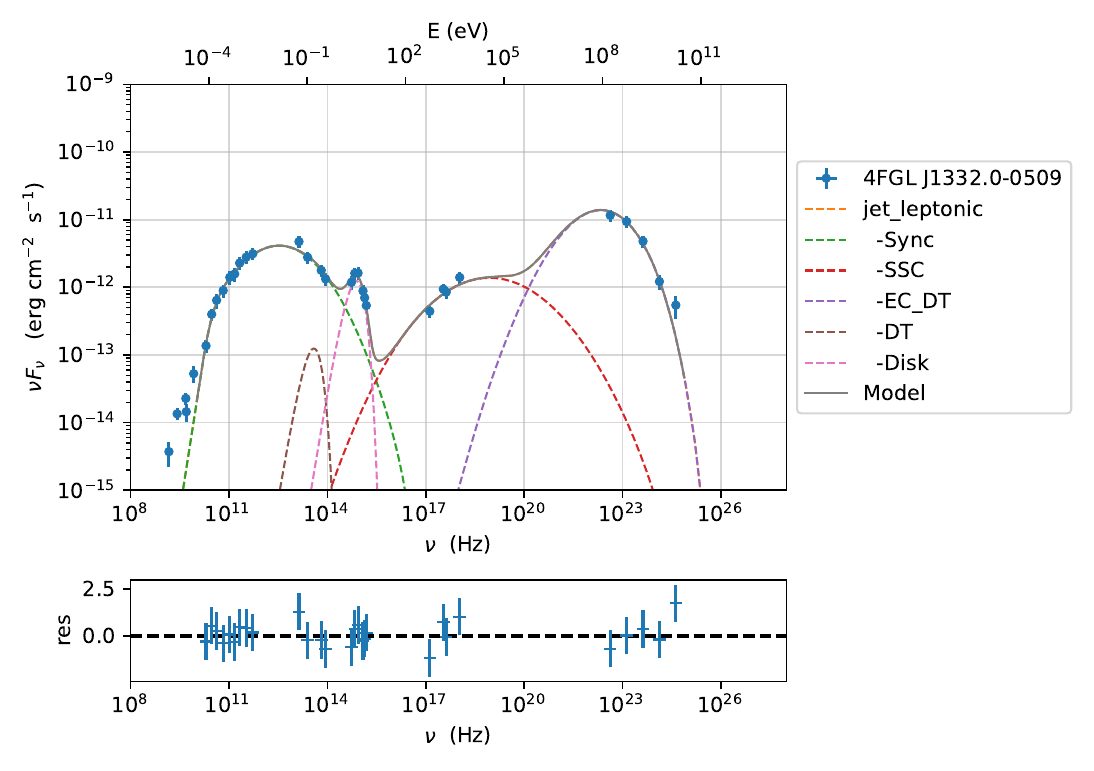}
\includegraphics[width=0.48\linewidth]{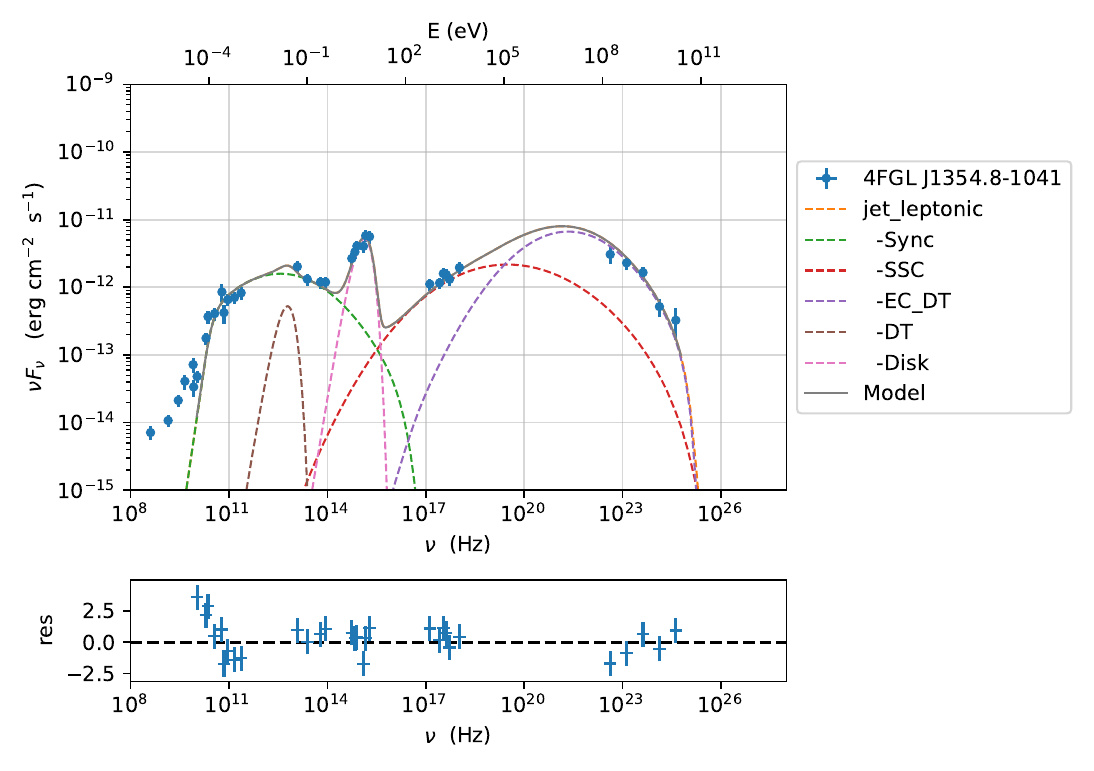}
\includegraphics[width=0.48\linewidth]{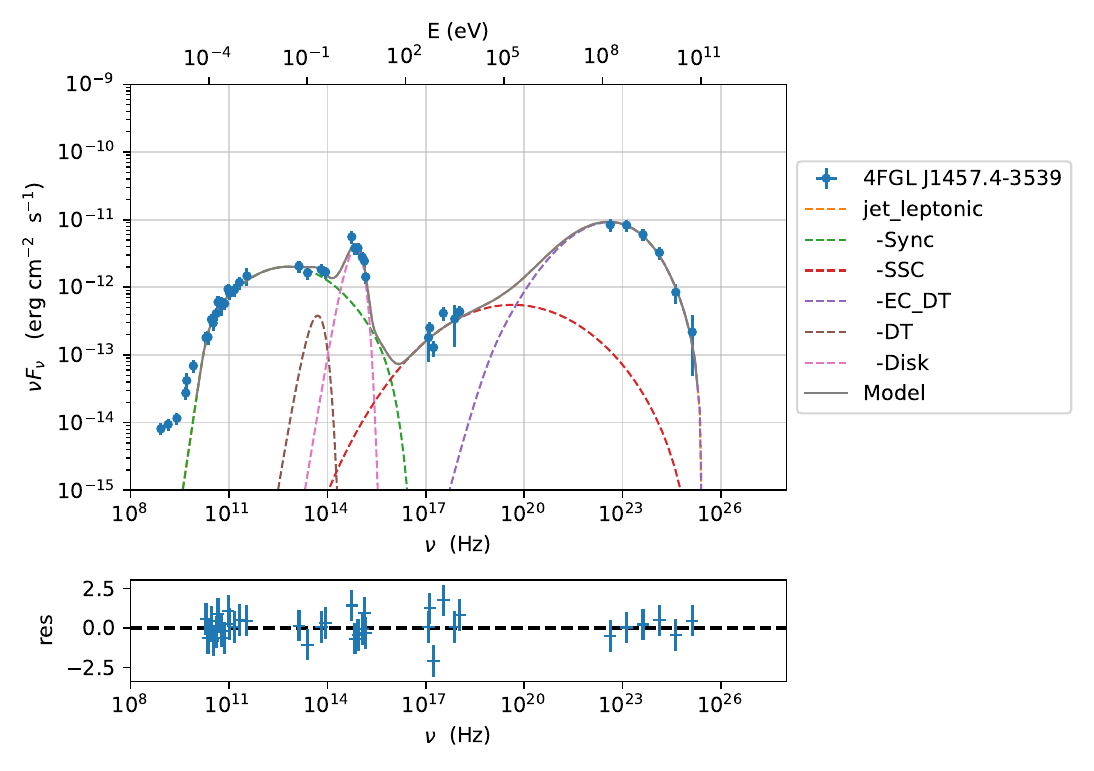}
\includegraphics[width=0.48\linewidth]{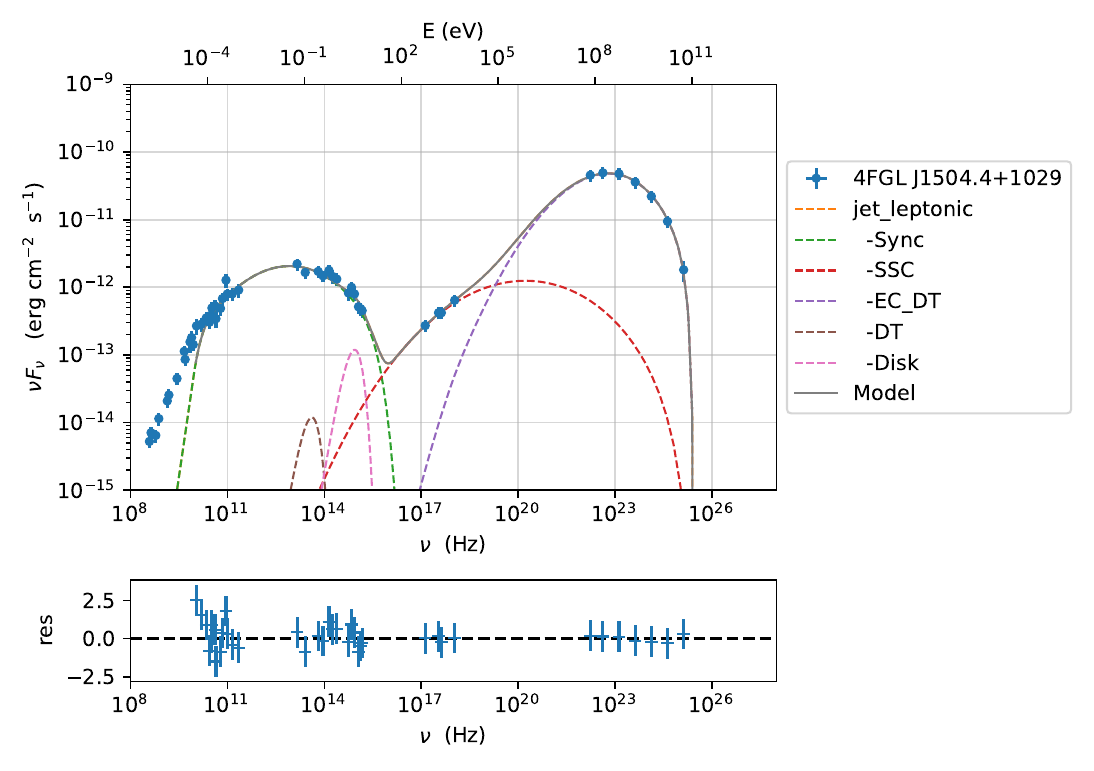}
\includegraphics[width=0.48\linewidth]{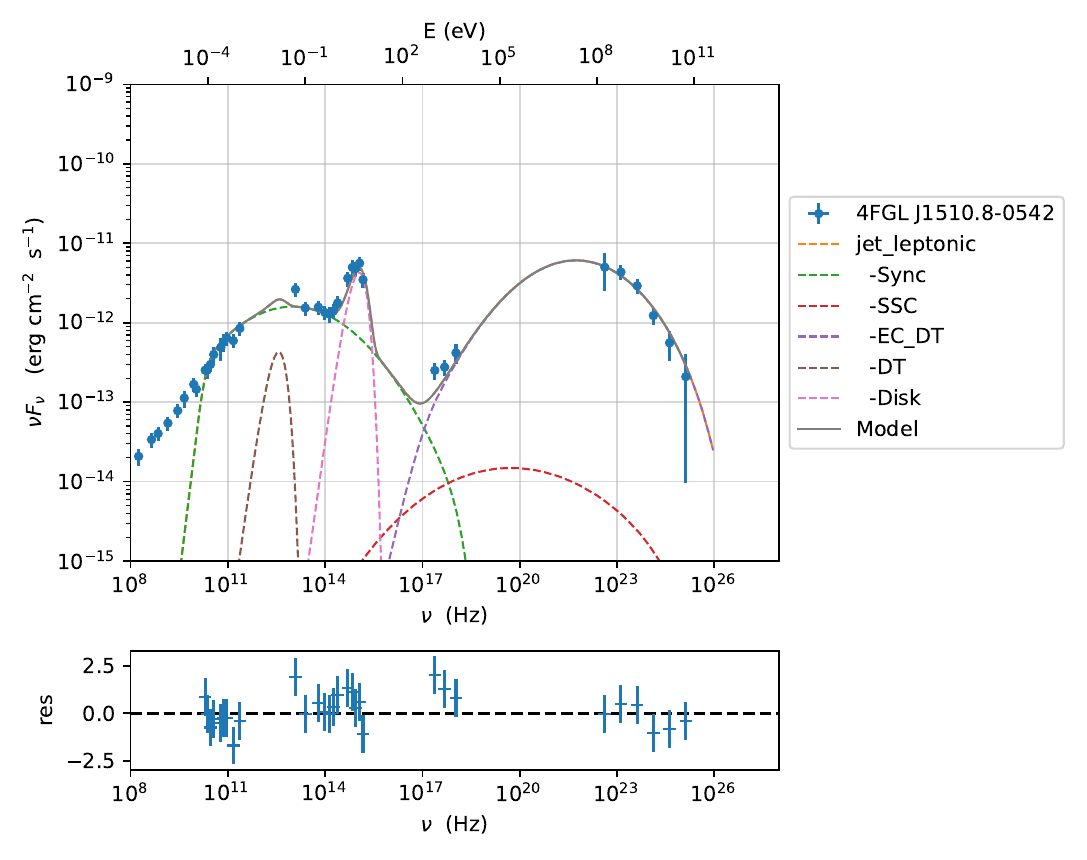}
\includegraphics[width=0.48\linewidth]{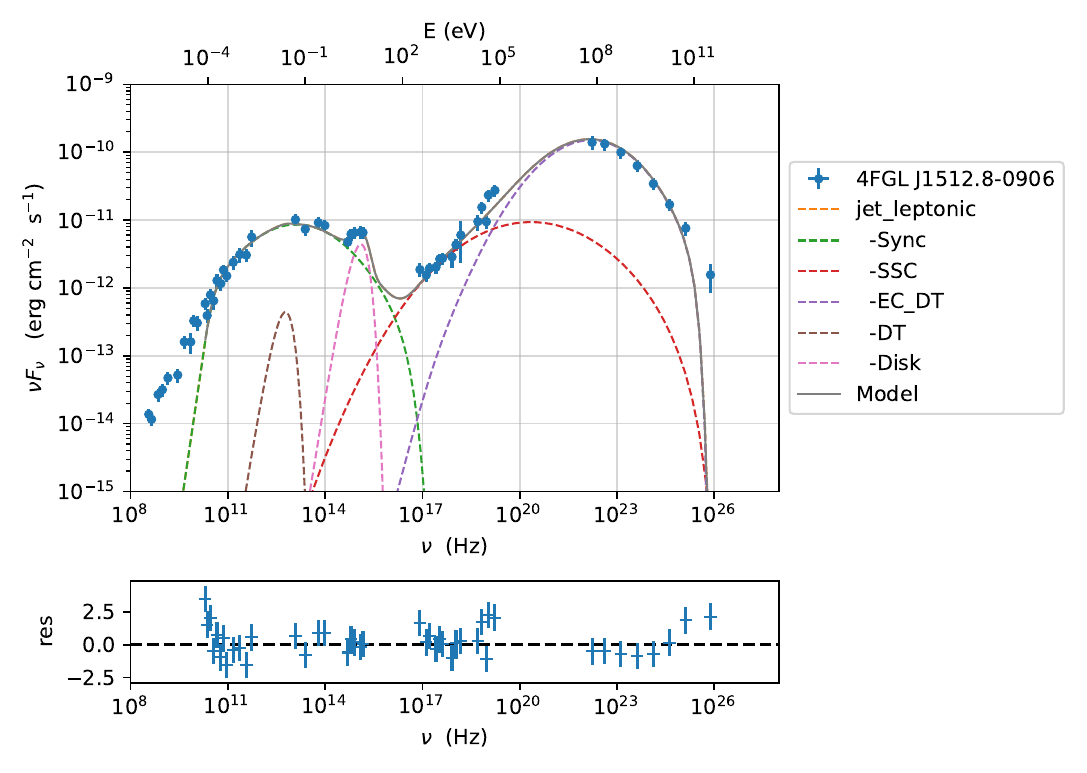}
\center{Figure \ref{fig:SED-EC} --- continued.}
\end{figure*}

\begin{figure*}
\centering
\includegraphics[width=0.48\linewidth]{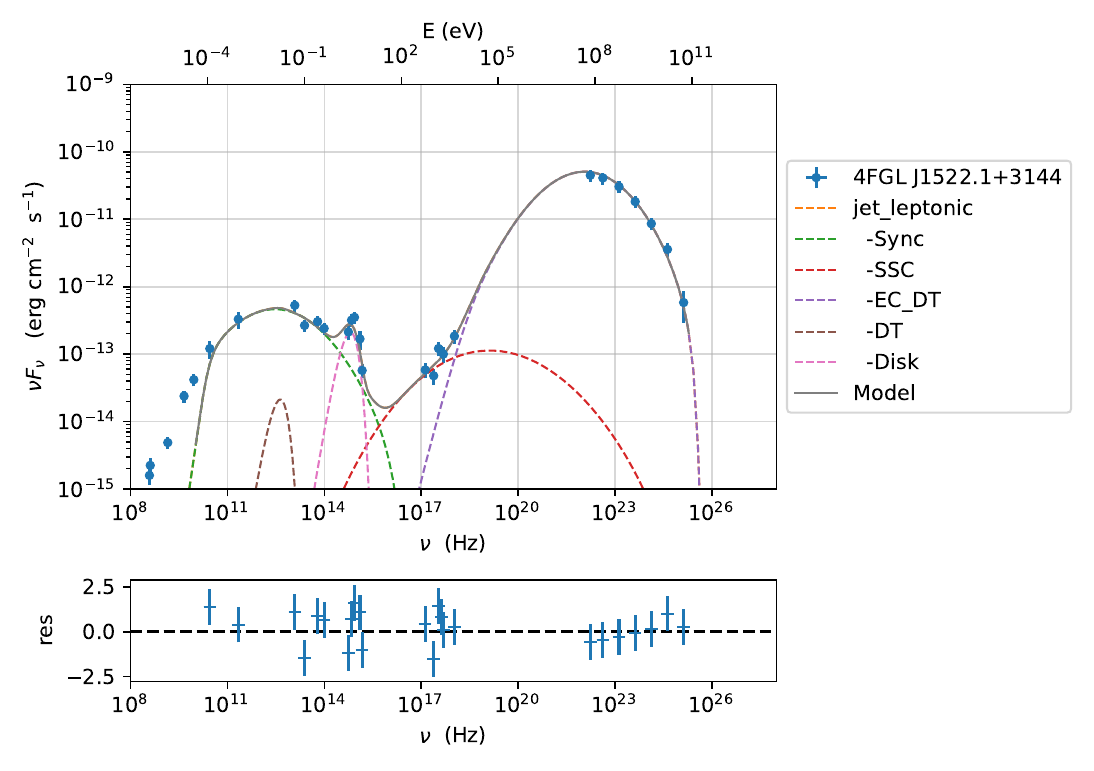}
\includegraphics[width=0.48\linewidth]{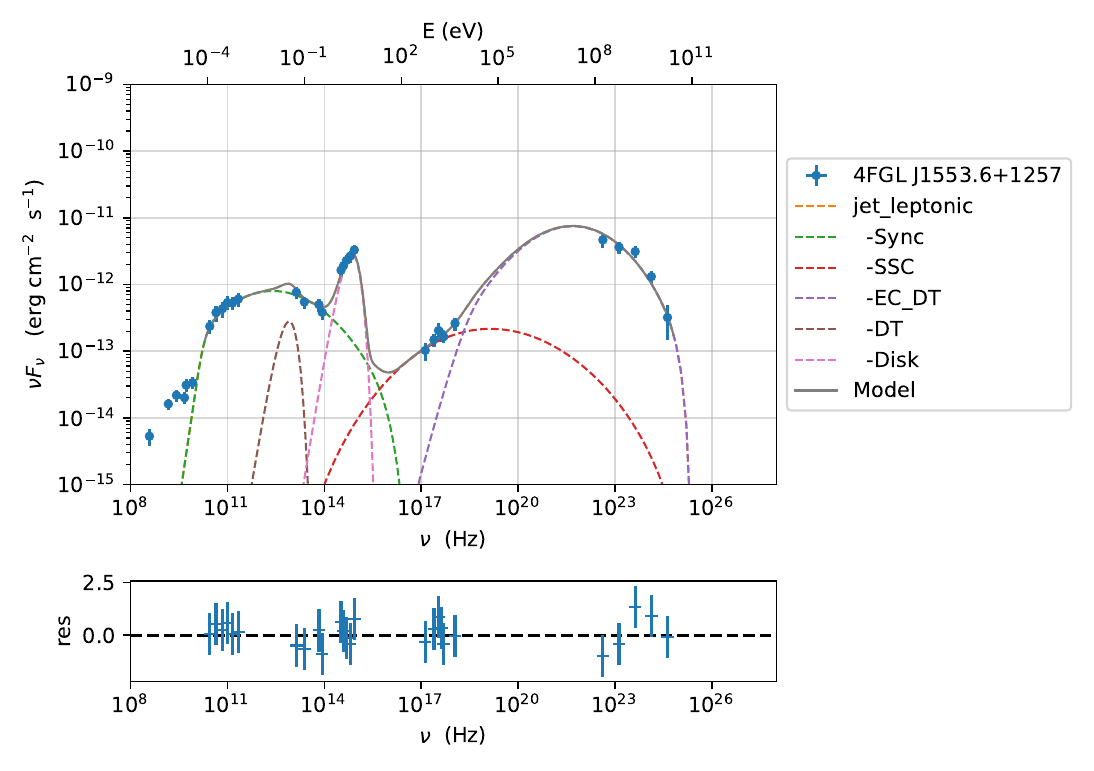}
\includegraphics[width=0.48\linewidth]{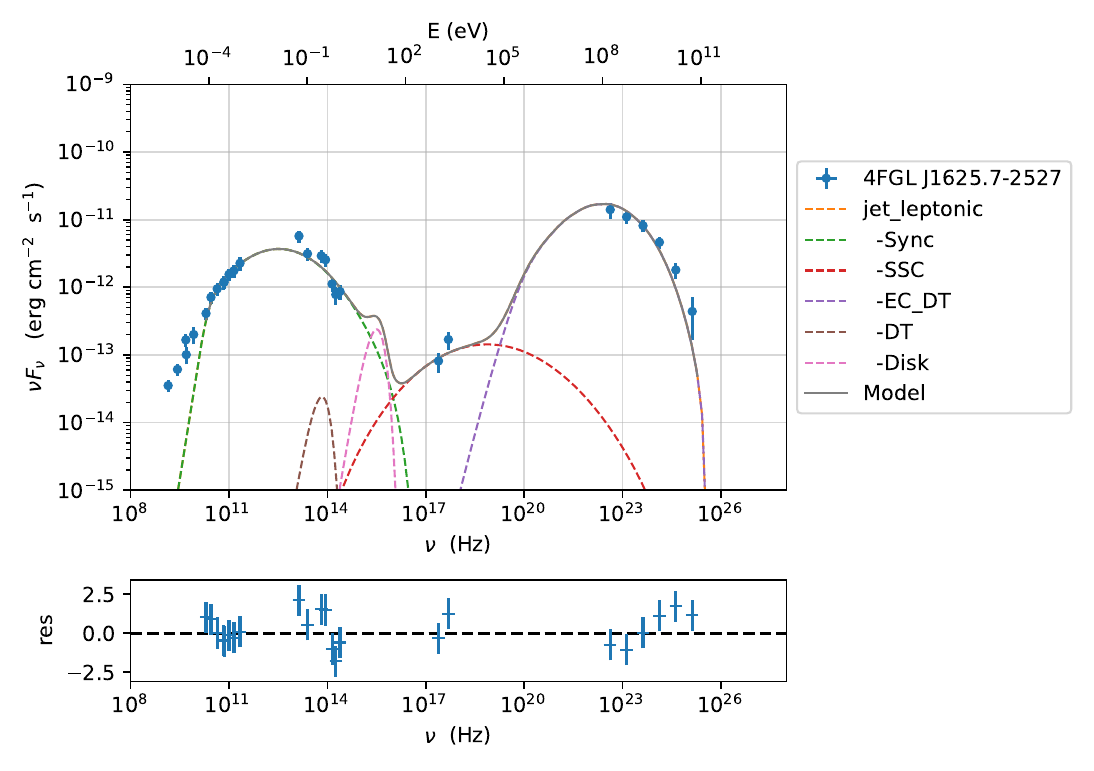}
\includegraphics[width=0.48\linewidth]{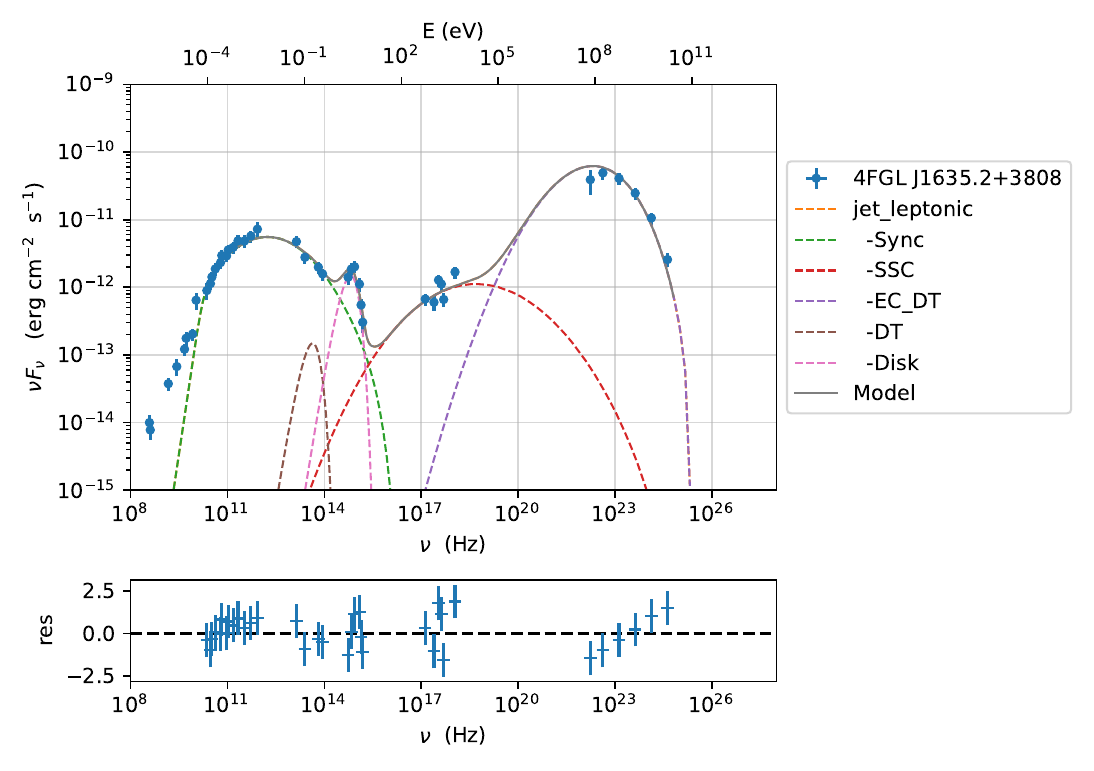}
\includegraphics[width=0.48\linewidth]{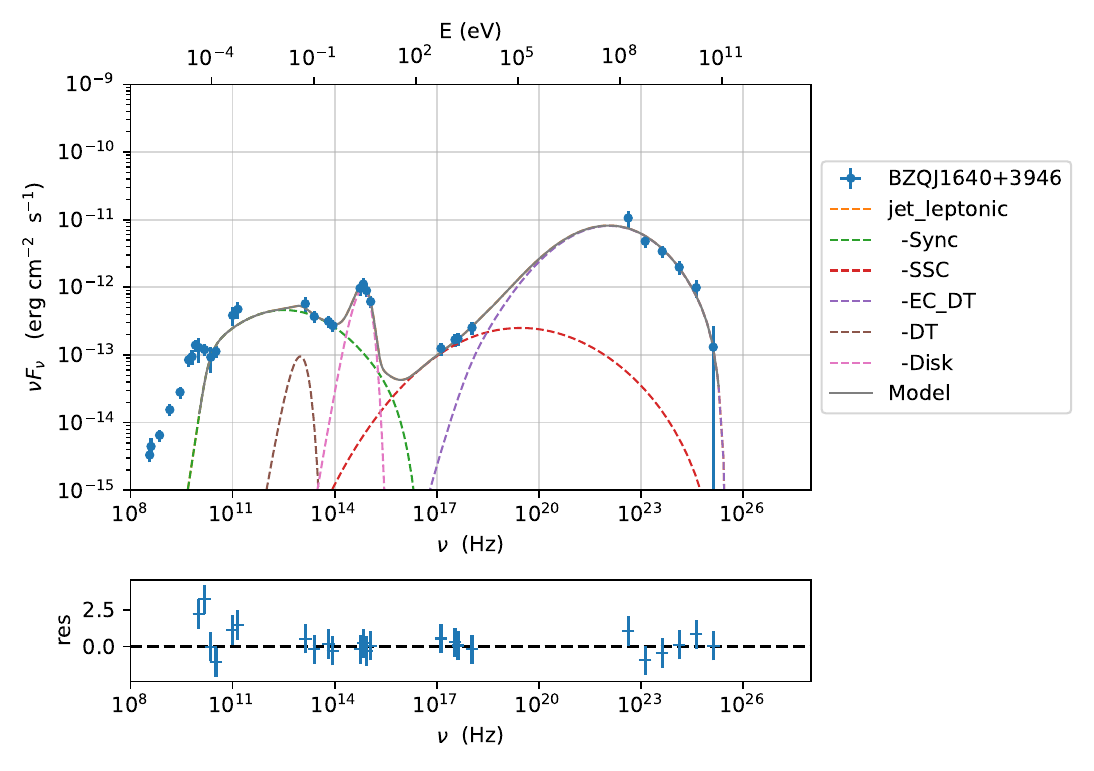}
\includegraphics[width=0.48\linewidth]{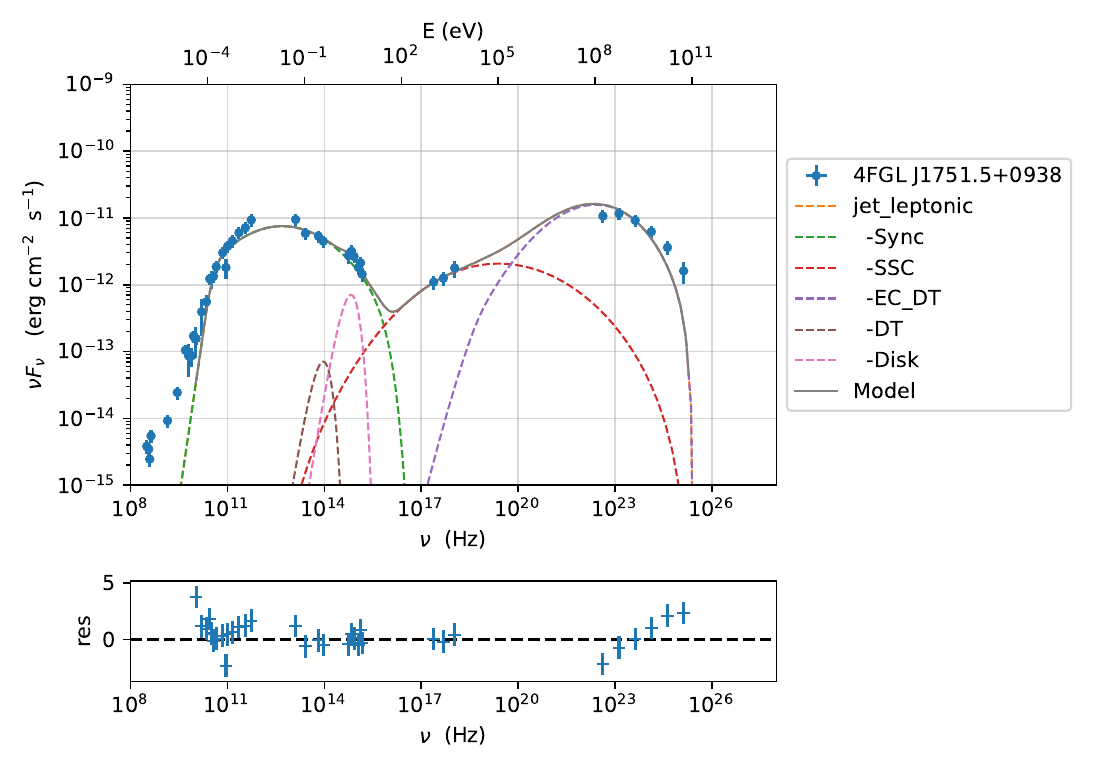}
\includegraphics[width=0.48\linewidth]{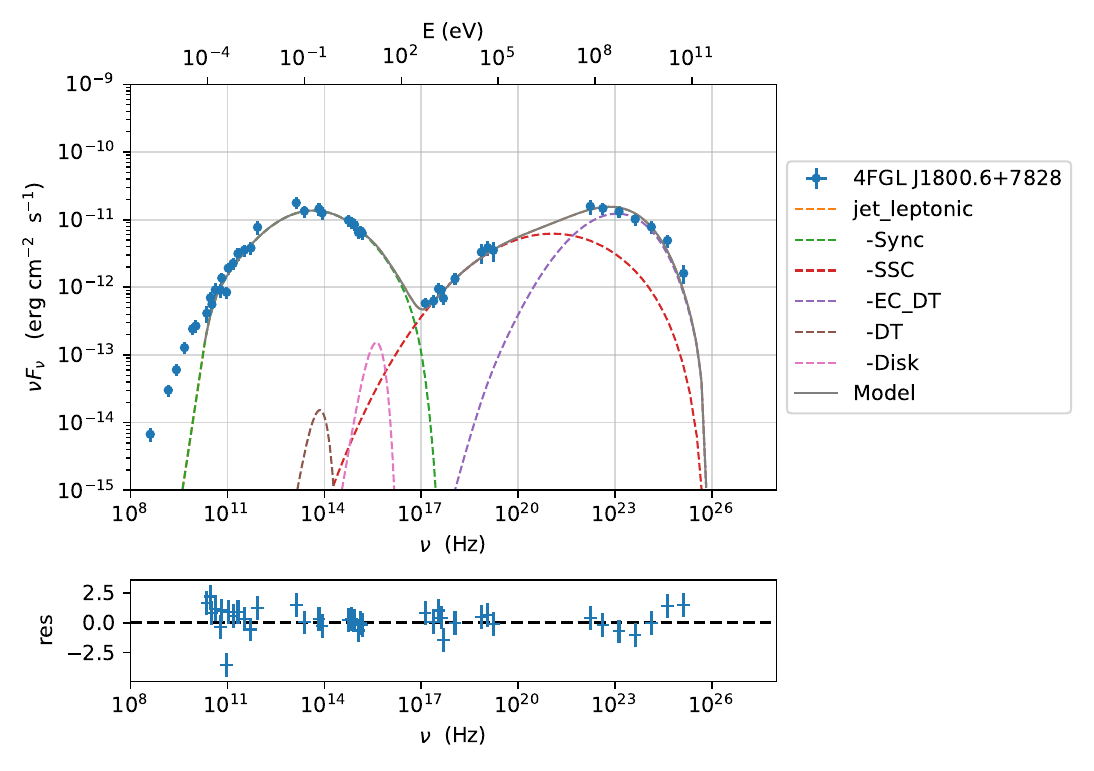}
\includegraphics[width=0.48\linewidth]{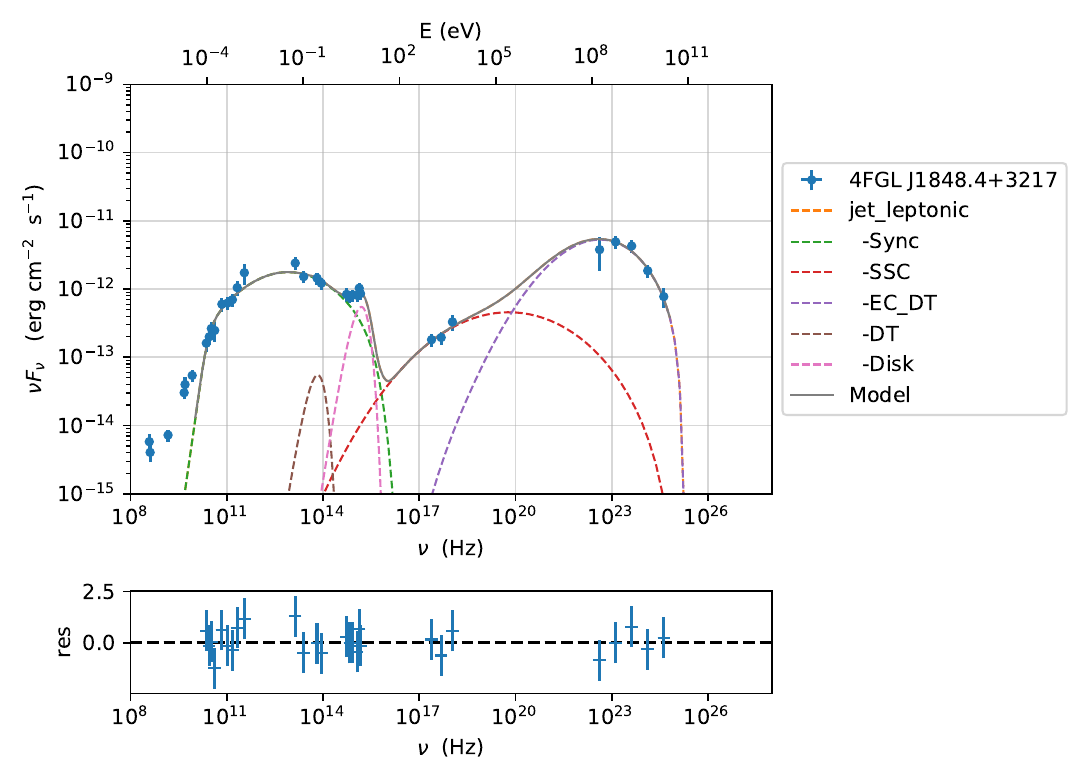}
\center{Figure \ref{fig:SED-EC} --- continued.}
\end{figure*}

\begin{figure*}
\centering
\includegraphics[width=0.48\linewidth]{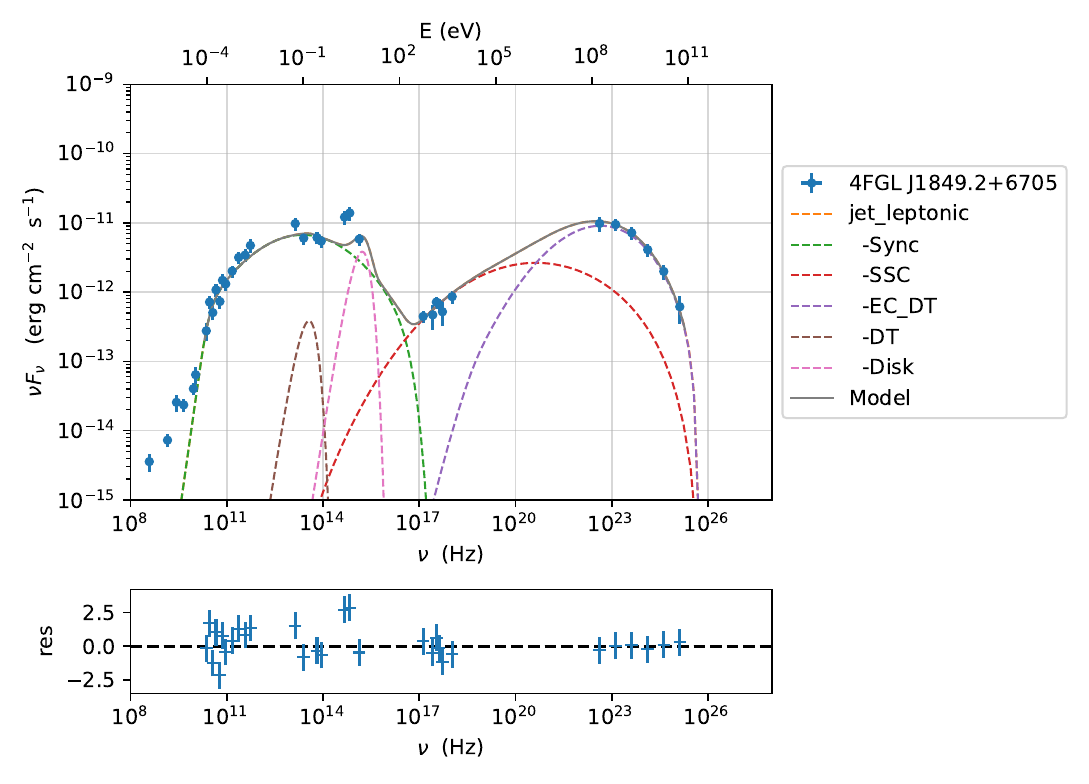}
\includegraphics[width=0.48\linewidth]{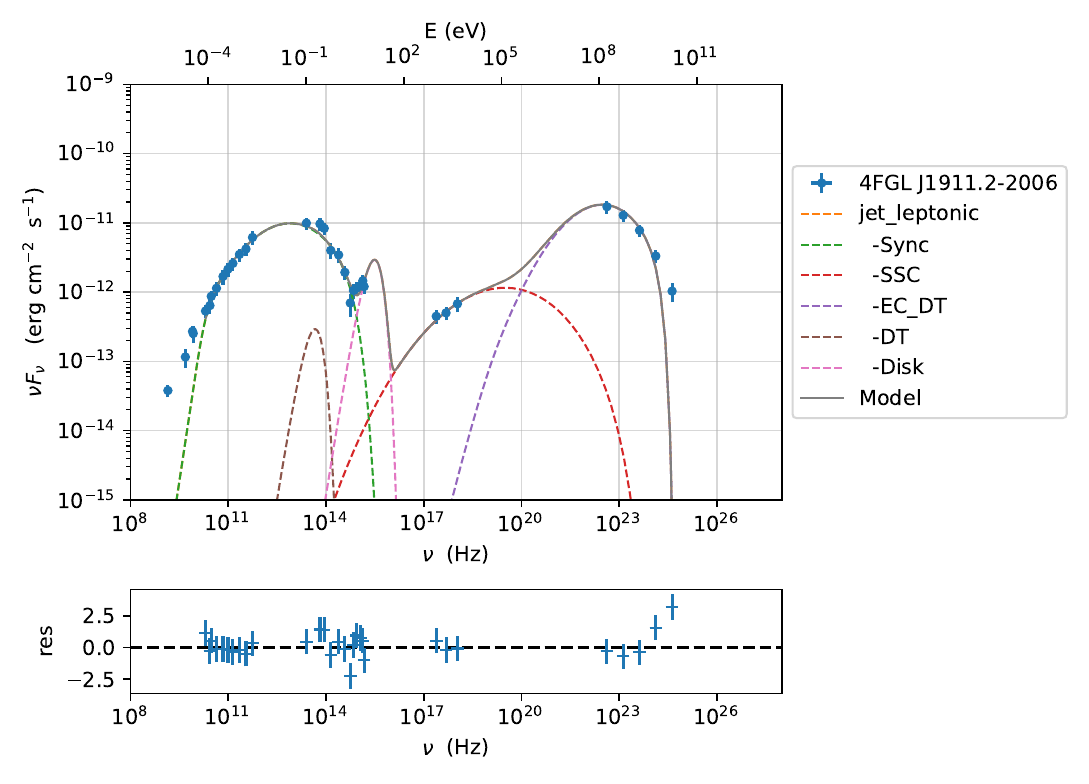}
\includegraphics[width=0.48\linewidth]{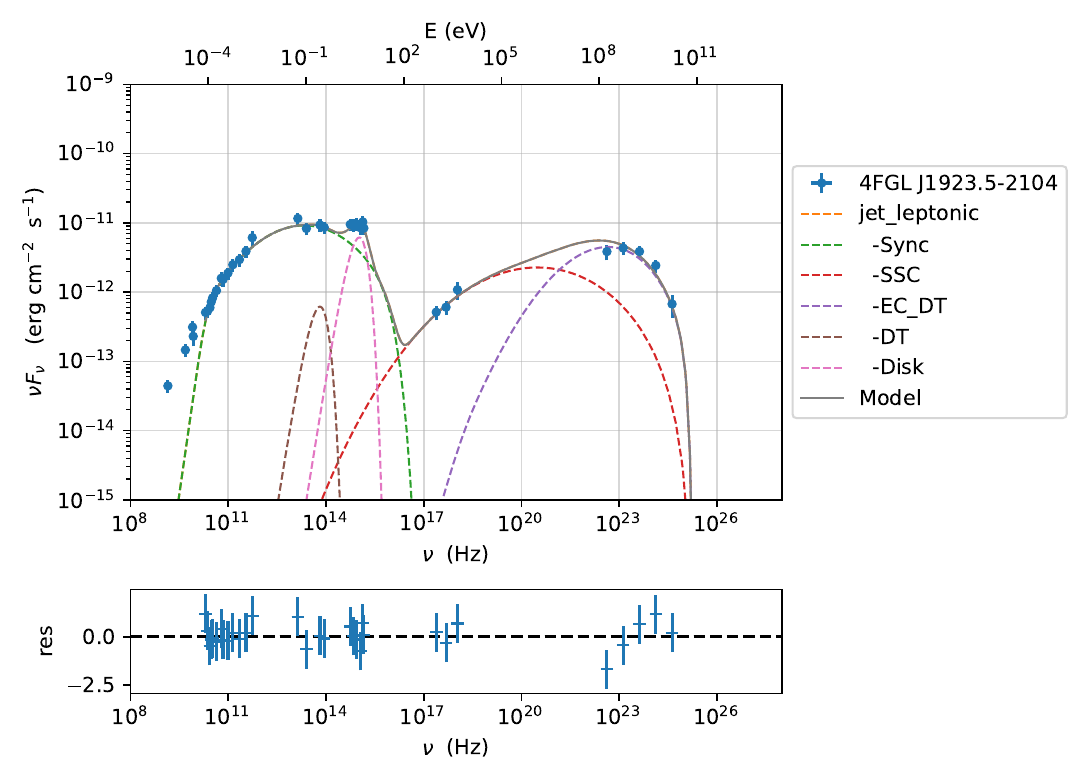}
\includegraphics[width=0.48\linewidth]{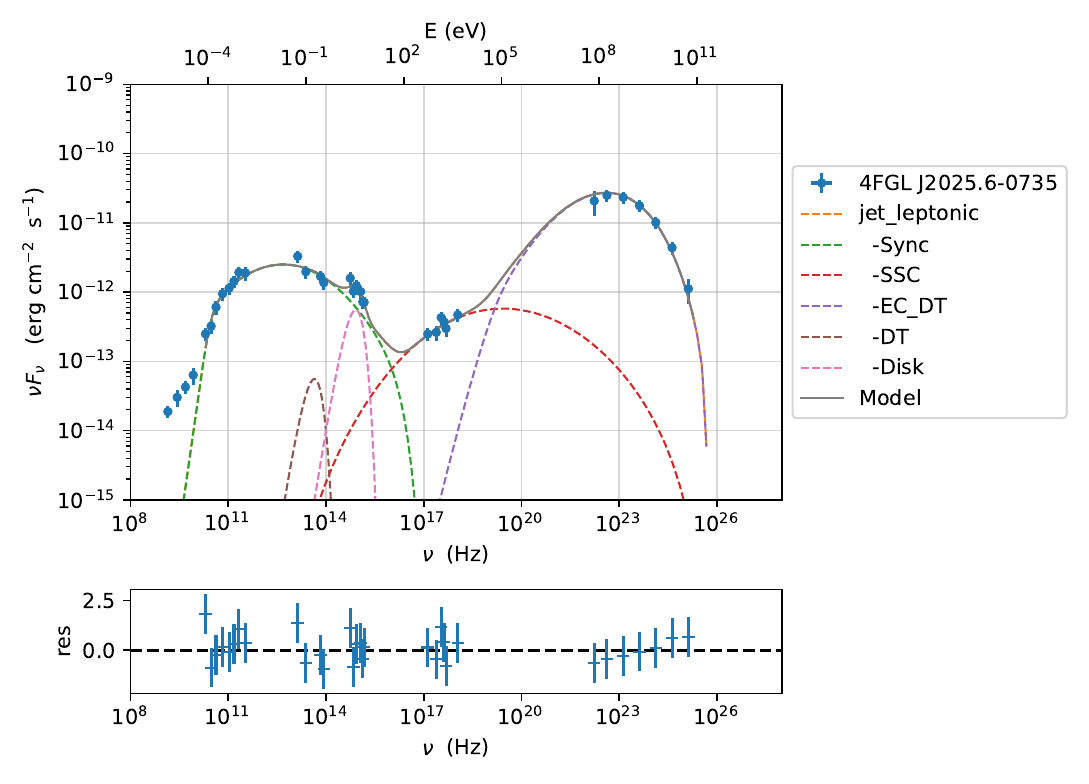}
\includegraphics[width=0.48\linewidth]{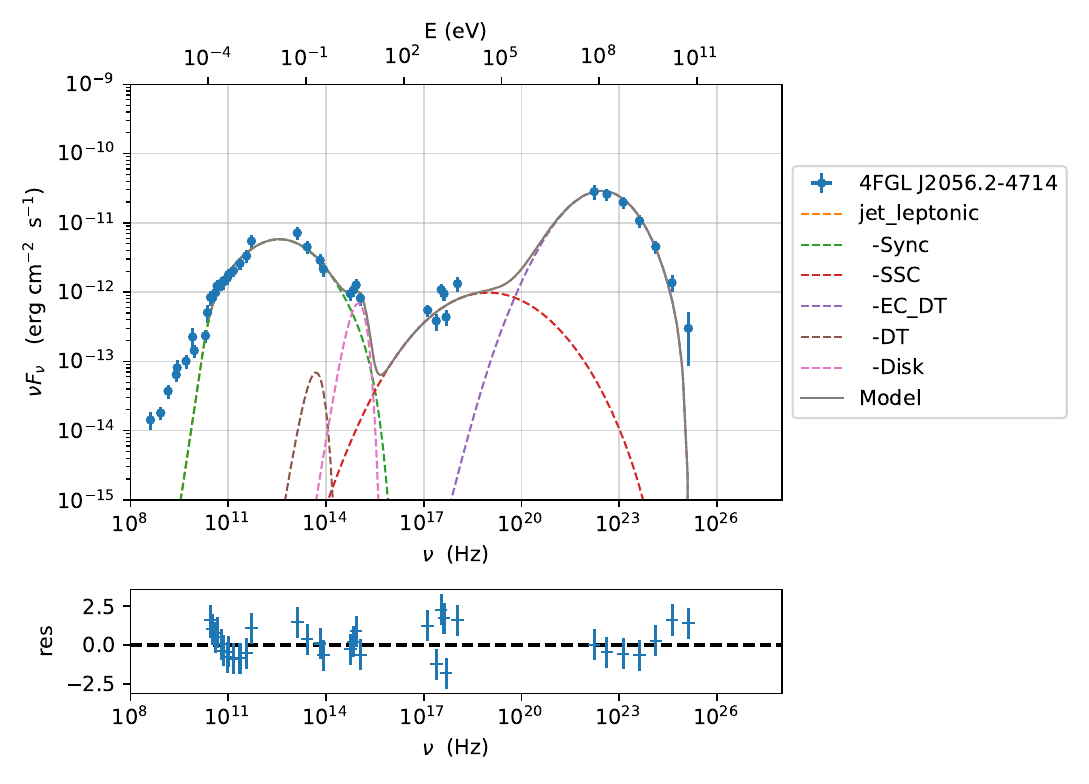}
\includegraphics[width=0.48\linewidth]{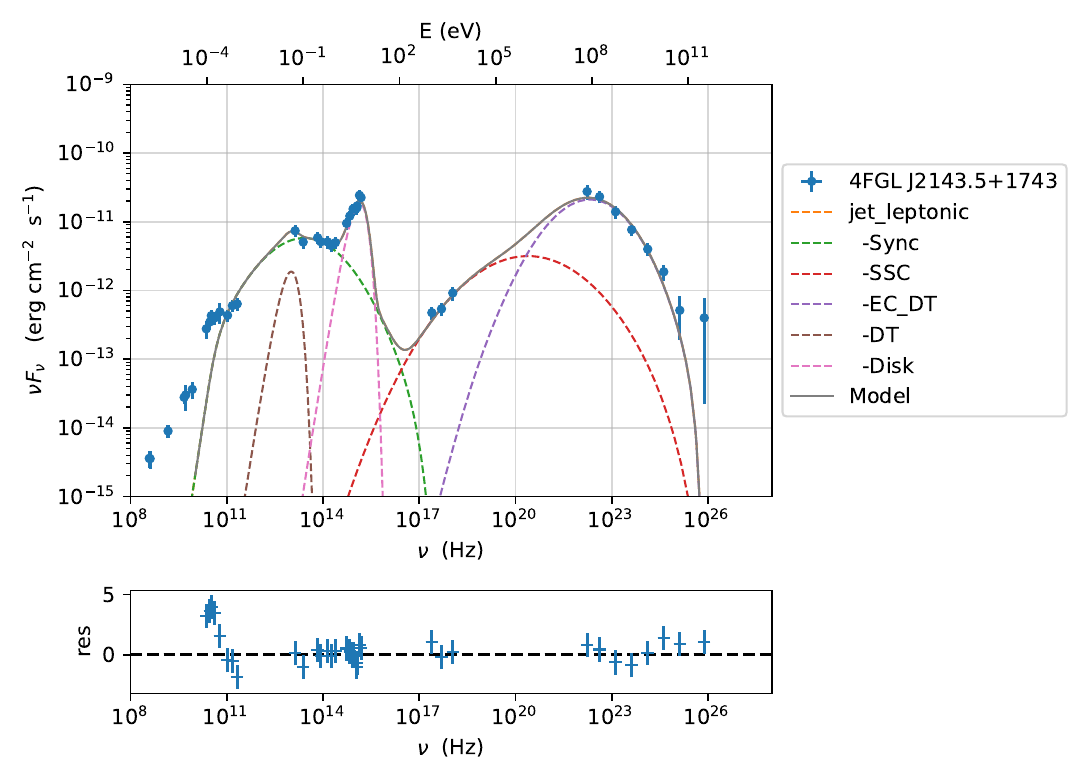}
\includegraphics[width=0.48\linewidth]{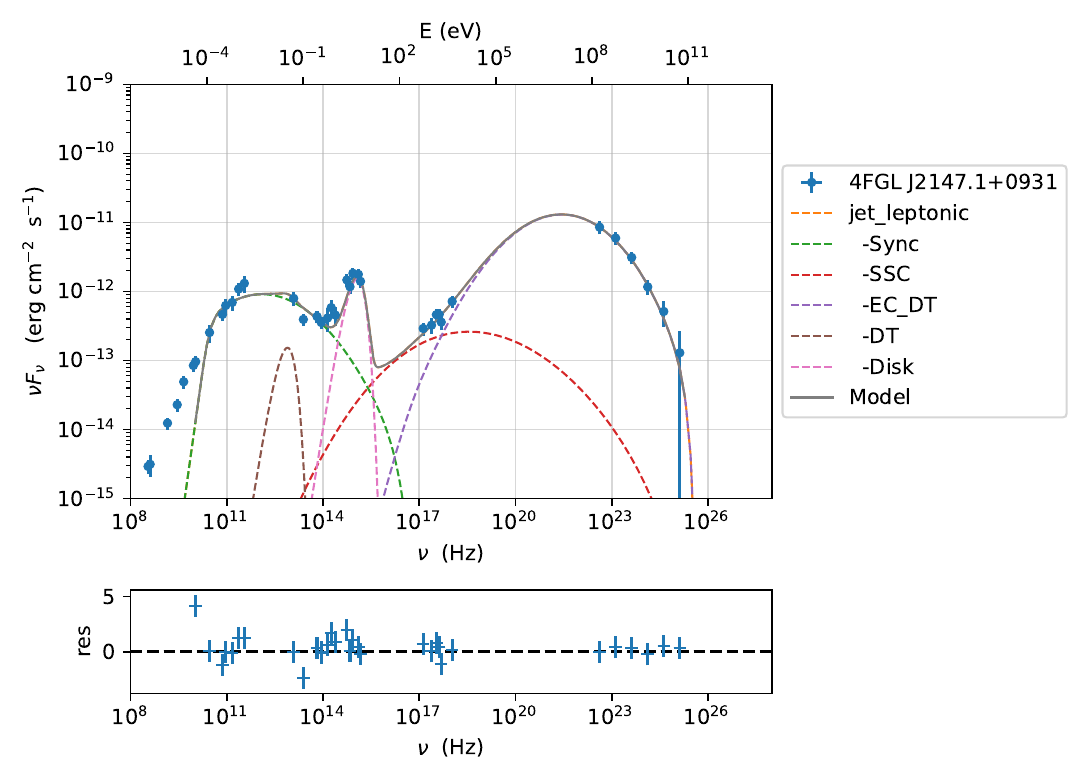}
\includegraphics[width=0.48\linewidth]{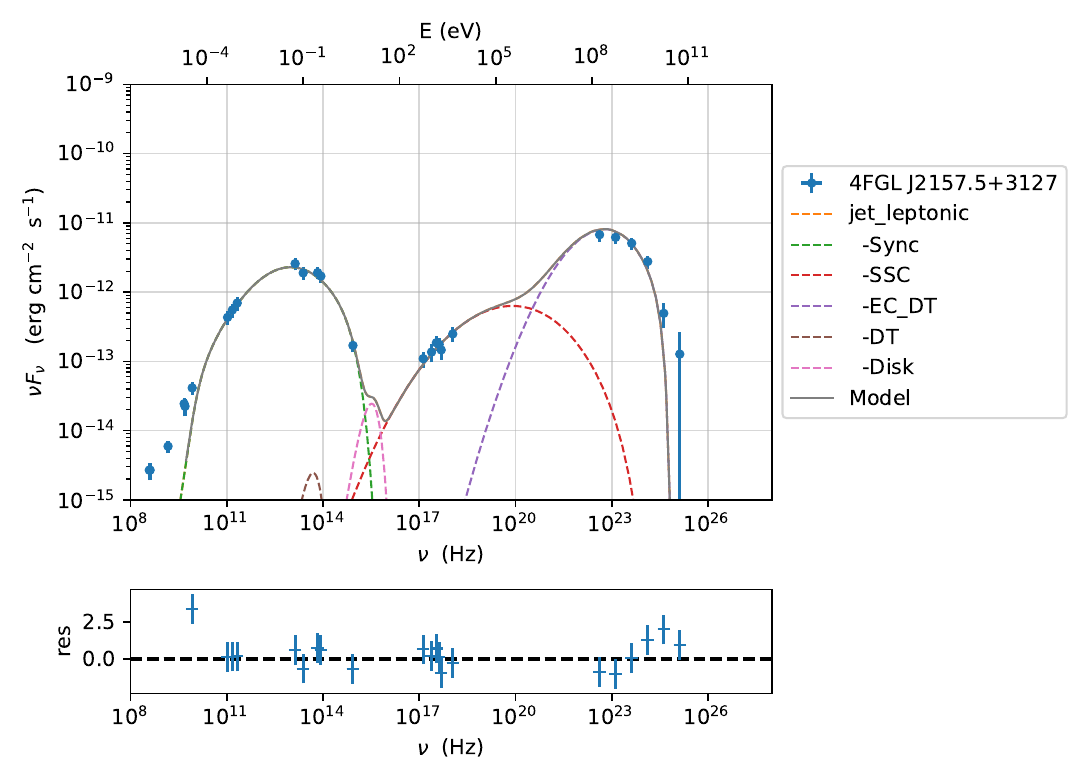}
\center{Figure \ref{fig:SED-EC} --- continued.}
\end{figure*}

\begin{figure*}
\centering
\includegraphics[width=0.48\linewidth]{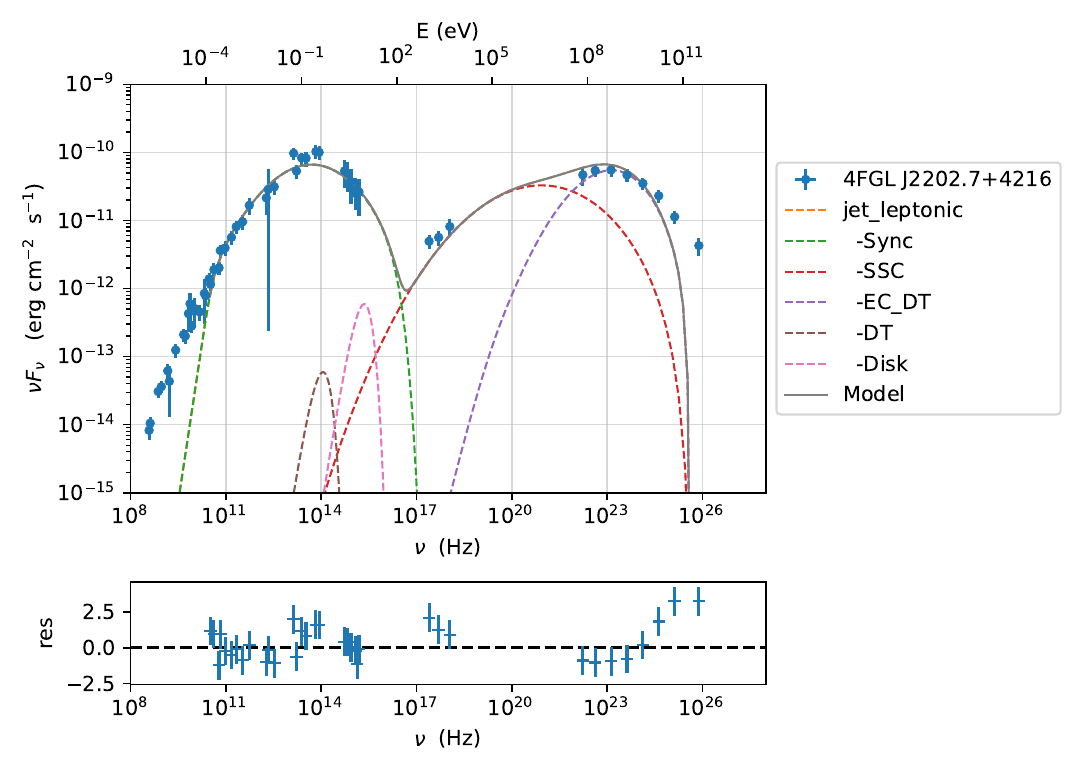}
\includegraphics[width=0.48\linewidth]{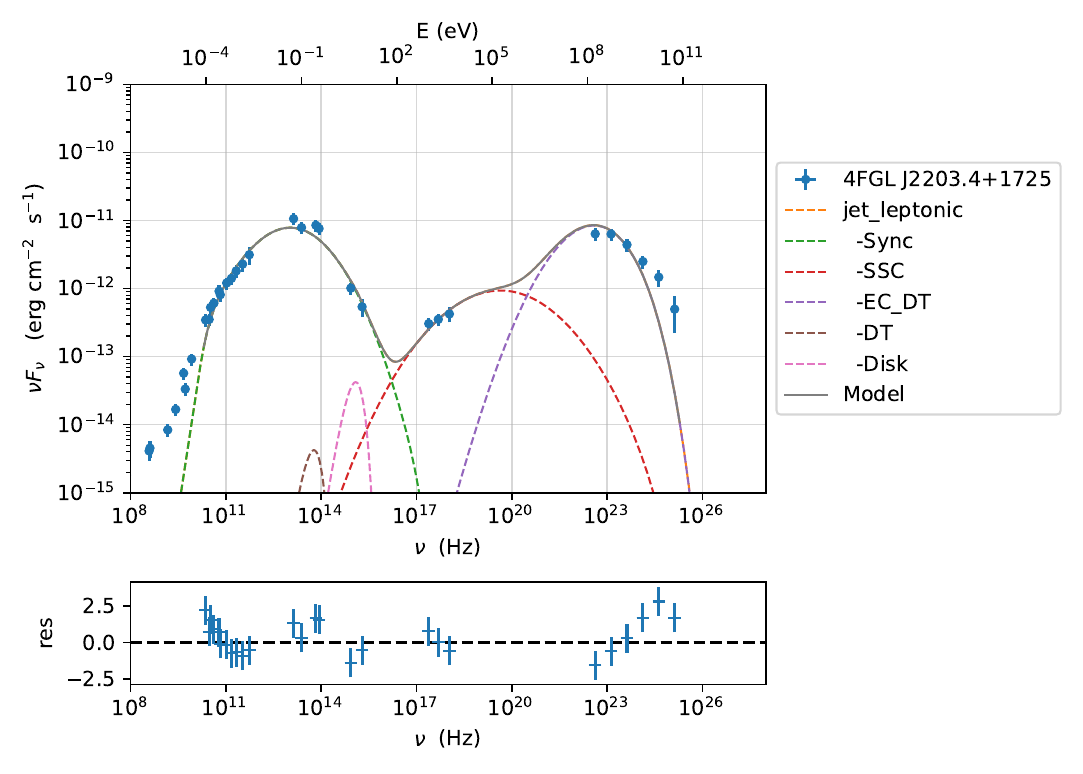}
\includegraphics[width=0.48\linewidth]{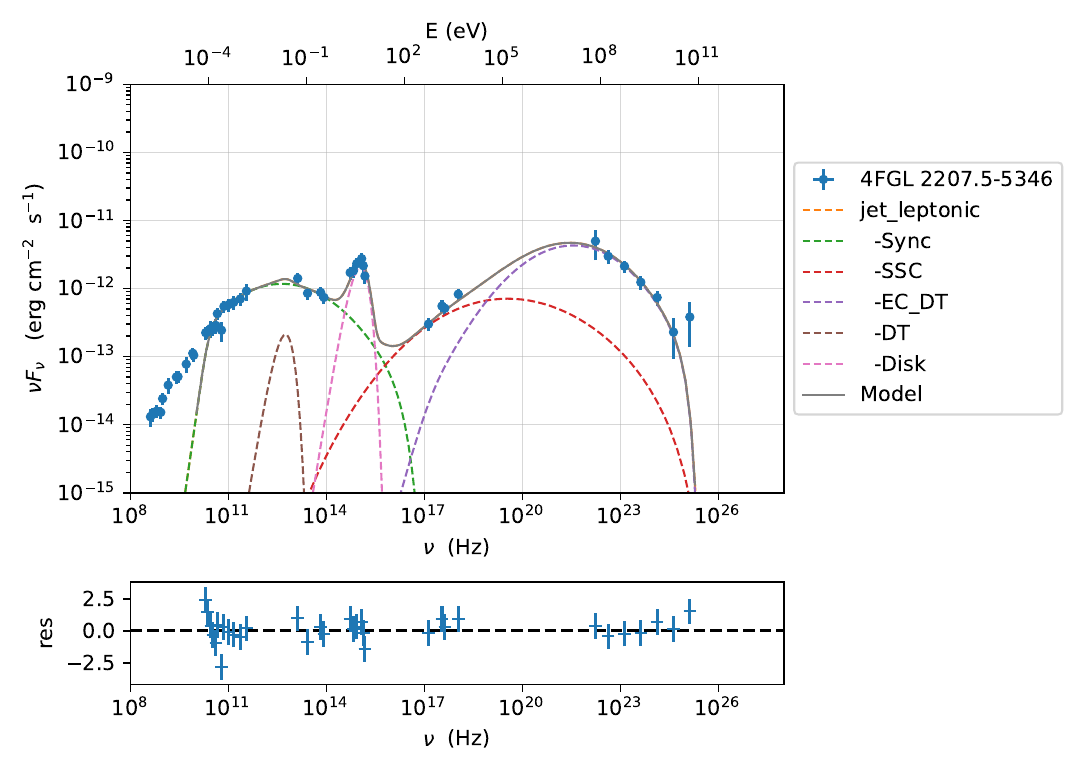}
\includegraphics[width=0.48\linewidth]{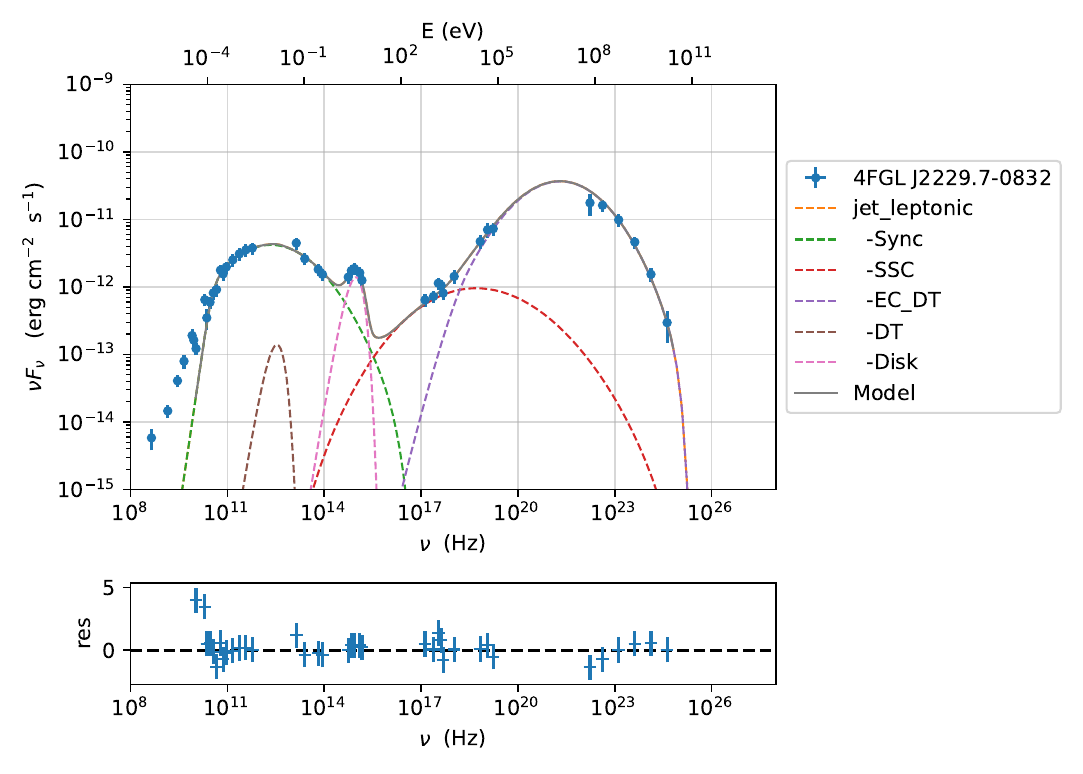}
\includegraphics[width=0.48\linewidth]{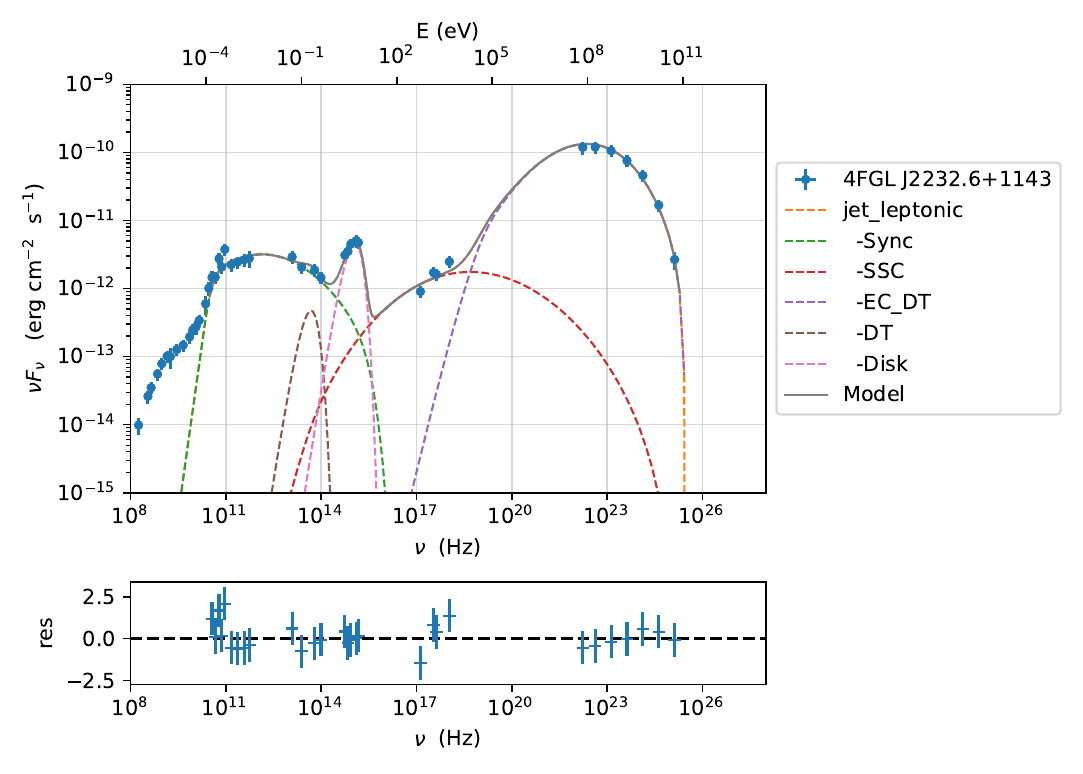}
\includegraphics[width=0.48\linewidth]{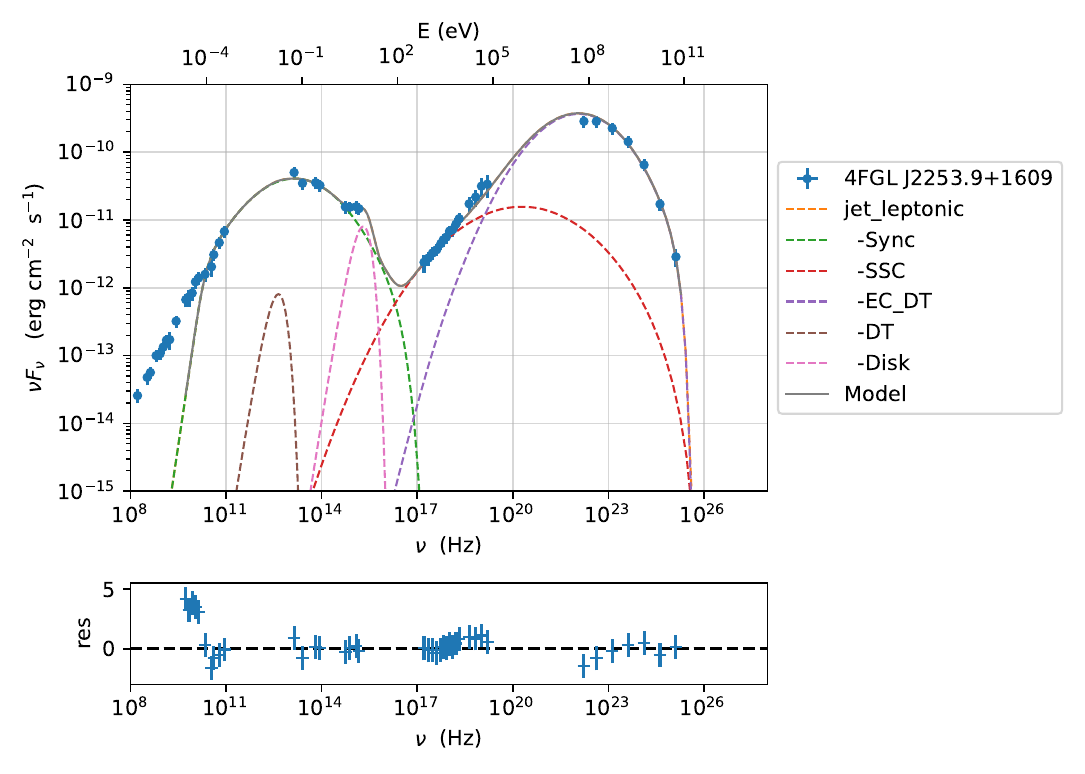}
\includegraphics[width=0.48\linewidth]{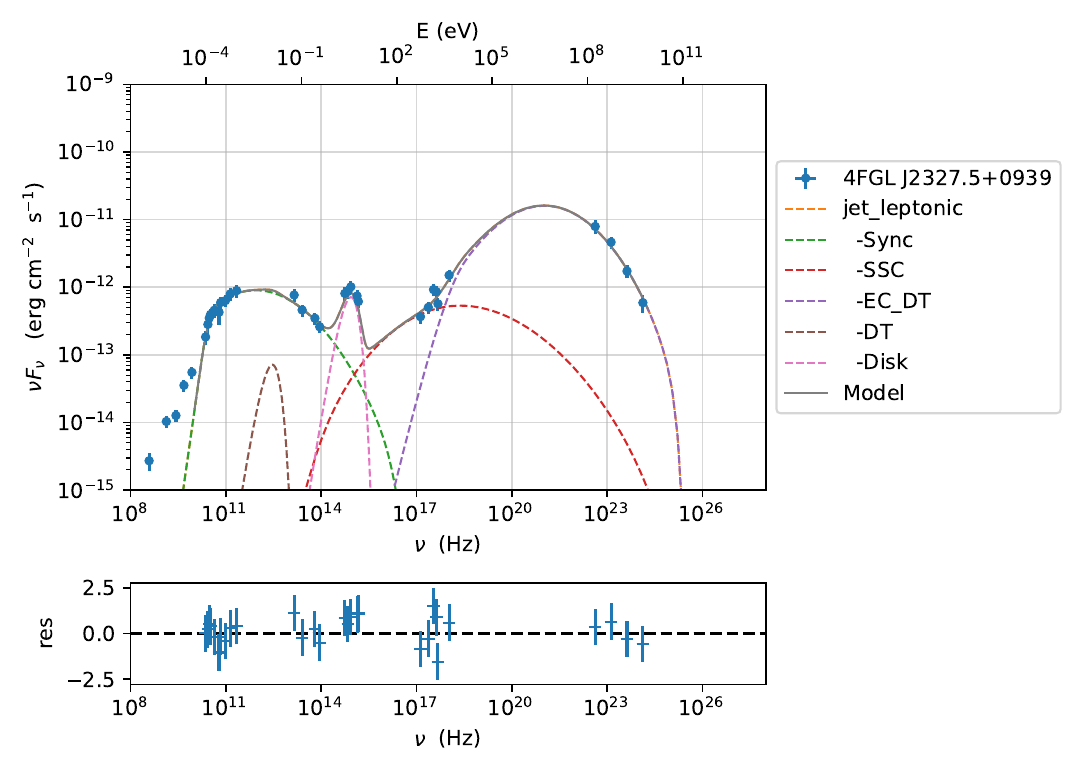}
\includegraphics[width=0.48\linewidth]{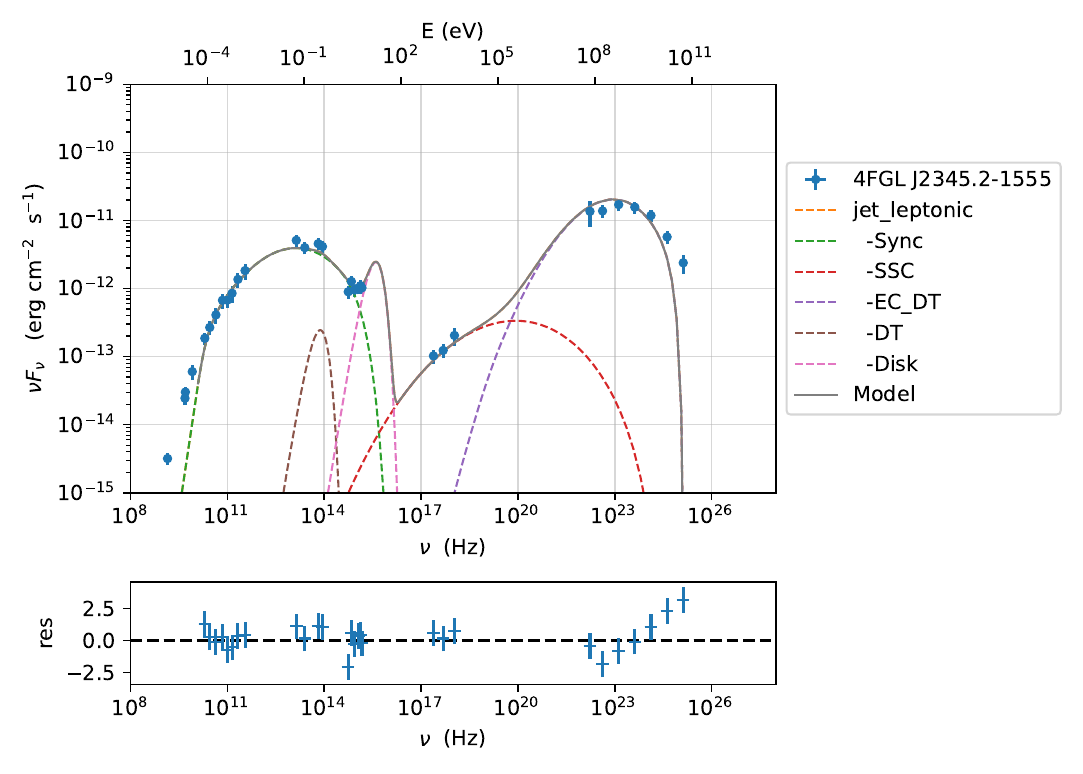}
\center{Figure \ref{fig:SED-EC} --- continued.}
\end{figure*}


\end{document}